\documentclass[12pt,a4paper]{article}

\pdfoutput=1

\usepackage{amsmath,amssymb,bm}
\usepackage{booktabs}
\usepackage[a4paper,includehead,includefoot,ignorehead,left=3cm,right=3cm,
  top=2.4cm,bottom=2.4cm,marginparwidth=2.2cm,marginparsep=0.5cm]{geometry}
\usepackage{graphicx}
\usepackage[numbers,sort&compress]{natbib}
\usepackage[onehalfspacing]{setspace}
\usepackage[bookmarks=true,linktocpage]{hyperref}
\hypersetup{pdftitle={CSE6SSM},
            pdfauthor={P. Athron, D. Harries, R. Nevzorov, A. G. Williams},
            pdfkeywords={supersymmetry,E6,dark matter}}

\newcommand{\FS}{FlexibleSUSY}
\newcommand{\FSv}{FlexibleSUSY-1.1.0}  
\newcommand{\SARAH}{SARAH}
\newcommand{\SARAHv}{SARAH-4.5.6}  
\newcommand{\SUSYHD}{SUSYHD}
\newcommand{\SUSYHDv}{SUSYHD-1.0.2}
\newcommand{\MICROMEGAS}{micrOMEGAs}
\newcommand{\MICROMEGASv}{micrOMEGAs-4.1.8}
\newcommand{\CALCHEP}{CalcHEP}
\newcommand{\checkmate}{Check\textsc{mate}}
\newcommand{\madanalysis}{MadAnalysis}
\newcommand{\smodels}{SModelS}
\newcommand{\fastlim}{Fastlim}
\newcommand{\lux}{LUX}
\newcommand{\xenon}{XENON1T}
\newcommand{\textoverline}[1]{$\overline{\mbox{#1}}$}
\newcommand{\DRbar}{\textoverline{DR}}
\newcommand{\appref}[1]{Appendix~\ref{#1}}
\newcommand{\figref}[1]{\figurename~\ref{#1}}
\newcommand{\secref}[1]{Section~\ref{#1}}
\newcommand{\tabref}[1]{\tablename~\ref{#1}}

\DeclareMathOperator{\Imag}{Im}
\DeclareMathOperator{\Real}{Re}
\DeclareMathOperator{\Tr}{Tr}

\newcommand{\trace}[1]{\Tr \Bigl ( #1 \Bigr )}

\allowdisplaybreaks

\begin{document}

\begin{titlepage}

  \begin{flushright}
    ADP--16--37/T993\\
    CoEPP--MN--16--23
  \end{flushright}

  \begin{center}
    {\Large\textbf{Dark Matter in a Constrained $E_6$ Inspired SUSY Model}}
    \\[8mm]
    P.~Athron$^{a}$\footnote{E-mail: \texttt{peter.athron@coepp.org.au}},
    D.~Harries$^{b}$\footnote{E-mail: \texttt{dylan.harries@adelaide.edu.au};
      ORCID: 0000--0002--2476--6989},
    R.~Nevzorov$^{b}$\footnote{E-mail: \texttt{roman.nevzorov@adelaide.edu.au}}
    \, and
    A.G.~Williams$^{b}$
    \footnote{E-mail: \texttt{anthony.williams@adelaide.edu.au};
      ORCID: 0000--0002--1472--1592}
    \\[3mm]
      {\small\textit{$^a$ ARC Centre of Excellence for Particle Physics at the
          Terascale,}}\\
      {\small\textit{School of Physics, Monash University, Melbourne,
          Victoria 3800, Australia}}\\[3mm]
      {\small\textit{$^b$ ARC Centre of Excellence for Particle Physics at the
          Terascale,}}\\
      {\small\textit{Department of Physics, The University of Adelaide,
          Adelaide, South Australia 5005, Australia}}\\
  \end{center}

  \vspace*{0.75cm}

  \begin{abstract}
    \noindent We investigate dark matter in a constrained $E_6$
    inspired supersymmetric model with an exact custodial symmetry and
    compare with the CMSSM.  The breakdown of $E_6$ leads to an
    additional $U(1)_N$ symmetry and a discrete matter parity.  The
    custodial and matter symmetries imply there are two stable dark
    matter candidates, though one may be extremely light and
    contribute negligibly to the relic density.  We demonstrate that a
    predominantly Higgsino, or mixed bino-Higgsino, neutralino can
    account for all of the relic abundance of dark matter, while
    fitting a $125$ GeV SM-like Higgs and evading LHC limits on new
    states.  However we show that the recent \lux\ 2016 limit on direct
    detection places severe constraints on the mixed bino-Higgsino
    scenarios that explain all of the dark matter.  Nonetheless we
    still reveal interesting scenarios where the gluino, neutralino
    and chargino are light and discoverable at the LHC, but the full
    relic abundance is not accounted for.  At the same time we also
    show that there is a huge volume of parameter space, with a
    predominantly Higgsino dark matter candidate that explains all the
    relic abundance, that will be discoverable with \xenon.  Finally
    we demonstrate that for the $E_6$ inspired model the exotic
    leptoquarks could still be light and within range of future LHC
    searches.
  \end{abstract}

\end{titlepage}

\setcounter{page}{2}
\setcounter{footnote}{0}

\section{Introduction}
A plethora of astrophysical and cosmological observations provide
strong evidence for the presence of non--baryonic, non--luminous
matter, so called dark matter (DM), that constitutes about $25\%$ of
the energy density of the Universe \cite{Agashe:2014kda}.  So far its
microscopic composition remains unknown.  However it is clear that
dark matter can not consist of any standard model (SM) particles.
Therefore its existence represents the strongest piece of evidence for
physics beyond the SM.

Models with softly broken supersymmetry (SUSY) are currently the best motivated
extensions of the SM.  Within these models the quadratic divergences,
which give rise to the destabilization of the electroweak (EW) scale, get
cancelled \cite{Witten:1981nf,Sakai:1981gr,Dimopoulos:1981zb,Kaul:1981hi}.
Models with softly broken SUSY also provide an attractive framework for the
incorporation of the gravitational interactions.  Indeed, a partial unification
of the SM gauge interactions with gravity can be attained within models
based on the $(N=1)$ local SUSY (supergravity).  Nevertheless $(N=1)$
supergravity (SUGRA) is a non--renormalizable theory.  The $(N=1)$ SUGRA models
can arise from ten dimensional $E_8 \times E_8^\prime$ heterotic string theory
\cite{Green:1987sp}.  The compactification of the extra dimensions in this
theory results in breaking $E_8 \to E_6$ \cite{delAguila:1985cb,
  Kaplunovsky:1993rd,Brignole:1993dj}.  The remaining $E_8^\prime$ constitutes
a hidden sector that gives rise to spontaneous breakdown of local SUSY.  The
hidden sector and visible sectors interact only gravitationally, which allows
for the breaking of local SUSY in the hidden sector to be communicated to the
visible sector and results in a set of soft SUSY breaking interactions.

When $R$--parity is conserved the lightest SUSY particle (LSP) in the models
with softly broken SUSY is stable and therefore can play the role of dark matter
\cite{Jungman:1995df}.  Moreover in the simplest SUSY extension of the SM,
i.e., the minimal supersymmetric standard model (MSSM), the SM gauge couplings
extrapolated to high energies using the renormalization group (RG) equations
(RGEs) converge to a common value at some high energy scale
$M_X\sim 10^{16}$ GeV \cite{Ellis:1990zq,Ellis:1990wk,Amaldi:1991cn,
  Langacker:1991an}.  This permits to embed the SM gauge
group into Grand Unified Theories (GUTs) \cite{Georgi:1974sy} based on $E_6$
or its subgroups such as $SU(5)$ and $SO(10)$.

In this context it is especially important to explore the implications
for dark matter and collider phenomenology within well motivated $E_6$
inspired SUSY extensions of the SM.  The breakdown of $E_6$ may lead
to a variety of SUSY models at low energies.  In particular, a set of
the simplest $E_6$ inspired SUSY extensions of the SM includes
supersymmetric models based on the SM gauge group, like the MSSM, as
well as extensions of the MSSM with an extra $U(1)$ gauge symmetry.
Within the class of the $E_6$ inspired $U(1)$ extensions of the MSSM,
there is a unique choice of Abelian $U(1)_{N}$ gauge symmetry that
allows zero charges for right-handed neutrinos and this is the $U(1)^\prime$
that appears in the exceptional supersymmetric standard model
(E$_6$SSM) \cite{King:2005jy, King:2005my}.  This choice ensures that
the right--handed neutrinos can be superheavy, so that a high scale
see-saw mechanism can be used to generate the mass hierarchy in the
lepton sector, providing a comprehensive understanding of the neutrino
oscillations data.  Successful leptogenesis is also a distinctive
feature of the E$_6$SSM because the heavy Majorana right-handed
neutrinos may decay into final states with lepton number $L=\pm 1$,
creating a lepton asymmetry in the early Universe
\cite{Hambye:2000bn,King:2008qb}.  Since sphalerons violate $B+L$ but
conserve $B-L$, this lepton asymmetry gets converted into the observed
baryon asymmetry of the Universe through the EW phase transition.  In
this case substantial values of the CP--asymmetries can be generated
even for the lightest right--handed neutrino masses $M_1 \sim 10^6$
GeV so that successful thermal leptogenesis may be achieved without
encountering a gravitino problem \cite{King:2008qb}.

To ensure anomaly cancellation the matter content of the E$_6$SSM is extended
to include three $\bm{27}$ representations of $E_6$.  In addition the low energy
spectrum can be supplemented by a $SU(2)_W$ doublet $L_4$ and anti-doublet
$\overline{L}_4$ from extra $\bm{27}'$ and $\overline{\bm{27}'}$ to preserve the
unification of the SM gauge couplings at high energies \cite{King:2007uj}.
Thus the E$_6$SSM contains extra exotic matter beyond the MSSM.  Over the last
ten years, several variants of the E$_6$SSM have been proposed
\cite{King:2005jy,King:2005my,Howl:2007hq,Howl:2007zi,Howl:2008xz,Howl:2009ds,
  Athron:2010zz,Hall:2011zq,Callaghan:2012rv,Nevzorov:2012hs,Callaghan:2013kaa,
  Athron:2014pua,King:2016wep}.  The $E_6$ inspired SUSY models with an extra
$U(1)_N$ gauge symmetry have been thoroughly investigated as well.  For
example, the possibility of mixing between doublet and singlet neutrinos
\cite{Ma:1995xk}, the effects of $Z - Z^\prime$ mixing \cite{Suematsu:1997au},
the neutralino sector \cite{Keith:1997zb,Suematsu:1997au,Keith:1996fv}, the
implications of the exotic states for the dark matter \cite{Hall:2009aj}, the
renormalization group flow \cite{Keith:1997zb,King:2007uj} and
EW symmetry breaking (EWSB) in the model \cite{Suematsu:1994qm,Keith:1997zb,
  Daikoku:2000ep} have all been studied.  More recently, the RG flow of the
Yukawa couplings and the theoretical upper bound on the lightest Higgs boson
mass were explored in the vicinity of the quasi--fixed point
\cite{Nevzorov:2013ixa,Nevzorov:2015iya} that appears as a result of the
intersection of the invariant and quasi--fixed lines \cite{Nevzorov:2001vj}.
Detailed studies of the E$_6$SSM have established that the additional exotic
matter and $Z^\prime$ in the model would lead to distinctive LHC signatures
\cite{King:2005jy,King:2005my,King:2006vu,
  Accomando:2006ga,King:2006rh,Howl:2007zi,Athron:2010zz,Athron:2011wu,
  Belyaev:2012si,Belyaev:2012jz}, as well as result in non-standard Higgs
decays for sufficiently light exotics \cite{Nevzorov:2013tta,Hall:2010ix,
  Hall:2010ny,Hall:2011au,Athron:2014pua,Nevzorov:2014sha,Nevzorov:2015iya,
  Athron:2016usd}.  In this SUSY model the particle spectrum has been examined
in Refs.~\cite{Athron:2008np,Athron:2009ue,Athron:2009bs,Athron:2012sq},
including the effects of threshold corrections from heavy states
\cite{Athron:2012pw}.  The renormalization of the vacuum expectation values
(VEVs) that lead to EWSB in the model has also been calculated
\cite{Sperling:2013eva,Sperling:2013xqa}, and the fine tuning in the model has
been studied \cite{Athron:2013ipa,Athron:2015tsa}.

Although the presence of exotic matter in the E$_6$SSM may lead to spectacular
collider signatures it also gives rise to non-diagonal flavor transitions and
rapid proton decay.  In principle, an approximate $Z_2^H$ symmetry can be
imposed to suppress flavor changing processes in these $U(1)$ extensions of
the MSSM while the most dangerous baryon and lepton number violating operators
can be forbidden by another exact $Z_2$ symmetry which plays a similar role to
the $R$--parity in the MSSM \cite{King:2005jy,King:2005my}.  Using the method
proposed in \cite{Hesselbach:2007te,Hesselbach:2007ta,Hesselbach:2008vt} it
was shown that the LSP and next--to--lightest SUSY particle (NLSP) in the
E$_6$SSM have masses below $60-65$ GeV \cite{Hall:2010ix}.  As a
consequence these states can give rise to unacceptably large branching ratios
of the exotic decays of the SM--like Higgs boson into the LSP and NLSP.  In
order to suppress such exotic Higgs decays and to prevent the decays of the
lightest MSSM--like neutralino into the LSP and NLSP in models with approximate
$Z_2^H$ symmetry an additional $Z^S_2$ symmetry needs to be postulated
\cite{Hall:2011zq}.  All discrete symmetries mentioned above do not commute
with $E_6$ and the imposition of such symmetries to ameliorate
phenomenological problems, which generically arise because of the presence of
the exotic matter at low energies, is an undesirable feature of the models
under consideration.

Here we focus on the investigation of the $U(1)_{N}$ extension of the MSSM
(SE$_6$SSM) in which a single discrete $\tilde{Z}^{H}_2$ symmetry forbids
tree-level flavor-changing transitions and the most dangerous operators that
violate baryon and lepton numbers \cite{Nevzorov:2012hs,Nevzorov:2013ixa,
  Athron:2014pua}.  In a recent letter \cite{Athron:2015vxg} we specified a
set of benchmark points representing scenarios with a $125$ GeV SM--like
Higgs, which are consistent with the LHC limits on SUSY particles and measured
dark matter abundance, within the constrained version of the above SE$_6$SSM
(CSE$_6$SSM).  As in any other constrained SUSY model, the soft SUSY--breaking
scalar masses, gaugino masses, the trilinear and bilinear scalar couplings in
the CSE$_6$SSM are each assumed to be universal at the scale $M_X$, where all
gauge couplings coincide, i.e., $m^2_i(M_X)=m_0^2$, $M_i(M_X)=M_{1/2}$,
$A_i(M_X)=A_0$ and $B_i(M_X)=B$.  The benchmark scenarios presented in
Ref.~\cite{Athron:2015vxg} lead to large spin-independent (SI) dark
matter-nucleon scattering cross section observable soon at \xenon\ experiment
and new physics signatures that may be observable at the $13$ TeV LHC.  These
new signatures should allow to distinguish the SUSY model under consideration
from the simplest SUSY extensions of the SM.  At the same time in this letter we
did not examine the CSE$_6$SSM parameter space thoroughly and did not provide
full details of our calculations.  We also did not include the full set of the
two--loop RGEs which were used in our analysis.

In this article we present the results of the comprehensive analysis of the
CSE$_6$SSM parameter space which is consistent with the $125$ GeV
SM--like Higgs, measured dark matter density and present LHC limits on
sparticle masses.  As in the MSSM the matter parity in the SE$_6$SSM is
preserved.  Therefore in both models the lightest $R$--parity odd state, i.e.,
LSP, is absolutely stable.  In most scenarios that have been explored within
the MSSM and its extensions the LSP is the lightest neutralino.  In the CMSSM
the lightest neutralino state is predominantly a linear superposition of the
Higgsino and bino.  Since the lightest neutralinos are heavy weakly interacting
massive particles (WIMPs) they explain well the large scale structure of the
Universe \cite{Blumenthal:1984bp} and can provide the correct relic abundance
of dark matter as long as the mass of the lightest neutralino is below the TeV
scale \cite{Jungman:1995df}.  The conservation of $\tilde{Z}^{H}_2$ symmetry and
matter parity in the SE$_6$SSM results in the lightest neutralino as well as
the lightest exotic state being stable.  In the simplest phenomenologically
viable scenarios the lightest exotic states have masses substantially lower
than $1$ eV forming hot dark matter in the Universe.  The results of our
analysis indicate that in this case the lightest neutralino in the CSE$_6$SSM,
which is either mostly Higgsino or a mixed bino-Higgsino state, can account for
all or some of the observed cold dark matter relic density.

We perform a scan of the parameter space of the CSE$_6$SSM enforcing
successful EW symmetry breaking and imposing theoretical and
low energy experimental constraints mentioned above.  We also compute
the dark matter density and SI neutralino-nucleon scattering cross
section as well as examine their dependence on the parameters of the
CSE$_6$SSM.  The obtained results are compared with the corresponding
ones in the CMSSM.  We show that present \lux\ bounds set sufficiently
stringent constraints on the mixing between bino and Higgsino states,
for cases where they give a substantial contribution to the observed
dark matter density.  We therefore find that if the relic density is
to be explained with Higgsino dark matter in either the CMSSM or
CSE$_6$SSM, then the lightest neutralino must be a relatively pure
Higgsino state with a highly restricted level of bino mixing, and this
is what we find in most of the allowed parameter space.  As a
consequence the observed dark matter abundance can be reproduced only
if the mass of lightest neutralino is relatively close to $1$ TeV.  In
this scenario all sparticles are so heavy that it won't be possible to
discover these states at the LHC.  If the lightest neutralino is
considerably lighter than $1$ TeV then this state can account for only
a small fraction of the measured dark matter density in the allowed
part of parameter space within both the CSE$_6$SSM and CMSSM.  At the
same time we argue that the scenarios with relatively small masses of
lightest neutralino and low relic dark matter abundance can still lead
to the spectrum of SUSY particles that may be observed at the $13$ TeV
LHC.  In the CSE$_6$SSM the set of states detectable at the LHC
includes gluino, chargino and neutralino states as well as exotic
fermions.  In the most part of the allowed CSE$_6$SSM parameter space
the lightest neutralino has sufficiently large direct detection cross
section which should be observable soon at the \xenon\ experiment.

The paper is organized as follows.  In the next section we briefly
review the $E_6$ inspired $U(1)_N$ extension of the MSSM with exact
custodial $\tilde{Z}^{H}_2$ symmetry and define the CSE$_6$SSM.  In
\secref{sec:ewsb} we consider the breakdown of gauge symmetry within this SUSY
model.  In \secref{sec:particle-spectrum} the analytical expressions for
the mass matrices and masses of all new states that appear in the SE$_6$SSM are
specified.  In \secref{sec:results} we discuss the implications of the
SUSY model under consideration for dark matter and summarize the
results of our studies.  \secref{sec:conclusions} is reserved for our
conclusions. \appref{app:rges} contains the complete system of the two--loop
RGEs that we use in our analysis.

\section{The SE$_6$SSM} \label{sec:se6ssm}
In orbifold GUT models the breakdown of $E_6$ gauge symmetry can lead to the
SM gauge group along with two additional $U(1)^\prime$ factors
\cite{Nevzorov:2012hs}, i.e.,
\begin{equation} \label{eq:e6-rank-6}
  E_6 \to SU(3)_C \times SU(2)_L \times U(1)_Y \times U(1)_\chi
  \times U(1)_\psi ,
\end{equation}
where $U(1)_\psi$ and $U(1)_\chi$ are associated with the subgroups
$E_6 \supset SO(10) \times U(1)_\psi \supset SU(5) \times U(1)_\chi \times
U(1)_\psi$.  Further symmetry breaking can then result in a low-energy model
with a single additional $U(1)'$ that is a linear combination of
$U(1)_\psi$ and $U(1)_\chi$,
\begin{equation} \label{eq:e6-rank-5}
  U(1)^\prime = U(1)_\chi \cos \theta_{E_6} + U(1)_\psi \sin \theta_{E_6} .
\end{equation}
In this case, the value of the mixing angle $\theta_{E_6}$ characterizes the
resulting $U(1)^\prime$ at low-energies, and several choices of symmetry
breaking pattern have been considered (for reviews, see for example
Refs.~\cite{Hewett:1988xc,Langacker:2008yv,London:1986dk}).  In $U(1)$
extensions with $E_6$ inspired charges, gauge anomalies automatically cancel
provided that the low-energy matter content fills in complete representations
of $E_6$.  The SM particle content can be accommodated if each generation is
embedded within a fundamental $\bm{27}$-plet of $E_6$, which requires the
introduction of extra matter to form complete multiplets.  In addition to the
SM fermions, each of these $\bm{27}$-plets ($\bm{27}_i$, $i = 1, 2, 3$)
contains a pair of $SU(2)_L$ doublets, $H_u$, $H_d$, a pair of color triplets,
$D_i$, $\overline{D}_i$, a right-handed neutrino, $N_i^c$, and a SM singlet,
$S_i$.  In general, both $N_i^c$ and $S_i$ carry non-zero $U(1)^\prime$ charges.
The doublets $H^u_i$ and $H^d_i$ may be identified as Higgs or inert Higgs
doublets, the distinction being that the latter do not develop VEVs.  The
states $\overline{D}_i$ and $D_i$ have electric charge $\pm 1/3$ and carry
$B - L$ charge twice that of ordinary quarks, and therefore may either be
diquarks or leptoquarks.

The potential for interesting phenomenology associated with these exotic
states, along with at least one $Z^\prime$ boson, has provided substantial
motivation for studying $E_6$ inspired models \cite{Gunion:1989we,
  Binetruy:1985xm,Ellis:1986yg,Ibanez:1986si,Gunion:1986ky,Haber:1986gz,
  Ellis:1986ip,Drees:1987tp,Baer:1987eb,Gunion:1987jd}.  Possible signatures of
the exotic states at colliders have been studied \cite{Kang:2007ib}, as well
as limits on the $Z^\prime$ mass \cite{Accomando:2010fz}.  In addition to
observing these exotic states, an underlying $E_6$ GUT might leave identifiable
fingerprints on the ordinary MSSM mass spectrum, such as in the pattern of
first and second generation sfermion masses \cite{Miller:2012vn}.  Further
motivation for studying this class of models has come from the fact that
they are able to address several weaknesses of the MSSM.  The extended gauge
sector and the presence of additional singlets, some of which may get VEVs,
allows for the solution of the MSSM $\mu$-problem \cite{Kim:1983dt} in a way
similar to in the next-to-MSSM (NMSSM) \cite{Ellwanger:2009dp,Maniatis:2009re}.
These same features also lead to a theoretical upper bound on the lightest
Higgs boson mass that is larger than can be achieved in the MSSM, and indeed in
the NMSSM \cite{Accomando:2006ga,Daikoku:2000ep,Barger:2006dh,Barger:2007ay}.
The accompanying enlarged Higgs \cite{Barger:2006dh,Barger:2007ay,King:2005jy}
and neutralino \cite{Keith:1997zb,Suematsu:1997tv,GutierrezRodriguez:2006hb,
  Suematsu:1997qt,Suematsu:1997au,Keith:1996fv,Hesselbach:2001ri,Barger:2005hb,
  Choi:2006fz,Barger:2007nv,Gherghetta:1996yr,Barger:2007ay} sectors have been
extensively studied.  It has been proposed that the extra $D$-terms
could also solve the tachyon problems encountered in anomaly mediated SUSY
breaking scenarios \cite{Asano:2008ju}, while the inclusion of appropriate
family symmetries could provide an explanation for the hierarchy of fermion
masses and mixings \cite{Stech:2008wd,Howl:2008xz,Howl:2009ds,
  Camargo-Molina:2016bwm}.  Many further implications of these models have
also been considered, including for EWSB
\cite{Suematsu:1994qm,Cvetic:1995rj,Cvetic:1996mf,Cvetic:1997ky,Keith:1997zb,
  Langacker:1998tc,Daikoku:2000ep}, neutrino physics \cite{Kang:2004ix,
  Ma:1995xk}, leptogenesis \cite{Hambye:2000bn,King:2008qb} and EW
baryogenesis \cite{Ma:2000jf,Kang:2004pp}, the muon anomalous magnetic moment
\cite{Grifols:1986vr,Morris:1987fm}, electric dipole moments
\cite{Suematsu:1997tv,GutierrezRodriguez:2006hb}, lepton flavor violating
processes \cite{Suematsu:1997qt} and the possibility of CP--violation
in the extended Higgs sector \cite{Ham:2008fx}.

As was noted above, in the rank-5 models described by Eq.~(\ref{eq:e6-rank-5})
both the singlets $S_i$ and the right-handed neutrinos $N_i^c$ are charged
under the additional $U(1)^\prime$ in general.  However, for the choice
of $\theta_{E_6} = \arctan \sqrt{15}$, the right-handed neutrinos are uncharged
under the resulting $U(1)^\prime$, denoted $U(1)_N$.  In this case, a large
Majorana mass is allowed for the $N_i^c$ and a see-saw mechanism can be used to
explain the observed neutrino masses, while also allowing for an explanation of
the baryon asymmetry via leptogenesis \cite{Hambye:2000bn,King:2008qb}.

In this article we study a $U(1)_{N}$ extension of the MSSM in which
tree-level flavor-changing transitions and the most dangerous baryon and
lepton number violating operators are forbidden by a single discrete
$\tilde{Z}^{H}_2$ symmetry \cite{Nevzorov:2012hs,Nevzorov:2013ixa,
  Athron:2014pua}.  In the SUSY models under consideration
\cite{Nevzorov:2012hs}, $E_6$ is assumed to be broken directly to
$SU(3)_C \times SU(2)_L \times U(1)_Y \times U(1)_\chi \times U(1)_\psi$ at or
near the GUT scale, $M_X$, which can be achieved in 5 or 6 dimensional
orbifold GUT models.  The additional $U(1)_\chi \times U(1)_\psi$ is then
broken near $M_X$ to $U(1)_N \times Z_2^M$, where the matter parity $Z_2^M$ is
defined by
\begin{equation} \label{eq:matter-parity}
  Z_2^M = (-1)^{3(B - L)} .
\end{equation}
Below $M_X$, the three complete $\bm{27}$-plets of $E_6$ are taken to be
accompanied by a set of pairs of multiplets $M_l$, $\overline{M}_l$, coming
from incomplete $\bm{27}^\prime$ and $\overline{\bm{27}^\prime}$
representations, respectively.  Note that anomalies still cancel, since the
fields from $M_l$ and $\overline{M}_l$ carry opposite $U(1)$ charges.  A
single exact $\tilde{Z}_2^H$, commuting with $E_6$, may then be imposed under
which all components of the $\bm{27}$-plets are odd, thereby forbidding both
interactions that generate large flavor changing neutral currents (FCNCs) and
those that would lead to rapid proton decay.  Doing so precludes any of the
components of the $\bm{27}$-plets from getting VEVs to break EW symmetry, so
that, for example, all of the $\bm{27}$-plet Higgs states $H^u_i$, $H^d_i$ are
inert and cannot be identified with the usual MSSM Higgs doublets.  But, at
the same time the multiplets $M_l$ and $\overline{M}_l$ may be either even or
odd under $\tilde{Z}_2^H$, allowing some of them to get VEVs for spontaneous
symmetry breaking.  In the model considered here we include two pairs of
$SU(2)_L$ doublets, $H_u$ and $\overline{H}_u$, $H_d$ and $\overline{H}_d$, as
well as a pair of singlets $S$ and $\overline{S}$.  The fields $H_u$, $H_d$,
$S$ and $\overline{S}$ are postulated to be even under $\tilde{Z}_2^H$
symmetry and are responsible for the breaking of $SU(2)_L \times U(1)_Y \times
U(1)_N \to U(1)_{\text{em}}$ at the TeV scale\footnote{The initial breaking of
  $U(1)_\psi \times U(1)_\chi \to U(1)_N \times Z_2^M$ can be achieved with the
  VEVs of a multiplet pair $N_H^c$ and $\overline{N}_H^c$ with the quantum
  numbers of right-handed neutrinos.  These VEVs may also be responsible for
  the generation of Majorana masses for the $\bm{27}$-plet right-handed
  neutrinos; the full details of the construction can be found in
  Ref.~\cite{Nevzorov:2012hs}.}.  The doublets $\overline{H}_u$ and
$\overline{H}_d$ are odd under $\tilde{Z}_2^H$, so that they can mix with a
combination of the $\bm{27}$-plet states, defined to be the third generation
$H^u_3$, $H^d_3$.  In this case they may form vectorlike states with masses of
order $M_X$, and so may be integrated out of the low-energy spectrum.

With only this set of multiplets, the imposed $\tilde{Z}_2^H$ would forbid any
renormalizable operators allowing the exotic quarks to decay.  Such
long-lived exotics would be produced in the early Universe and would lead
to estimated concentrations \cite{Wolfram:1978gp,Dover:1979sn} in excess of
the observed limits on heavy isotopes \cite{Rich:1987jd,Smith:1988ni,
  Hemmick:1989ns}.  To avoid this, a pair of $\tilde{Z}_2^H$ even $SU(2)_L$
doublets $L_4$ and $\overline{L}_4$ with the quantum numbers of leptons are
also included at the TeV scale that couple to the exotic $D_i$, $\overline{D}_i$
and allow the exotic quarks to decay.  This choice also implies that $D_i$
and $\overline{D}_i$ are leptoquarks in this scenario.

In addition to the above sets of multiplets, in the model considered here we
also include a pure singlet superfield $\hat{\phi}$ in the spectrum below the
GUT scale, which is uncharged under all of the gauge symmetries
\cite{Athron:2014pua}.  This superfield is likewise taken to be even under
$\tilde{Z}_2^H$ so that the superpotential may contain a term proportional
to $\hat{\phi} \hat{S} \hat{\overline{S}}$, to stabilize the scalar potential,
and the scalar component of $\hat{\phi}$ is allowed to develop a non-zero VEV.
The fields $H_u$, $H_d$, $S$, $\overline{S}$ and $\hat{\phi}$ are all expected
to get masses at or below the TeV scale.  Thus after integrating out
superheavy states the low-energy matter content in this model, which we refer
to as the SE$_6$SSM, consists of the superfields shown in
\tabref{tab:matter-content}.
\begin{table*}[ht]
  \centering
  \begin{tabular}{ccccccccccccccccc}
    \toprule
    & $\hat{Q}_i$ & $\hat{u}_i^c$ & $\hat{d}_i^c$ & $\hat{L}_i$ & $\hat{e}_i^c$
    & $\hat{D}_i$ & $\hat{\overline{D}}_i$ & $\hat{S}_i$ & $\hat{H}_{\alpha}^u$
    & $\hat{H}_{\alpha}^d$ & $\hat{H}_u$ & $\hat{H}_d$ & $\hat{S}$
    & $\hat{\overline{S}}$ & $\hat{L}_4$ & $\hat{\overline{L}}_4$ \\
    \midrule
    $SU(3)_C$ & $\bm{3}$ & $\overline{\bm{3}}$ & $\overline{\bm{3}}$ & $\bm{1}$
    & $\bm{1}$ & $\bm{3}$ & $\overline{\bm{3}}$ & $\bm{1}$ & $\bm{1}$
    & $\bm{1}$ & $\bm{1}$ & $\bm{1}$ & $\bm{1}$ & $\bm{1}$ & $\bm{1}$
    & $\bm{1}$ \\
    $SU(2)_L$ & $\bm{2}$ & $\bm{1}$ & $\bm{1}$ & $\bm{2}$ & $\bm{1}$ & $\bm{1}$
    & $\bm{1}$ & $\bm{1}$ & $\bm{2}$ & $\bm{2}$ & $\bm{2}$ & $\bm{2}$ & $\bm{1}$
    & $\bm{1}$ & $\bm{2}$ & $\overline{\bm{2}}$ \\
    $\sqrt{\frac{5}{3}} Q_i^Y$ & $\frac{1}{6}$ & $-\frac{2}{3}$ & $\frac{1}{3}$
    & $-\frac{1}{2}$ & $1$ & $-\frac{1}{3}$ & $\frac{1}{3}$ & $0$
    & $\frac{1}{2}$ & $-\frac{1}{2}$ & $\frac{1}{2}$ & $-\frac{1}{2}$ & $0$
    & $0$ & $-\frac{1}{2}$ & $\frac{1}{2}$ \\
    $\sqrt{40} Q_i^N$ & $1$ & $1$ & $2$ & $2$ & $1$ & $-2$ & $-3$ & $5$
    & $-2$ & $-3$ & $-2$ & $-3$ & $5$ & $-5$ & $2$ & $-2$ \\
    $\tilde{Z}_2^H$ & $-$ & $-$ & $-$ & $-$ & $-$ & $-$ & $-$ & $-$ & $-$
    & $-$ & $+$ & $+$ & $+$ & $+$ & $+$ & $+$ \\
    $Z_2^M$ & $-$ & $-$ & $-$ & $-$ & $-$ & $+$ & $+$ & $+$ & $+$ & $+$ & $+$
    & $+$ & $+$ & $+$ & $-$ & $-$ \\
    $Z_2^E$ & $+$ & $+$ & $+$ & $+$ & $+$ & $-$ & $-$ & $-$ & $-$ & $-$
    & $+$ & $+$ & $+$ & $+$ & $-$ & $-$ \\
    \bottomrule
  \end{tabular}
  \caption{Summary of the chiral superfields present at low-energies,
    showing their representations and charges under the gauge symmetries
    and their transformation properties under the discrete symmetries
    defined in the text.  Here and throughout this paper, the generation
    index $i = 1, 2, 3$, while $\alpha = 1, 2$.  Note that the pure singlet
    field $\hat{\phi}$ is omitted from the table, as it transforms trivially
    under all of the symmetries.}
  \label{tab:matter-content}
\end{table*}
At low-energies and neglecting suppressed non-renormalizable interactions,
the superpotential can then be written
\begin{align}
  W &= \lambda \hat{S} \hat{H}_d \cdot \hat{H}_u - \sigma \hat{\phi} \hat{S}
  \hat{\overline{S}} + \frac{\kappa_\phi}{3} \hat{\phi}^3 + \frac{\mu_\phi}{2}
  \hat{\phi}^2 + \Lambda_F \hat{\phi} + \tilde{\lambda}_{\alpha \beta} \hat{S}
  \hat{H}_\alpha^d \cdot \hat{H}_\beta^u + \kappa_{ij} \hat{S} \hat{D}_i
  \hat{\overline{D}}_j \nonumber \\
  & \quad {} + \tilde{f}_{i\alpha} \hat{S}_i \hat{H}_u \cdot
  \hat{H}_\alpha^d + f_{i\alpha} \hat{S}_i \hat{H}_\alpha^u \cdot \hat{H}_d
  - g_{ij}^D \hat{Q}_i \cdot \hat{L}_4 \hat{\overline{D}}_j - h_{i\alpha}^E
  \hat{e}_i^c \hat{H}_\alpha^d \cdot \hat{L}_4 + \mu_L \hat{L}_4 \cdot
  \hat{\overline{L}}_4 \nonumber \\
  & \quad {} + \tilde{\sigma} \hat{\phi} \hat{\overline{L}}_4 \cdot
  \hat{L}_4 + y_{ij}^U \hat{u}_i^c \hat{H}_u \cdot \hat{Q}_j
  + y_{ij}^D \hat{d}_i^c \hat{Q}_j \cdot \hat{H}_d + y_{ij}^E \hat{e}_i^c
  \hat{L}_j \cdot \hat{H}_d. \label{eq:superpotential}
\end{align}
We denote superfields with hats, and adopt the convention
$\hat{A} \cdot \hat{B} \equiv \epsilon_{\alpha \beta} \hat{A}^\alpha
\hat{B}^\beta = \hat{A}^2 \hat{B}^1 - \hat{A}^1 \hat{B}^2$ for the $SU(2)$ dot
product.  The exact $\tilde{Z}_2^H$ symmetry forbids all terms of the form
$\bm{27} \times \bm{27} \times \bm{27}$, so that the allowed trilinear
interactions involving non-singlet fields are of the form
$\bm{27}^\prime \times \bm{27}^\prime \times \bm{27}^\prime$ or $\bm{27}^\prime
\times \bm{27} \times \bm{27}$.  By making appropriate rotations of the
superfields $(\hat{H}_\alpha^d, \hat{H}_\alpha^u)$ and $(\hat{D}_i,
\hat{\overline{D}}_i)$, the trilinear couplings $\tilde{\lambda}_{\alpha \beta}$
and $\kappa_{ij}$ are chosen to be flavor diagonal, while the other new
couplings are $\tilde{f}_{i\alpha}$, $f_{i\alpha}$, $g_{ij}^D$ and
$h_{i\alpha}^E$ are not, in general.  The superpotential also contains several
bilinear terms, such as those of the form $\bm{27}^\prime \times
\overline{\bm{27}^\prime}$.  The corresponding couplings, for example $\mu_L$,
may be generated\footnote{For example, in a SUGRA model this term can be
  induced after the breakdown of local SUSY if the K\"{a}hler potential contains
  an extra term of the form $\left [ Z_L (\hat{L}_4 \hat{\overline{L}_4}) + h.c.
    \right ]$ \cite{Casas:1992mk}} through the Giudice-Masiero mechanism
\cite{Giudice:1988yz}.

As well as being invariant under the single imposed $\tilde{Z}_2^H$ symmetry,
the superpotential is also invariant under the residual $Z_2^M$ symmetry
resulting from the breakdown of $U(1)_\psi \times U(1)_\chi \to U(1)_N \times
Z_2^M$.  The presence of multiple $Z_2$ symmetries suggests that it is not
unreasonable to expect multiple stable states that may play the role of DM.
For our analysis, it is convenient to define a combination of these two $Z_2$
symmetries by $\tilde{Z}_2^H = Z_2^M \times Z_2^E$.  The transformation
properties of each field under this $Z_2^E$ symmetry are also shown in
\tabref{tab:matter-content}, and henceforth we shall refer to states that are
odd under $Z_2^E$ as exotics.  Since the Lagrangian is separately invariant
under $\tilde{Z}_2^H$ and $Z_2^M$, it is also the case that transformations
under $Z_2^E$ leave the Lagrangian invariant.  In particular, this means that
the lightest $Z_2^E$-odd, exotic state is absolutely stable and so can
potentially be a DM candidate.  The automatically conserved matter parity
$Z_2^M$, meanwhile, is equivalent to $R$--parity and also implies the existence
of a stable state, as in the MSSM.  Examination of the possible cases shows
that these two states are in fact distinct, so that the model contains two DM
candidates.  In the case that the stable, lightest $Z_2^E$ odd state is
$R$--parity even\footnote{Or, more precisely, if it is not the lightest
  $R$--parity odd state.}, then the lightest $R$--parity odd state must be
stable, as usual.  Conversely, if the lightest $Z_2^E$ odd state is also the
lightest $R$-parity odd state, then either the lightest $R$--parity even,
$Z_2^E$ odd state or the lightest $R$--parity odd, $Z_2^E$ even state
(depending on which is lighter) is absolutely stable.

By applying the method described in
Ref.~\cite{Hesselbach:2007te,Hesselbach:2007ta,Hesselbach:2008vt}, it has
previously been found that the lightest inert neutralinos can have masses
no larger than $60 - 65$ GeV \cite{Hall:2010ix,Hall:2010ny,Hall:2011au}.
These states then tend to be the lightest exotic states in the spectrum,
and are predominantly combinations of the fermionic components of
the inert singlet superfields $\hat{S}_i$.  Substantial masses for these inert
singlinos, of more than $\sim 1$ eV, are ruled out by measurements of the
SM-like Higgs branching ratios and the DM relic density.  The simplest viable
solution is instead for the inert singlino masses to be much lighter than 1 eV,
which can be achieved provided that the couplings $\tilde{f}_{i\alpha},
f_{i\alpha} \lesssim 10^{-6}$.  This results in the inert singlinos forming hot
dark matter, giving a negligible contribution to the observed relic
density\footnote{The presence of very light neutral fermions in the particle
  spectrum may also lead to some interesting implications for the neutrino
  physics (see, for example, Ref.~\cite{Frere:1996gb}).}.

In this case, the second DM candidate should account fully or partially
for the DM density, with the latter possibility requiring either
additional DM candidates or a non-standard thermal history of the Universe to
be consistent with measurements.  The sub-eV inert singlinos are both the
lightest exotic and lightest $R$--parity odd states in the spectrum.  This
implies that the lightest $R$--parity even exotic state or the lightest
$R$--parity odd, $Z_2^E$ even state is a possible second DM candidate.  As can
be read from \tabref{tab:matter-content}, the possible exotic candidates are
the exotic squarks arising from the superfields
$(\hat{D}_i, \hat{\overline{D}}_i)$, the inert Higgs scalars coming from
the mixing of $(\hat{S}_i, \hat{H}_\alpha^u, \hat{H}_\alpha^d)$, or the
fermionic components of $(\hat{L}_4, \hat{\overline{L}}_4)$.  The masses
of these states are required to be sufficiently heavy to have evaded detection
to date.  In particular, for large values of the SUSY breaking scale $M_S$
the scalars receive large soft SUSY breaking masses and can be of similar
mass to the ordinary squarks.  The fermionic components of $(\hat{L}_4,
\hat{\overline{L}}_4)$, meanwhile, receive a supersymmetric mass contribution
from the superpotential bilinear term $\mu_L \hat{L}_4 \cdot
\hat{\overline{L}}_4$, which is not constrained by the requirement of successful
EWSB and need not be small\footnote{The principal constraints on the value of
  $\mu_L$ come from requiring that gauge unification still occurs, which
  restricts $\mu_L \lesssim 100$ TeV \cite{King:2007uj}, and that the states
  associated with $\hat{L}_4$ and $\hat{\overline{L}}_4$ are light enough so
  that the exotic leptoquarks $D_i$, $\overline{D}_i$ decay sufficiently
  quickly.}.  In the model studied here this means that the lightest $R$--parity
odd, $Z_2^E$ even state tends to be the stable state, corresponding to the
lightest neutralino with $Z_2^E = +1$.  Depending on the composition of this
state, it may then account for some or all of the DM relic density, as in the
MSSM.  In the following we shall focus on cases where the lightest neutralino is
a mixed bino-Higgsino state, or pure Higgsino; we will find that this leads to
a DM candidate that is also MSSM-like in its interactions and predictions for
the DM relic density.

As usual in low-energy SUSY models, the relevant masses and mixings of interest
in the neutralino sector are governed both by the superpotential interactions in
Eq.~(\ref{eq:superpotential}) as well as a subset of the soft SUSY breaking
interactions.  Including the standard set of soft scalar masses, soft
trilinears, and soft gaugino masses, the full set of soft SUSY breaking terms
that we consider is
\begin{align}
  -\mathcal{L}_{\text{soft}} &= m_{H_u}^2 |H_u|^2
    + m_{H_d}^2 |H_d|^2 + m_S^2 |S|^2 + m_{\overline{S}}^2 |\overline{S}|^2
    + m_{\Sigma_{ij}}^2 S_i^\dagger S_j + m_{\phi}^2 |\phi|^2 \nonumber \\
  & \quad {} + m_{H_{2,\alpha \beta}}^2 (H^u_{\alpha})^\dagger H^u_{\beta} +
    m_{H_{1,\alpha\beta}}^2 (H^d_{\alpha})^\dagger H^d_{\beta}
    + m_{D_{ij}}^2 D_i^\dagger D_j + m_{\overline{D}_{ij}}^2
    \overline{D}_i^\dagger \overline{D}_j + m_{L_4}^2 |L_4|^2
    \nonumber \\
  & \quad {} + m_{\overline{L}_4}^2 |\overline{L}_4|^2
    + m_{Q_{ij}}^2 \tilde{Q}_i^\dagger \tilde{Q}_j + m_{u^c_{ij}}^2
    (\tilde{u}^c_i)^\dagger \tilde{u}^c_j + m_{d^c_{ij}}^2
    (\tilde{d}^c_i)^\dagger \tilde{d}^c_j + m_{L_{ij}}^2 \tilde{L}_i^\dagger
    \tilde{L}_j \nonumber \\
  & \quad {} + m_{e^c_{ij}}^2 (\tilde{e}^c_i)^\dagger \tilde{e}^c_j
    + \biggl ( \mu_L B_L L_4 \cdot \overline{L}_4 +
    \frac{\mu_\phi B_\phi}{2} \phi^2 + \Lambda_S \phi + h.c. \biggr )
    \nonumber \\
  & \quad {} +  \biggl ( T_\lambda S H_d \cdot H_u
    - T_\sigma \phi S \overline{S} + T^\kappa_{ij} S D_i \overline{D}_j
    + T^U_{ij}\tilde{u}_i^c H_{u} \cdot \tilde{Q}_j + T^D_{ij} \tilde{d}_i^c
    \tilde{Q}_j \cdot H_{d} \nonumber \\
    & \quad {} + T^E_{ij} \tilde{e}_i^c \tilde{L} \cdot H_{d}
    + T^{\tilde{\lambda}}_{\alpha \beta} S H^d_{\alpha} \cdot
    H^u_{\beta} + T^{\tilde{f}}_{i\alpha} S_i H_u \cdot H^d_\alpha
    + T^f_{i\alpha} S_i H^u_\alpha \cdot H_d
    \nonumber \\
  & \quad {} + T_{\tilde{\sigma}} \phi \overline{L}_4 \cdot L_4 +
    \frac{T_{\kappa_\phi}}{3}\phi^3
    - T^{g^D}_{ij} \tilde{Q}_i \cdot L_4 \overline{D}_j -
    T^{h^E}_{i\alpha} \tilde{e}^c_{i} H^d_{\alpha} \cdot L_4 + h.c. \biggr )
    \nonumber \\
  & \quad {}  + \frac{1}{2} \biggl ( M_1 \tilde{B} \tilde{B}
      + M_2 \tilde{W} \tilde{W} + M_3 \tilde{G} \tilde{G}
      + M_1^\prime \tilde{B}^\prime \tilde{B}^\prime + 2 M_{11}
      \tilde{B} \tilde{B}^\prime + h.c. \biggr ) .
    \label{eq:soft-terms}
\end{align}
The general soft SUSY breaking Lagrangian, in which all of the soft parameters
are treated as independent, introduces a large number of additional free
parameters on top of the extra couplings already present in the
superpotential.  The number of free parameters can be much reduced by
considering a constrained model in which certain relations are assumed to hold
between the soft parameters at some high scale.

The CSE$_6$SSM is defined by imposing boundary conditions at the GUT scale
$M_X$ where all gauge couplings coincide.  In the SE$_6$SSM, since all of the
low-energy matter content can be placed in complete $SU(5)$ multiplets with
the exception of the doublets $\hat{L}_4$ and $\hat{\overline{L}}_4$, gauge
coupling unification still occurs at the two-loop level for any value of
$\alpha_3 (M_Z)$, the strong coupling evaluated at the scale $M_Z$, consistent
with the measured value \cite{King:2007uj,Nevzorov:2012hs}.  Therefore, at the
GUT scale $M_X$ we take
\begin{equation} \label{eq:gut-scale}
  g_1(M_X) \approx g_1^\prime(M_X) \approx g_2(M_X) \approx g_3(M_X),
\end{equation}
where $g_1$, $g_1^\prime$, $g_2$ and $g_3$ are the GUT-normalized $U(1)_Y$,
$U(1)_N$, $SU(2)_L$ and $SU(3)_C$ gauge couplings, respectively.  This allows
for the $U(1)_N$ gauge coupling $g_1^\prime$ to be fixed.  The presence
of multiple $U(1)$ symmetries implies the possibility of kinetic mixing between
the $U(1)$ field strengths \cite{Holdom:1985ag,Babu:1997st}.  In practice, this
mixing can be handled by working in a rotated basis for the $U(1)$ gauge fields
where the mixing leads instead to non-zero off-diagonal gauge couplings, i.e.,
in covariant derivatives one finds terms of the form $Q_{\hat{\Phi}}^T G A_\mu$
with
\begin{equation} \label{eq:off-diagonal-gauge-couplings}
  Q_{\hat{\Phi}} = \begin{pmatrix} Q_{\hat{\Phi}}^Y \\ Q_{\hat{\Phi}}^N
  \end{pmatrix}, \quad
  G = \begin{pmatrix} g_1 & g_{11} \\ 0 & g_1^\prime \end{pmatrix}, \quad
  A_\mu = \begin{pmatrix} B_\mu \\ B^\prime_\mu \end{pmatrix} .
\end{equation}
This field redefinition is also responsible for the appearance of the mixed
gaugino soft mass, $M_{11}$, in the last bracketed term of
Eq.~(\ref{eq:soft-terms}).  It is natural to expect that at $M_X$, the
kinetic mixing should vanish so that $g_{11}(M_X) = 0$, $M_{11}(M_X) = 0$.
However, even if this holds at $M_X$, in general non-zero mixing terms will be
generated at low-energies by RG running
\cite{delAguila:1987st,delAguila:1988jz}.  Previous analyses
\cite{Rizzo:1998ut,King:2007uj} suggest that in this particular model, provided
that the off-diagonal gauge coupling vanishes at $M_X$, it remains very small
at all scales below $M_X$ as well, $g_{11} \sim 0.02 \ll g_1, g_1^\prime$.
Therefore in our analysis we neglect the effects of gauge kinetic mixing,
setting $g_{11}(M_X) = 0$, $M_{11}(M_X) = 0$ and taking them to vanish at scales
below this.  Nevertheless, it is important to note that in general the effects
of this kinetic mixing can be non-negligible \cite{Rizzo:1998ut,Krauss:2012ku,
  Rizzo:2012rf}; it is small here as the only non-vanishing contribution to
the mixing comes from the $(\hat{L}_4, \hat{\overline{L}}_4)$ multiplet pair.

The remaining soft masses satisfy high-scale relations analogous to those
applied in the CMSSM.  The soft scalar masses squared are taken to be
flavor diagonal with diagonal elements set to the common value $m_0^2$ at
$M_X$, and similarly the gaugino masses (with the exception of $M_{11}$, as
noted above) are assumed to unify to the value $M_{1/2}$ at this scale.  The
values of the soft breaking trilinears are related to a single common trilinear
parameter $A_0$ by
\begin{equation}
  \begin{aligned}
    & T_\lambda(M_X) = \lambda(M_X) A_0 \, , \quad T_\sigma(M_X) =
      \sigma(M_X) A_0 \, , \\
    & T^\kappa_{ij}(M_X) = \kappa_{ij}(M_X) A_0 \, , \quad
      T^U_{ij}(M_X) = y^U_{ij}(M_X) A_0 \, , \\
    & T^D_{ij}(M_X) = y^D_{ij}(M_X) A_0 \, , \quad
      T^E_{ij}(M_X) = y^E_{ij}(M_X) A_0 \, , \\
    & T^{\tilde{\lambda}}_{\alpha \beta}(M_X) =
      \tilde{\lambda}_{\alpha \beta}(M_X) A_0 \, , \quad
      T^{\tilde{f}}_{i\alpha}(M_X) = \tilde{f}_{i\alpha}(M_X) A_0 \, , \\
    & T^f_{i\alpha}(M_X) = f_{i\alpha}(M_X) A_0 \, , \quad
      T_{\tilde{\sigma}}(M_X) = \tilde{\sigma}(M_X) A_0 \, , \\
    & T_{\kappa_\phi}(M_X) = \kappa(M_X) A_0 \, , \quad
      T^{g^D}_{ij}(M_X) = g^D_{ij}(M_X) A_0 \, , \\
    & T^{h^E}_{i\alpha}(M_X) = h^E_{i\alpha}(M_X) A_0 \, .
  \end{aligned} \label{eq:soft-trilinear-bcs}
\end{equation}
Similarly, the soft breaking bilinears are assumed to unify,
$B_L(M_X) = B_\phi(M_X) = B_0$.  The parameter $B_0$ is taken to be independent
of $A_0$; to do so we assume that these soft terms are also generated via a
Giudice-Masiero term, as used to produce the superpotential bilinears.
The soft breaking tadpole $\Lambda_S$ is not required to be related to other
soft parameters by the high-scale boundary condition.

With this choice of boundary conditions, the remaining
unfixed parameters in the CSE$_6$SSM consist of the new superpotential
couplings, namely $\lambda(M_X)$, $\sigma(M_X)$, $\kappa_\phi(M_X)$,
$\mu_\phi(M_X)$, $\Lambda_F(M_X)$, $\tilde{\lambda}_{\alpha \beta}(M_X)$,
$\kappa_{ij}(M_X)$, $\tilde{f}_{i\alpha}(M_X)$, $f_{i\alpha}(M_X)$,
$g_{ij}^D(M_X)$, $h_{i\alpha}^E(M_X)$, $\mu_L(M_X)$ and $\tilde{\sigma}(M_X)$,
and the soft breaking parameters $m_0$, $M_{1/2}$, $A_0$, $B_0$ and $\Lambda_S$.
To simplify our analysis, in the following we assume that all of these
parameters are real.  Once these high-scale parameters, together with the
MSSM gauge and Yukawa couplings are specified, the model at low-energies is
studied by integrating the RGEs given in \appref{app:rges} from $M_X$ to the
EWSB scale.

\section{Gauge Symmetry Breaking} \label{sec:ewsb}
The Higgs fields $H_u$, $H_d$, $S$, $\overline{S}$ and $\phi$ develop non-zero
VEVs breaking $SU(2)_L \times U(1)_Y\times U(1)_N \to U(1)_{\text{em}}$.
The relevant part of the scalar potential reads
\begin{align}
  V &= \lambda^2 |S|^2 ( |H_d|^2 + |H_u|^2 ) +
  \sigma^2 |\phi|^2 |S|^2 + | \lambda H_d \cdot H_u - \sigma \phi
  \overline{S} |^2 \nonumber \\
  & \quad {} + |\kappa_\phi \phi^2 + \mu_\phi \phi + \Lambda_F - \sigma S
  \overline{S}|^2 + \frac{\bar{g}^2}{8} ( |H_u|^2 - |H_d|^2 )^2
  + \frac{g_2^2}{2} |H_d^\dagger H_u|^2 \nonumber \\
  & \quad {} + \frac{g_1'^2}{2} ( Q_{H_d} |H_d|^2
  + Q_{H_u} |H_u|^2 + Q_S |S|^2 - Q_S |\overline{S}|^2
  )^2 \nonumber \\
  & \quad {} + m_{H_d}^2 |H_d|^2 + m_{H_u}^2 |H_u|^2 + m_S^2 |S|^2
  + m_{\overline{S}}^2 |\overline{S}|^2 + m_\phi^2 |\phi|^2 \nonumber \\
  & \quad {} + \biggl ( \frac{T_{\kappa_\phi}}{3} \phi^3 + \frac{\mu_\phi}{2}
  B_\phi \phi^2 + \Lambda_S \phi + T_\lambda S H_d \cdot H_u - T_\sigma \phi S
  \overline{S} + h.c. \biggr ) + \Delta V, \label{eq:higgs-potential}
\end{align}
where $\bar{g}^2 = g_2^2 + 3 g_1^2 /5$ and $\Delta V$ contains the loop
corrections to the effective potential.  We denote by $Q_\Phi$ the $U(1)_N$
charge of the field $\Phi$.  In the presence of kinetic mixing these charges
should be replaced by effective $U(1)_N$ charges \cite{King:2005jy}.

At the physical minimum of this potential, the VEVs of the Higgs fields are
taken to be of the form
\begin{equation}
  \begin{aligned}
    &\langle H_d \rangle = \frac{1}{\sqrt{2}} \begin{pmatrix} v_1 \\ 0
    \end{pmatrix} , \quad
    \langle H_u \rangle = \frac{1}{\sqrt{2}} \begin{pmatrix} 0 \\ v_2
    \end{pmatrix} \\
    &\langle S \rangle = \frac{s_1}{\sqrt{2}} , \quad
    \langle \overline{S} \rangle = \frac{s_2}{\sqrt{2}} , \quad
    \langle \phi \rangle = \frac{\varphi}{\sqrt{2}} .
  \end{aligned} \label{eq:HiggsVEVs}
\end{equation}
The corresponding conditions for these non-zero VEVs to be a stationary point
of the potential are\footnote{Here we are using the shorthand
  $\partial V / \partial \langle \Phi \rangle \equiv \partial V / \partial
  \Phi |_{\Phi = \langle \Phi \rangle}$.},
\begin{subequations} \label{eq:ewsb-conditions}
  \begin{align}
    \frac{\partial V}{\partial v_1} &= m_{H_d}^2 v_1 - \frac{T_\lambda}
         {\sqrt{2}} s_1 v_2 + \frac{\lambda^2}{2} ( v_2^2 + s_1^2 ) v_1
         + \frac{\lambda \sigma}{2} v_2 s_2 \varphi + \frac{\bar{g}^2}{8}
         (v_1^2 - v_2^2) v_1 \nonumber \\
         & \quad {} + \frac{g_1'^2}{2} ( Q_{H_d} v_1^2 + Q_{H_u} v_2^2
         + Q_S s_1^2 - Q_S s_2^2 ) Q_{H_d} v_1
         + \frac{\partial \Delta V}{\partial v_1} = 0 , \label{eq:EWSBCond1} \\
    \frac{\partial V}{\partial v_2} &= m_{H_u}^2 v_2 - \frac{T_\lambda}
         {\sqrt{2}} s_1 v_1 + \frac{\lambda^2}{2} ( v_1^2 + s_1^2 ) v_2
         + \frac{\lambda \sigma}{2} v_1 s_2 \varphi - \frac{\bar{g}^2}{8}
         ( v_1^2 - v_2^2 ) v_2 \nonumber \\
         & \quad {} + \frac{g_1'^2}{2} (Q_{H_d} v_1^2 + Q_{H_u} v_2^2
         + Q_S s_1^2 - Q_S s_2^2 ) Q_{H_u} v_2 +
         \frac{\partial \Delta V}{\partial v_2} = 0 , \label{eq:EWSBCond2} \\
    \frac{\partial V}{\partial s_1} &= m_S^2 s_1 - \frac{T_\lambda}
         {\sqrt{2}} v_1 v_2 - \frac{T_\sigma}{\sqrt{2}} \varphi s_2
         + \frac{\sigma^2}{2} \varphi^2 s_1 + \sigma s_2 \biggl (
         \frac{\sigma}{2} s_1 s_2 - \frac{\kappa_\phi}{2} \varphi^2 -
         \frac{\mu_\phi}{\sqrt{2}} \varphi - \Lambda_F \biggr ) \nonumber \\
         & \quad {} + \frac{\lambda^2}{2}
         ( v_1^2 + v_2^2 ) s_1 + \frac{g_1'^2}{2} ( Q_{H_d} v_1^2 +
         Q_{H_u} v_2^2 + Q_S s_1^2 - Q_S s_2^2 ) Q_S
         s_1 + \frac{\partial \Delta V}{\partial s_1} = 0 ,
         \label{eq:EWSBCond3} \\
    \frac{\partial V}{\partial s_2} &= m_{\overline{S}}^2 s_2 - \frac{
         T_\sigma}{\sqrt{2}} \varphi s_1 + \frac{\sigma^2}{2} \varphi^2 s_2
         + \frac{\lambda \sigma}{2} \varphi v_1 v_2 + \sigma s_1 \biggl (
         \frac{\sigma}{2} s_1 s_2 - \frac{\kappa_\phi}{2} \varphi^2
         - \frac{\mu_\phi}{\sqrt{2}} \varphi - \Lambda_F \biggr ) \nonumber \\
         & \quad {} - \frac{g_1'^2}{2} ( Q_{H_d}
         v_1^2 + Q_{H_u} v_2^2 + Q_S s_1^2 - Q_S s_2^2
         ) Q_S s_2 + \frac{\partial \Delta V}{\partial s_2} = 0 ,
         \label{eq:EWSBCond4} \\
    \frac{\partial V}{\partial \varphi} &= m_\phi^2 \varphi -
    \frac{T_\sigma}{\sqrt{2}} s_1 s_2 + \mu_\phi B_\phi \varphi
         + \sqrt{2} \Lambda_S + \frac{T_{\kappa_\phi}}{\sqrt{2}} \varphi^2
         + \frac{\sigma^2}{2} ( s_1^2 + s_2^2 ) \varphi + \frac{\lambda
         \sigma}{2} s_2 v_1 v_2 \nonumber \\
         & \quad {} - 2 \biggl ( \frac{\sigma}{2} s_1 s_2
         - \frac{\kappa_\phi}{2} \varphi^2 - \frac{\mu_\phi}{\sqrt{2}} \varphi
         - \Lambda_F \biggr ) \biggl ( \kappa_\phi \varphi + \frac{\mu_\phi}
         {\sqrt{2}} \biggr ) + \frac{\partial \Delta V}{\partial \varphi}
         = 0 . \label{eq:EWSBCond5}
  \end{align}
\end{subequations}
Of the 14 degrees of freedom associated with this set of Higgs fields, after
EWSB four massless Goldstone modes are swallowed to generate masses for the
physical $W^\pm$, $Z$ and $Z^\prime$ bosons.  The masses of the charged gauge
bosons remain the same as in the MSSM.  The neutral gauge boson masses are
rather different, since the fields $H_u^0$ and $H_d^0$ are charged under
both $U(1)$ groups and therefore there is $Z - Z^\prime$ mixing even when gauge
kinetic mixing is neglected.  It is convenient to define the combinations of
the VEVs,
\begin{equation} \label{eq:vev-combinations}
  v^2 = v_1^2 + v_2^2 , \quad \tan\beta = \frac{v_2}{v_1} , \quad
  s^2 = s_1^2 + s_2^2 , \quad \tan\theta = \frac{s_2}{s_1} .
\end{equation}
The tree-level masses $M_{Z_1}$, $M_{Z_2}$ of the physical $Z$ and $Z^\prime$
bosons are then found by diagonalizing the squared mass matrix
\begin{equation} \label{eq:ZZprime-mass-matrix}
  M_{Z Z^\prime}^2 = \begin{pmatrix} M_Z^2 & \Delta^2 \\
  \Delta^2 & M_{Z^\prime}^2 \end{pmatrix} ,
\end{equation}
where $M_Z^2 = \bar{g}^2 v^2 /4$ and
\begin{align*}
  M_{Z^\prime}^2 &= g_1'^2 v^2 \left ( Q_{H_d}^2 \cos^2 \beta +
  Q_{H_u}^2 \sin^2 \beta \right ) + g_1'^2 Q_S^2 s^2 , \\
  \Delta^2 &= \frac{\bar{g} g_1^\prime}{2} v^2 \left ( Q_{H_d}
  \cos^2 \beta - Q_{H_u} \sin^2 \beta \right ) .
\end{align*}
The mixing between the two gauge bosons is strongly constrained by EW
precision measurements \cite{Erler:2009jh}, while LHC searches currently
place lower bounds on the mass of the extra $Z^\prime$ in $U(1)_N$ models of
$M_{Z_2} \gtrsim 3.4$ TeV \cite{ATLAS:2016cyf}.  The physical $Z^\prime$ mass
can be made acceptably large provided that the combination of the SM singlet
VEVs is large, $s \gtrsim 9$ TeV.  This leads to negligible mixing between the
physical states $Z_1$ and $Z_2$, with a mixing angle $\lesssim 10^{-4}$, so
that the light state $Z_1$ is approximately the SM $Z$ boson with $M_{Z_1}
\approx M_Z = \bar{g} v / 2$ and $v \approx 246$ GeV, while the heavier gauge
boson has its mass set by the singlet VEVs with $M_{Z_2} \approx M_{Z^\prime}
\approx g_1^\prime Q_S s$.

The presence of the singlet fields involved in EWSB means that the
set of EWSB conditions, Eq.~(\ref{eq:ewsb-conditions}), is somewhat larger than
in the MSSM.  In the MSSM, there are two such conditions, which read
\begin{subequations} \label{eq:mssm-ewsb-conditions}
  \begin{align}
    \frac{\partial V}{\partial v_1} &= \bigl ( |\mu|^2 + m_{H_d}^2
    \bigr ) v_1 + \frac{\bar{g}^2}{8} \bigl ( v_1^2 - v_2^2 \bigr ) v_1
    - \mu B v_2 + \frac{\partial \Delta V}{\partial v_1} = 0,
    \label{eq:mssm-ewsb-cond-1} \\
    \frac{\partial V}{\partial v_2} &= \bigl ( |\mu|^2 + m_{H_u}^2
    \bigr ) v_2 - \frac{\bar{g}^2}{8} \bigl ( v_1^2 - v_2^2 \bigr ) v_2
    - \mu B v_1 + \frac{\partial \Delta V}{\partial v_2} = 0 .
    \label{eq:mssm-ewsb-cond-2}
  \end{align}
\end{subequations}
Imposing the EWSB conditions allows for a subset of the model parameters to
be fixed.  Conventionally in the CMSSM, the two parameters fixed by
Eq.~(\ref{eq:mssm-ewsb-conditions}) are chosen to be $\mu$ and $B$.  However,
this choice is not unique, nor is it always the most convenient.  In particular,
when studying scenarios for dark matter, it is ideal to be able to vary
$\mu$ directly, as this controls the Higgsino masses and therefore permits the
composition of the lightest neutralino to be directly chosen.  In all of our
results below, in both models we allow $\mu_{(\text{eff})}$ to remain free and
instead fix $m_0$ using the EWSB conditions.  This can be done by expressing
the soft masses in terms of the GUT scale parameters using semi-analytic
solutions to the RGEs, as detailed in \secref{sec:results} below.  In the MSSM,
the remaining EWSB condition can be used to fix $B_0$, while in the SE$_6$SSM
there are still four conditions available.

In this paper we primarily examine the part of the parameter space where all
SUSY particles are considerably lighter than $M_{Z^\prime}$.  This corresponds
to $s_1$, $s_2$ and $\varphi$ being much larger than the SUSY breaking scale
$M_S$.  These VEVs are fixed using two of the EWSB conditions to determine
$\tan\theta$ and $\varphi$, with the value of $s$ being a free input parameter.
The remaining two conditions can be used to fix the GUT scale parameters
$\Lambda_F(M_X)$ and $\Lambda_S(M_X)$.  The appropriate stationary points of
the scalar potential in Eq.~(\ref{eq:higgs-potential}) arise if $\Lambda_F
\gg M_S^2$ and $\Lambda_S \gg M_S^3$.  In this case the structure of the
potential is further simplified if the dimensionless couplings $\kappa_{\phi}$
and $\sigma$ are small.  Then in the leading approximation the quartic part of
the scalar potential in Eq.~(\ref{eq:higgs-potential}) is just given by
\begin{equation} \label{eq:leading-quartic-term}
 \frac{g_1'^2}{2} \tilde{Q}_S^2 (|S|^2 - |\overline{S}|^2)^2
\end{equation}
so that in the limit $|\langle S \rangle|, |\langle \overline{S} \rangle |
\to \infty$ the SM singlet VEVs tend to lie approximately along the $D$-flat
direction $s_1 \approx s_2$.  The inclusion of non-zero couplings $\sigma$ and
$\kappa_{\phi}$ stabilize the potential along this direction resulting in large
SM singlet VEVs, i.e.,
\begin{equation} \label{eq:singlet-vevs-magnitude}
  |\varphi| \sim |s_1| \approx |s_2| \sim \sqrt{\frac{2 \Lambda_F}{\sigma}}\,.
\end{equation}
For the ratio of the SM singlet VEVs $s_2/s_1$ one can obtain a more accurate
estimate using the minimization conditions Eq.~(\ref{eq:EWSBCond3}) and
Eq.~(\ref{eq:EWSBCond4}),
\begin{equation} \label{eq:tree-level-tantheta}
  \tan^2\theta \simeq \frac{m_S^2 + \frac{\sigma^2}{2} \varphi^2
    + \frac{g_1'^2}{2} \tilde{Q}_S^2 s^2}{m_{\overline{S}}^2
    + \frac{\sigma^2}{2} \varphi^2 + \frac{g_1'^2}{2} \tilde{Q}_S^2 s^2} .
\end{equation}
If the VEVs of the SM singlets $\varphi$, $s_1$ and $s_2$ are rather large
due to the large values of parameters $\Lambda_F$ and $\Lambda_S$ then
$M_{Z'}\gg M_S$ and from Eq.~(\ref{eq:tree-level-tantheta}) it follows that
$\tan\theta \simeq 1$.  This is in marked difference to the situation in the
simplest variants of the E$_6$SSM where the EWSB conditions imply that
$M_{Z^\prime} \sim M_S$, forcing the SUSY spectrum to be substantially heavier
than is required, for example, in the MSSM by collider searches, due to the
large lower bound on $M_{Z^\prime}$.  In our numerical studies we take
advantage of this behavior to search for solutions with a heavy $Z^\prime$ with
a mass well above current limits and a somewhat lighter SUSY scale than
could be achieved in the simplest $E_6$ inspired extensions of the MSSM.

After fixing the parameters $m_0$, $\tan\theta$, $\varphi$, $\Lambda_F(M_X)$
and $\Lambda_S(M_X)$, the remaining parameters listed after
Eq.~(\ref{eq:soft-trilinear-bcs}) are still free, up to the constraint of
requiring a viable mass spectrum.  In our analysis we mostly focus on the
scenarios with small Yukawa couplings $\lambda$, $\sigma$,
$\tilde{\lambda}_{\alpha \beta}$, $\kappa_{ij}$, $\tilde{f}_{i\alpha}$ and
$f_{i\alpha}$ that can lead to a set of relatively light exotic fermions which
might be discovered at the LHC.  Consequently for the high-scale boundary
condition $m_S^2(M_X) = m_{\overline{S}}^2(M_X) = m_0^2$, the running of $m_S^2$
and $m_{\overline{S}}^2$ is such that at the EWSB scale $m_S^2 \simeq
m_{\overline{S}}^2$.  Thus the value of $\tan\theta$ is always extremely close
to unity.

\section{Particle Spectrum} \label{sec:particle-spectrum}
The extension of the Higgs sector responsible for the breaking of $U(1)_N$ and
EW symmetry also modifies the masses of the physical states in the spectrum
compared to those found in the simplest variants of the E$_6$SSM.  The masses
of the MSSM sfermions are almost unchanged.  The smallness of the first and
second generation Yukawa couplings leads to negligible mixing between the
left- and right-handed states, so that their masses may be summarized as
\cite{Athron:2009bs}
\begin{align}
  \left ( m_{\tilde{d}_{L\alpha}}^{\text{\DRbar}} \right )^2 &\approx
    m_{Q_{\alpha\alpha}}^2 + \left (-\frac{1}{2}
    + \frac{1}{3} \sin^2 \theta_W \right ) M_Z^2 \cos 2\beta + \Delta_Q ,
    \label{eq:left-handed-sdown-mass} \\
  \left ( m_{\tilde{d}_{R\alpha}}^{\text{\DRbar}} \right )^2 &\approx
    m_{d^c_{\alpha\alpha}}^2 -\frac{1}{3} M_Z^2 \sin^2 \theta_W \cos 2\beta
    + \Delta_{d^c} ,
    \label{eq:right-handed-sdown-mass} \\
  \left ( m_{\tilde{u}_{L\alpha}}^{\text{\DRbar}} \right )^2 &\approx
    m_{Q_{\alpha\alpha}}^2 + \left (\frac{1}{2}
    - \frac{2}{3} \sin^2 \theta_W \right ) M_Z^2 \cos 2\beta + \Delta_Q ,
    \label{eq:left-handed-sup-mass} \\
  \left ( m_{\tilde{u}_{R\alpha}}^{\text{\DRbar}} \right )^2 &\approx
    m_{u^c_{\alpha\alpha}}^2 + \frac{2}{3}
    M_Z^2 \sin^2 \theta_W \cos 2\beta + \Delta_{u^c} ,
    \label{eq:right-handed-sup-mass} \\
  \left ( m_{\tilde{e}_{L\alpha}}^{\text{\DRbar}} \right )^2 &\approx
    m_{L_{\alpha\alpha}}^2 + \left (-\frac{1}{2}
    + \sin^2 \theta_W \right ) M_Z^2 \cos 2\beta + \Delta_L ,
    \label{eq:left-handed-selectron-mass} \\
  \left ( m_{\tilde{e}_{R\alpha}}^{\text{\DRbar}} \right )^2 &\approx
    m_{e^c_{\alpha\alpha}}^2 - M_Z^2 \sin^2 \theta_W \cos 2\beta
    + \Delta_{e^c} ,
    \label{eq:right-handed-selectron-mass} \\
  \left ( m_{\tilde{\nu}_i}^{\text{\DRbar}} \right )^2 &\approx m_{L_{ii}}^2
    + \frac{1}{2} M_Z^2 \cos 2\beta + \Delta_{L} ,
    \label{eq:sneutrino-mass}
\end{align}
The only differences appear in the form of the $U(1)_N$ $D$-term contributions
$\Delta_\Phi$, which now have a contribution from the extra singlet
$\overline{S}$ and read
\begin{equation} \label{eq:d-term-contribution}
  \Delta_\Phi = \frac{g_1'^2}{2} Q_\Phi v^2 \left ( Q_{H_d}
  \cos^2 \beta + Q_{H_u} \sin^2 \beta \right ) + \frac{g_1'^2}{2}
  Q_\Phi Q_S s^2 \cos 2\theta .
\end{equation}
Compared to the E$_6$SSM, this $D$-term contribution is significantly smaller
at large $s$, due to the suppression by $\cos 2\theta$, while the sign of the
contribution to the masses remains the same.

The same is true of the third generation squarks and sleptons.  Due to the
large third generation Yukawa couplings, the left-right mixing is in general
non-negligible so that the third generation sfermion masses follow from
diagonalizing $2 \times 2$ mass matrices (in the absence of flavor off-diagonal
soft terms as considered here).  The stop, sbottom and stau masses are found to
be
\begin{align}
  \left ( m_{\tilde{t}_{1,2}}^{\text{\DRbar}} \right )^2 &= \frac{1}{2} \Bigg \{
  m_{Q_{33}}^2 + m_{u_{33}^c}^2 + \frac{1}{2} M_Z^2 \cos 2\beta
    + \Delta_{Q} + \Delta_{u^c} + 2 \left ( m_t^{\text{\DRbar}} \right )^2
    \nonumber \\
    & \quad {} \mp \sqrt{\left [ m_{Q_{33}}^2 -
        m_{u_{33}^c}^2 + \left (\frac{1}{2} - \frac{4}{3} \sin^2 \theta_W
        \right ) M_Z^2 \cos 2\beta
        + \Delta_{Q} - \Delta_{u^c} \right ]^2 + 4 X_t^2} \Bigg \} ,
    \label{eq:stop-masses} \\
  \left ( m_{\tilde{b}_{1,2}}^{\text{\DRbar}} \right )^2 &= \frac{1}{2} \Bigg \{
    m_{Q_{33}}^2 + m_{d_{33}^c}^2 - \frac{1}{2} M_Z^2 \cos 2\beta
    + \Delta_{Q} + \Delta_{d^c} + 2 \left ( m_b^{\text{\DRbar}} \right )^2
    \nonumber \\
    & \quad {} \mp \sqrt{\left [ m_{Q_{33}}^2 -
        m_{d_{33}^c}^2 + \left (-\frac{1}{2} + \frac{2}{3} \sin^2 \theta_W
        \right ) M_Z^2 \cos 2\beta
        + \Delta_{Q} - \Delta_{d^c} \right ]^2 + 4 X_b^2} \Bigg \} ,
    \label{eq:sbottom-masses} \\
  \left ( m_{\tilde{\tau}_{1,2}}^\text{\DRbar} \right )^2 &= \frac{1}{2}
    \Bigg \{ m_{L_{33}}^2 + m_{e_{33}^c}^2 - \frac{1}{2} M_Z^2 \cos 2\beta
    + \Delta_{L} + \Delta_{e^c} + 2 \left ( m_\tau^{\text{\DRbar}} \right )^2
    \nonumber \\
    & \quad {} \mp \sqrt{\left [ m_{L_{33}}^2 -
        m_{e_{33}^c}^2 + \left (-\frac{1}{2} + 2 \sin^2 \theta_W
        \right ) M_Z^2 \cos 2\beta
        + \Delta_{L} - \Delta_{e^c} \right ]^2 + 4 X_\tau^2} \Bigg \} ,
  \label{eq:stau-masses}
\end{align}
where the top, bottom and tau running masses are $m_t^{\text{\DRbar}} =
y^U_{33} v \sin \beta / \sqrt{2}$, $m_b^{\text{\DRbar}} = y^D_{33} v \cos\beta
/ \sqrt{2}$, $m_\tau^{\text{\DRbar}} = y^E_{33} v \cos\beta / \sqrt{2}$ and the
mixing parameters are given by $X_t = \frac{T^U_{33} v}{\sqrt{2}} \sin \beta
- \frac{\lambda y^U_{33} v s}{2} \cos\beta \cos\theta$, $X_b =
\frac{T^D_{33} v}{\sqrt{2}} \cos \beta - \frac{\lambda y^D_{33} v s}{2}
\sin\beta \cos\theta$ and $X_\tau = \frac{T^E_{33} v}{\sqrt{2}} \cos \beta
- \frac{\lambda y^E_{33} v s}{2} \sin\beta \cos\theta$.  Mixing between the
left- and right-handed states allows the third generation sfermions to be
lighter than their first and second generation counterparts, as usual.  To be
phenomenologically viable the squarks and sleptons cannot be too light, so that
we require the SUSY breaking scale $M_S \gtrsim 1$ TeV with $M_S \gg M_Z$.

These formulas, as well as those in the following sections, give the
running \DRbar\ masses with all parameters appearing in them evaluated at a
renormalization scale $Q$; the above formulas also assume no significant
flavor mixing.  They are useful for gaining an analytical understanding of the
spectrum, but it should be emphasized that in our numerical calculations we
make use of the general tree-level mass matrices for all states.  To calculate
the physical spectrum, we also include the full one-loop self-energy corrections
to all of the mass matrices; further details about our numerical procedure are
given in \secref{sec:results}.  Such corrections are particularly
important for accurately estimating the physical gluino mass,
\begin{equation}
  m_{\tilde{g}} = M_3(M_S) + \Delta^{\tilde{g}}(M_S) ,
\end{equation}
for which the one-loop corrections $\Delta^{\tilde{g}}$ can be quite large,
of up to $20$\%--$30$\%.  Pair production of gluinos would lead to a significant
enhancement in $p\,p \to q \bar{q} q \bar{q} + E_T^{miss} + X$, with $X$
denoting any number of light quark or gluon jets \cite{Athron:2009bs}.
This signature can be used to discover the model when
$m_{\tilde{g}}$ is within the LHC reach, or exclude regions of SE$_6$SSM
parameter space where this is the case.  As the SE$_6$SSM contains the same
colored states as in the E$_6$SSM, the form of these radiative corrections
$\Delta^{\tilde{g}}$ is unchanged between the two models.

\subsection{The Chargino and Neutralino Sector}
On the other hand, the predictions for the masses of some other remaining
states, that is, the neutralino sector, the exotic states and the Higgs
sector, are rather different in the SE$_6$SSM compared to the E$_6$SSM.  At
the same time because the supermultiplet of the $Z'$ boson and the additional
singlet superfields in the Higgs sector are electrically neutral, the fermion
components of these superfields do not mix with chargino states,
$\tilde{\chi}_{1,2}^\pm$.  Therefore the tree-level chargino mass matrix  and
its eigenvalues are almost identical to the ones in the MSSM, the only
difference being that $\mu \to \mu_{\text{eff}}$, where
\begin{equation} \label{eq:mu-eff-defn}
  \mu_{\text{eff}} = \frac{\lambda s_1}{\sqrt{2}} = \frac{\lambda s}{\sqrt{2}}
  \cos \theta .
\end{equation}
By contrast, the neutral fermion components of $\hat{H}_u$, $\hat{H}_d$,
$\hat{S}$, $\hat{\overline{S}}$ and $\hat{\phi}$ as well as the neutral
gauginos may all mix, leading to a $Z_2^E = +1$ neutralino sector that is
twice as large as the MSSM neutralino sector.  The neutralino mass
eigenstates, $\tilde{\chi}_i^0$, $i = 1, \ldots, 8$, are linear combinations
of the neutral Higgsino and singlino fields $\tilde{H}_u^0$, $\tilde{H}_d^0$,
$\tilde{S}$, $\tilde{\overline{S}}$, $\tilde{\phi}$, the bino $\tilde{B}$,
the neutral $SU(2)_L$ gaugino $\tilde{W}_3$, and the $U(1)_N$ gaugino
$\tilde{B}^\prime$, and are obtained by diagonalizing the mass matrix
\begin{equation} \label{eq:neutralino-diagonalisation}
  \mathrm{diag}(m_{\tilde{\chi}_1^0}^{\text{\DRbar}},\ldots,
  m_{\tilde{\chi}_8^0}^{\text{\DRbar}}) = N^* M_{\tilde{\chi}^0} N^{\dagger} .
\end{equation}
The $8 \times 8$ tree-level mass matrix in the basis $(\tilde{H}_d^0,
\tilde{H}_u^0, \tilde{W}_3, \tilde{B}, \tilde{B}^\prime, \tilde{S} \cos \theta
- \tilde{\overline{S}} \sin \theta, \tilde{S} \sin \theta
+ \tilde{\overline{S}} \cos \theta, \tilde{\phi})$ can be written in terms
of $4 \times 4$ sub-matrices as
\begin{equation} \label{eq:full-neutralino-mass-matrix}
M_{\tilde{\chi}^0} = \begin{pmatrix} A & C^T \\ C & B \end{pmatrix} .
\end{equation}
The upper left sub-matrix has the same structure as the neutralino mass
matrix in the MSSM with $\mu \to \mu_{\text{eff}}$,
\begin{equation} \label{eq:mssm-neutralino-submatrix}
  A = \begin{pmatrix} 0 & -\mu_{\text{eff}} & \frac{g_2 v}{2} \cos\beta
    & -\frac{g_1 v}{2} \sqrt{\frac{3}{5}} \cos\beta \\
    -\mu_{\text{eff}} & 0 & -\frac{g_2 v}{2} \sin\beta &
    \frac{g_1 v}{2} \sqrt{\frac{3}{5}} \sin\beta \\
    \frac{g_2 v}{2} \cos\beta & -\frac{g_2 v}{2} \sin\beta & M_2 & 0 \\
    -\frac{g_1 v}{2} \sqrt{\frac{3}{5}} \cos\beta & \frac{g_1 v}{2}
    \sqrt{\frac{3}{5}} \sin\beta & 0 & M_1
    \end{pmatrix} .
\end{equation}
The remaining two sub-matrices then contain the mass terms for the additional
SM singlet neutralinos and their mixings with the MSSM-like neutralino
sector,
\begin{align}
  B &= \begin{pmatrix} M_1^\prime & g_1^\prime Q_S s & 0 & 0 \\
    g_1^\prime Q_S s & \frac{\sigma \varphi}{\sqrt{2}} \sin 2\theta
    & -\frac{\sigma \varphi}{\sqrt{2}} \cos 2\theta & 0 \\
    0 & -\frac{\sigma \varphi}{\sqrt{2}} \cos 2\theta & -\frac{\sigma
      \varphi}{\sqrt{2}} \sin 2\theta & -\frac{\sigma s}{\sqrt{2}} \\
    0 & 0 & -\frac{\sigma s}{\sqrt{2}} & \mu_\phi + \sqrt{2} \kappa_\phi \varphi
  \end{pmatrix} , \label{eq:singlet-neutralino-submatrix} \\
  C &= \begin{pmatrix} Q_{H_d} g_1^\prime v \cos\beta &
    Q_{H_u} g_1^\prime v \sin\beta & 0 & 0 \\
    -\frac{\lambda v}{\sqrt{2}} \sin\beta \cos\theta & -\frac{\lambda v}
    {\sqrt{2}} \cos\beta \cos\theta & 0 & 0 \\
    -\frac{\lambda v}{\sqrt{2}} \sin\beta \sin\theta & -\frac{\lambda v}
    {\sqrt{2}} \cos\beta \sin\theta & 0 & 0 \\
    0 & 0 & 0 & 0
  \end{pmatrix} . \label{eq:mixing-neutralino-submatrix}
\end{align}
As noted above, we neglect the mixed gaugino soft mass $M_{11}$ arising from
$U(1)$ mixing.

For general values of the parameters and VEVs, the neutralino mass matrix
of the SE$_6$SSM is clearly more complicated than its counterpart in the MSSM.
In the parameter space that we consider here, however, the mass matrix has a
rather simple structure so that the MSSM-like neutralinos and the states beyond
the MSSM tend not to mix.  Inspection of
Eq.~(\ref{eq:singlet-neutralino-submatrix}) shows that two of the neutralinos,
those that are a mixture of $\tilde{B}^\prime$ and $\tilde{S} \cos\theta -
\tilde{\overline{S}} \sin\theta$, have their masses set by the large value of
$M_{Z^\prime}$.  For large values of the singlet VEV, the value of
$\mu_{\text{eff}}$ would be similarly large unless $\lambda$ is taken to be
sufficiently small.  For large values of $\mu_{\text{eff}}$ the states that are
superpositions of $\tilde{H}_u^0$ and $\tilde{H}_d^0$ become very heavy, leading
to two very heavy pure Higgsino neutralinos that cannot account for the relic
dark matter density.  Therefore we restrict our attention to small values of
$\lambda$ so that $\mu_{\text{eff}} \lesssim 1$ TeV.  When $\lambda\ll \sigma$
while $\sigma$ is rather small and $M_{Z^\prime} \gg M_S$ as implied by
Eq.~(\ref{eq:singlet-vevs-magnitude}), the aforementioned states with masses
set by $M_{Z^\prime}$ become very heavy and decouple from the rest of the
spectrum.  For very large $s$, Eq.~(\ref{eq:tree-level-tantheta}) implies
that $\tan\theta \approx 1$ to high precision, and we find this is indeed the
case in our numerical results below, allowing us to express the masses of these
states approximately as
\begin{equation} \label{eq:heaviest-neutralino-masses}
  m_{\tilde{\chi}_{7,8}^0}^{\text{\DRbar}} \approx M_{Z^\prime} \left [
    \frac{\sqrt{2} M_1^\prime + \sigma \varphi}{2 \sqrt{2} M_{Z^\prime}} \pm
    \sqrt{1 + \frac{\left (\sqrt{2} M_1^\prime - \sigma \varphi
        \right )^2}{8 M_{Z^\prime}^2}}
    \right ] \sim M_{Z^\prime} .
\end{equation}
When $M_S \gg M_Z$ and $\lambda$ is small, the mixing of the remaining extra
states, which are a mixture of $\tilde{S} \sin\theta +
\tilde{\overline{S}} \cos\theta$ and $\tilde{\phi}$, and the MSSM-like
neutralinos is also highly suppressed.  The masses of these states are then
approximately given by
\begin{equation} \label{eq:susy-scale-neutralino-masses}
  m_{\tilde{\chi}_{5,6}^0}^{\text{\DRbar}} \approx \frac{1}{2} \left [ \mu_\phi
    + \frac{\varphi}{\sqrt{2}} \left ( 2\kappa_\phi - \sigma \right )
    \pm \sqrt{2 \sigma^2 s^2 + \left (\mu_\phi + \frac{\varphi}{\sqrt{2}}
      \left (2 \kappa_\phi + \sigma \right ) \right )^2} \right ] \sim M_S .
\end{equation}
For large values of $M_S \gtrsim 1$ TeV, these states will be heavy and,
due to the lack of significant mixing, can also be ignored in the first
approximation as far as determining the mass of the DM candidate goes.
Provided this is the case, the neutralino DM candidate is expected to be
predominantly MSSM-like, that is, a mixture of $\tilde{H}_d$, $\tilde{H}_u$,
$\tilde{W}_3$ and $\tilde{B}$, with mass given by the lightest eigenvalue of
the $4 \times 4$ sub-matrix in Eq.~(\ref{eq:mssm-neutralino-submatrix}).
In particular, since this matrix is identical to the MSSM neutralino mass
matrix (with $\mu \to \mu_{\text{eff}}$),  when $M_S \gg M_Z$ the masses of the
four lightest neutralinos are determined by $\mu_{\text{eff}}$, $M_1$ and
$M_2$ as they are in the MSSM.  In the CSE$_6$SSM, the condition of universal
gaugino masses at $M_X$ further implies that
\begin{equation} \label{eq:cse6ssm-gaugino-mass-relation}
  M_1 \approx 1.1 M_1^\prime \approx 0.5 M_2 \approx 0.3 M_3
  \approx 0.2 M_{1/2} ,
\end{equation}
so that the MSSM-like neutralino sector in our case depends only on the two
parameters $\mu_{\text{eff}}$ and $M_{1/2}$.  These values can also be compared
to the relations found in the CMSSM,
\begin{equation} \label{eq:cmssm-gaugino-mass-relation}
  M_1 \approx 0.5 M_2 \approx 0.15 M_3 \approx 0.4 M_{1/2} ,
\end{equation}
which are quite different due to the modified RG flow in the SE$_6$SSM.

\subsection{The Exotic Sector}
The states that are odd under $Z_2^E$ do not mix with the ordinary MSSM
states or the Higgs fields, forming a separate sector containing the
second DM candidate as well as additional exotic states, some of which may
generate spectacular collider signals.  As discussed above, the DM candidate
in this sector is expected to be an almost massless inert singlino, which is
the lightest of the inert neutralinos.  The inert neutralino sector is formed
by the fermion components ($\tilde{S}_i$, $\tilde{H}_\alpha^u$ and
$\tilde{H}_\alpha^d$) of the superfields $\hat{S}_i$, $\hat{H}_\alpha^u$ and
$\hat{H}_\alpha^d$.  The scalar components of the corresponding superfields also
mix to form a set of inert charged and neutral Higgs scalars.  The general
inert neutralino and neutral inert Higgs mass matrices are $7 \times 7$
matrices.  In the basis $((\tilde{H}^{d,0}_1 + \tilde{H}^{u,0}_1)/\sqrt{2},
(\tilde{H}^{u,0}_1 - \tilde{H}^{d,0}_1)/\sqrt{2}, (\tilde{H}^{d,0}_2 +
\tilde{H}^{u,0}_2)/\sqrt{2}, (\tilde{H}^{u,0}_2 - \tilde{H}^{d,0}_2)/\sqrt{2},
\tilde{S}_1, \tilde{S}_2, \tilde{S}_3)$, the inert neutralino mass matrix is of
the form
\begin{equation} \label{eq:inert-neutralino-mass-matrix}
  M_{\tilde{\chi}_I^0} = \begin{pmatrix} A_I & C_I^T \\ C_I & 0 \end{pmatrix} ,
\end{equation}
where
\begin{equation} \label{eq:inert-higgsino-masses}
  A_I = \mathrm{diag} \left ( -\mu_{\tilde{H}_{I1}^0}, \mu_{\tilde{H}_{I1}^0},
  -\mu_{\tilde{H}_{I2}^0}, \mu_{\tilde{H}_{I2}^0} \right )
\end{equation}
contains the tree-level masses of the inert Higgsinos,
$\mu_{\tilde{H}_{I\alpha}^0} = \tilde{\lambda}_{\alpha \alpha} s \cos\theta /
\sqrt{2}$, in the absence of mixing with the inert singlinos, while the mixing
is given by the $3 \times 4$ sub-matrix $C_I$ with elements
\begin{equation}
  \begin{aligned}
    (C_I)_{i1} &= \frac{v}{2} \left ( f_{i1} \cos\beta + \tilde{f}_{i1}
    \sin\beta \right ), \quad (C_I)_{i2} = \frac{v}{2} \left ( f_{i1}
    \cos\beta - \tilde{f}_{i1} \sin\beta \right ) \\
    (C_I)_{i3} &= \frac{v}{2} \left ( f_{i2} \cos\beta + \tilde{f}_{i2}
    \sin\beta \right ), \quad (C_I)_{i4} = \frac{v}{2} \left ( f_{i2}
    \cos\beta - \tilde{f}_{i2} \sin\beta \right ) . \\
  \end{aligned} \label{eq:inert-singlino-mixing}
\end{equation}
The couplings of the inert singlinos are required to satisfy $f_{i\alpha},
\tilde{f}_{i\alpha} \lesssim 10^{-6}$ to yield almost massless hot DM
candidates.  Then, provided that $\tilde{\lambda}_{\alpha \beta}
\gtrsim 10^{-6}$, the mixing between the inert Higgsinos and the inert singlinos
is entirely negligible, and the inert neutralinos correspond to two degenerate
pairs of inert Higgsinos with tree-level masses given by
Eq.~(\ref{eq:inert-higgsino-masses}) and three almost massless inert singlinos.
The inert charginos similarly have tree-level masses given by
$\mu_{\tilde{H}_{I\alpha}^\pm} = |\mu_{\tilde{H}_{I\alpha}^0}|$.

When the couplings $f_{i\alpha}$, $\tilde{f}_{i\alpha}$ are negligibly small,
the mass matrix associated with the scalar components of the superfields
$\hat{S}_i$, $\hat{H}_\alpha^u$ and $\hat{H}_\alpha^d$ also simplifies in a
similar fashion.  In this case, the mixing between the neutral inert Higgs
scalars ($H_\alpha^u$ and $H_\alpha^d$) and the inert singlets $S_i$ can be
ignored and the corresponding mass matrix decomposes into a
$3 \times 3$ singlet mass matrix and a $4 \times 4$ mass matrix for the inert
Higgs scalars\footnote{Strictly speaking, for non-zero $f$ and
  $\tilde{f}$ couplings, the inert neutral Higgs sector should actually be
  decomposed into CP--eigenstates.  This leads to 7 CP--even scalars and 7
  CP--odd scalars.  When the couplings $f_{i\alpha}$ and $\tilde{f}_{i\alpha}$
  are neglected, these states instead form 7 complex scalar mass
  eigenstates described by the mentioned $3 \times 3$ and $4 \times 4$ mass
  matrices.}.  The family-diagonal structure of the couplings
$\tilde{\lambda}_{\alpha \beta}$, as well as the fact that the off-diagonal soft
scalar masses vanish at the GUT scale, ensures that the mixing between
generations is very small.  Thus the mass matrix for the inert singlets is
approximately diagonal, with the tree-level masses for the inert singlet
scalars given by
\begin{equation} \label{eq:inert-singlet-masses}
  \left ( m_{S_{Ii}}^{\text{\DRbar}} \right )^2 = m_{\Sigma_{ii}}^2
  + \Delta_{S_i} .
\end{equation}
For $\tan\theta \approx 1$, the inert singlet masses are therefore $\sim M_S$,
and so are somewhat lighter than $M_{Z^\prime}$.  In the absence of generation
mixing, the inert Higgs mass matrix can be written as two $2 \times 2$ matrices,
yielding the tree-level masses
\begin{align}
  \left (m_{H_{\alpha_{1,2}}^0}^{\text{\DRbar}} \right )^2 &= \frac{1}{2}
  \Bigg [ m_{H_{1,\alpha\alpha}}^2 + m_{H_{2,\alpha\alpha}}^2
    + \Delta_{H_\alpha^d} + \Delta_{H_\alpha^u}
    + 2 \mu_{\tilde{H}_{I\alpha}^0}^2 \nonumber \\
    & \quad {} \mp \sqrt{\left ( m_{H_{1,\alpha\alpha}}^2 -
      m_{H_{2,\alpha\alpha}}^2 + M_Z^2 \cos 2\beta
      + \Delta_{H_\alpha^d} - \Delta_{H_\alpha^u} \right )^2
      + 4 X_{H_\alpha}^2} \Bigg ] ,
  \label{eq:inert-neutral-higgs-masses}
\end{align}
where $X_{H_\alpha} = \frac{T^{\tilde{\lambda}}_
  {\alpha \alpha} s}{\sqrt{2}} \cos\theta- \frac{
  \tilde{\lambda}_{\alpha\alpha}}{4} \left ( \lambda v^2 \sin 2\beta
+ 2 \sigma \varphi s \sin\theta \right )$.
The masses of the inert charged Higgs states are likewise
\begin{align}
  \left ( m_{H_{\alpha_{1,2}}^\pm}^{\text{\DRbar}} \right )^2 &= \frac{1}{2}
    \Bigg [ m_{H_{1,\alpha\alpha}}^2 + m_{H_{2,\alpha\alpha}}^2
    + \Delta_{H_\alpha^d} + \Delta_{H_\alpha^u}
    + 2 \mu_{\tilde{H}_{I\alpha}^\pm}^2 \nonumber \\
  & \quad {} \mp \sqrt{ \left ( m_{H_{1,\alpha\alpha}}^2 -
    m_{H_{2,\alpha\alpha}}^2 -  M_Z^2 \cos 2\theta_W \cos 2\beta
    + \Delta_{H_\alpha^d} - \Delta_{H_\alpha^u} \right )^2
    + 4 X_{H_\alpha}^2} \Bigg ] . \label{eq:inert-charged-higgs-masses}
\end{align}
The contribution to the mixing proportional to $\sigma \varphi s
\sim M_S M_{Z^\prime}$ can be of the order of the soft mass contributions
to the masses.  To prevent this potentially dangerous term from causing
tachyonic states, the inert Higgs couplings $\tilde{\lambda}_{\alpha \beta}$
cannot be too large.  In practice, in our numerical study we take these
couplings to be not much larger than $\lambda$, e.g.,
$\tilde{\lambda}_{\alpha \beta} \sim 10^{-3}$, to satisfy this
requirement.  Doing so implies that the mixing is rather small so that the
inert scalars tend to have masses of order $M_S$.  At the same time, small
values of the Yukawas $\tilde{\lambda}_{\alpha \beta}$ imply that the inert
Higgsinos and charginos can be light, with masses not much heavier than the
lightest $Z_2^E = +1$ neutralino, in which case they may be observable in LHC
searches.  The exact $\tilde{Z}_2^H$ symmetry forbids the Yukawa couplings of
the inert Higgs and singlet superfields to ordinary quark and lepton
superfields.  In the $E_6$ models with only an approximate $Z_2$ symmetry
responsible for suppressing FCNCs, such couplings in general are permitted
along with those for the ordinary Higgs fields, leading to the inert Higgsinos
and charginos decaying predominantly into third generation fermion-sfermion
pairs \cite{King:2005jy}.  The absence of these couplings in the SE$_6$SSM
due to $\tilde{Z}_2^H$ symmetry means that the decay channels of the inert
Higgsinos are rather different in this model.  Pair production of the
$Z_2^E$ and $R$--parity odd inert Higgsinos and charginos can occur through
off-shell $W$ and $Z$ bosons.  They then decay into an inert singlino and
an on-shell $W$ or $Z$ boson, or a $Z_2^E$ even Higgs boson, through the
mixing induced by the $f_{i\alpha}$ and $\tilde{f}_{i\alpha}$ superpotential
couplings.  When both of the produced states decay into gauge bosons it is
expected that they should lead to enhancements in the rates of $p\,p \to Z\,Z
+ E_T^{miss} + X$, $p\,p \to W\,Z + E_T^{miss} + X$ and $p\,p \to W\,W +
E_T^{miss} + X$.

The choice of flavor diagonal couplings $\kappa_{ij}$ also means that there is
no substantial mixing between generations of the exotic leptoquarks,
$D_i$ and $\overline{D}_i$.  The $6 \times 6$ mass matrix for the scalar
leptoquarks reduces to three $2 \times 2$ matrices, giving the tree-level
masses
\begin{align}
  \left ( m_{\tilde{D}_{i1,2}}^{\text{\DRbar}} \right )^2& = \frac{1}{2}
    \Bigg [ m_{D_{ii}}^2 + m_{\overline{D}_{ii}}^2 + \Delta_D
    + \Delta_{\overline{D}} + 2 \mu_{D_i}^2 \nonumber \\
    & \quad {} \mp \sqrt{\left ( m_{D_{ii}}^2 - m_{\overline{D}_{ii}}^2 +
      \frac{2}{3} M_Z^2 \sin^2 \theta_W \cos 2\beta + \Delta_D -
      \Delta_{\overline{D}} \right )^2 + 4 X_{D_i}^2 } \Bigg ] ,
  \label{eq:scalar-leptoquark-masses}
\end{align}
where $X_{D_i} = \frac{T^\kappa_{ii} s}{\sqrt{2}} \cos\theta
- \frac{\kappa_{ii}}{4} \left ( \lambda v^2 \sin 2\beta + 2\sigma\varphi s
\sin\theta \right )$ and the corresponding spin-$1/2$ leptoquark
masses are $\mu_{D_i} = \kappa_{ii} s \cos\theta / \sqrt{2}$.  The same
potentially dangerous contribution to the mixing that occurs in the inert Higgs
mass matrices is also present here.  To ensure that this does not lead to an
instability of the physical vacuum, we require the couplings $\kappa_{ij}$ to
be small as well, $\kappa_{ij} \sim 10^{-3}$.  As is the case for the inert
Higgs states, this leads to the scalar leptoquark $\tilde{D}_i$ being heavier,
with masses of the order of $M_S$, while the exotic fermions $D_i$ can be
light.  These exotic fermion states are colored and, once past threshold, can be
pair produced at the 13 TeV LHC.  They subsequently decay with missing energy
via a decay chain involving an initial decay into an ordinary squark (quark) and
an exotic $L_4$ fermion (scalar) component, through the couplings $g_{ij}^D$.
This is followed by a decay involving the couplings $h^E_{i\alpha}$ of the
exotic $L_4$ state into a lepton and inert Higgs or singlet (inert neutralino).
If a hierarchy exists in the sizes of the couplings $g_{ij}^D$ and
$h^E_{i\alpha}$ as is present in the SM Yukawas, then such a process leads to an
enhancement in signals with third generation final states, namely in
$p\,p \to t\,\bar{t}\,\tau^+\,\tau^-+E_t^{miss}+X$ and $p\,p \to b\,\bar{b}\,
\tau^+\,\tau^- + E_T^{miss} + X$.

For the branching ratio of these leptoquark decays to be significant, and also
for the lifetimes of the exotic leptoquarks to be sufficiently short, the states
associated with $\hat{L}_4$ and $\hat{\overline{L}}_4$ should not be too heavy.
The fermion and scalar components of $\hat{L}_4$ and $\hat{\overline{L}}_4$ form
a set of exotic lepton and slepton states that do not mix with the other exotic
fields.  The fermion components lead to a pair of charged and neutral states
$\tilde{L}_4^\pm$ and $\tilde{L}_{4,1}^0$, $\tilde{L}_{4,2}^0$ with degenerate
tree-level masses given by
\begin{equation} \label{eq:l4-fermion-mass}
  \mu_{\tilde{L}_4^\pm} = \mu_{\tilde{L}_4^0} = \mu_L -
  \frac{\tilde{\sigma} \varphi}{\sqrt{2}} .
\end{equation}
The tree-level masses of the neutral exotic sleptons are given by
\begin{align}
  \left ( m_{L_{4_{1,2}}^0}^{\text{\DRbar}} \right )^2 &= \frac{1}{2} \Bigg [
    m_{L_4}^2 + m_{\overline{L}_4}^2 + \Delta_{L_4} + \Delta_{\overline{L}_4}
    + 2 \mu_{\tilde{L}_4^0}^2 \nonumber \\
  & \quad {} \mp \sqrt{\left (m_{L_4}^2 - m_{\overline{L}_4}^2 +
    M_Z^2 \cos 2\beta + \Delta_{L_4} - \Delta_{\overline{L}_4} \right )^2
    + 4 X_{L_4}^2} \Bigg ] , \label{eq:l4-neutral-scalar-mass}
\end{align}
where the mixing parameter is
\begin{equation}
  X_{L_4} = \mu_L B_L - \frac{T_{\tilde{\sigma}} \varphi}{\sqrt{2}} +
  \tilde{\sigma} \left ( \frac{\sigma}{4} s^2 \sin 2\theta -
  \frac{\kappa_\phi}{2} \varphi^2 - \frac{\mu_\phi}{\sqrt{2}} \varphi
  - \Lambda_F \right ) ,
\end{equation}
while those for the charged exotic sleptons read
\begin{align}
  \left ( m_{L_{4_{1,2}}^\pm}^{\text{\DRbar}} \right )^2 &= \frac{1}{2}
    \Bigg [ m_{L_4}^2 + m_{\overline{L}_4}^2 + \Delta_{L_4}
    + \Delta_{\overline{L}_4} + 2 \mu_{\tilde{L}_4^\pm}^2
    \nonumber \\
  & \quad {} \mp \sqrt{\left (m_{L_4}^2 - m_{\overline{L}_4}^2 -
    M_Z^2 \cos 2\theta_W \cos 2\beta + \Delta_{L_4}
    - \Delta_{\overline{L}_4} \right )^2
    + 4 X^2_{L_4}} \Bigg ] . \label{eq:l4-charged-scalar-mass}
\end{align}
By tuning the above mixing parameter, the exotic sleptons could be made light
enough so that the exotic $D$ fermions decay rapidly enough.  Alternatively,
these states are allowed to be heavier than the spin-1/2 leptoquarks
provided that the couplings $g^D$ and $h^E$ are taken to be sufficiently
large.  In the numerical results below, we find that taking values for these
couplings of $\sim 10^{-2}$ lead to lifetimes of the exotic fermions short
enough to be consistent with constraints from Big Bang nucleosynthesis.  At
the same time, the impact of the couplings $g^D$ and $h^E$ on the mass spectrum
and DM predictions is negligible for these small values of the couplings.
Consequently they may be safely varied in this range without having any
substantial impact on the other sectors.

\subsection{The Higgs Sector}
The Higgs sector of the SE$_6$SSM is substantially different from the simplest
version of the E$_6$SSM, for which the spectrum of the Higgs bosons was explored
in Ref.~\cite{King:2005jy}.  In the simplest case the sector responsible
for the breakdown of the $SU(2)_L\times U(1)_Y\times U(1)_{N}$ gauge
symmetry includes just $H_u$, $H_d$ and $S$ resulting in three CP--even, one
CP--odd and two charged states.  One CP--even Higgs state, which is
predominantly SM singlet field, is always almost degenerate with the $Z'$ gauge
boson.  The qualitative pattern of the Higgs spectrum in the simplest variant
of the E$_6$SSM depends on the coupling $\lambda$ which is a coupling of
the SM singlet superfield $\hat{S}$ to the Higgs doublets $\hat{H}_u$ and
$\hat{H}_d$, i.e., $\lambda \hat{S} \hat{H}_u \hat{H}_d$, as in the SE$_6$SSM.
If $\lambda < g'_1$ the singlet dominated CP--even state is very heavy and
decouples which makes the rest of the Higgs spectrum indistinguishable from
the one in the MSSM.  When $\lambda\gtrsim g'_1$ the spectrum of the Higgs
bosons has a very hierarchical structure, which is similar to the one that
appears in the NMSSM with the approximate Peccei-Quinn (PQ) symmetry
\cite{Miller:2003ay,Miller:2003hm,Miller:2005qua,King:2014xwa,Nevzorov:2004ge}.
As a result the mass matrix of the CP--even Higgs sector can be diagonalized
using the perturbation theory \cite{Nevzorov:2004ge,Kovalenko:1998dc,
  Nevzorov:2000uv,Nevzorov:2001um}.  In this case the heaviest CP--even,
CP--odd and charged states are almost degenerate and lie beyond the multi-TeV
range whereas the mass of the second lightest CP--even Higgs state is set by
the $Z'$ boson mass.

As was mentioned before in the SE$_6$SSM the sector responsible for the
breakdown of gauge symmetry involves five multiplets of scalar fields  $H_u$,
$H_d$, $S$, $\overline{S}$ and $\phi$ that give rise to ten physical degrees of
freedom in the Higgs sector which form a set of charged and neutral Higgs
bosons.  The unbroken $U(1)_{\text{em}}$ symmetry ensures that the charged
components of $H_u$ and $H_d$ do not mix with the other Higgs and singlet
fields.  Two massive charged Higgs states are formed by the linear combination
\begin{equation} \label{eq:charged-higgs-state}
  H^+ = H_d^{-*} \sin\beta + H_u^+ \cos\beta ,
\end{equation}
with a mass given by
\begin{equation} \label{eq:charged-higgs-mass}
  \left ( m_{H^\pm}^{\text{\DRbar}} \right )^2 = \frac{\sqrt{2} s}{\sin 2\beta}
    \left ( T_\lambda \cos\theta - \frac{\lambda \sigma \varphi}{\sqrt{2}}
    \sin\theta \right ) - \frac{\lambda^2}{2} v^2 + \frac{g_2^2}{4} v^2 .
\end{equation}
The linear combination orthogonal to Eq.~(\ref{eq:charged-higgs-state})
constitutes the longitudinal degrees of freedom of the $W^\pm$ bosons.

In the absence of CP--violation in the Higgs sector, the real and imaginary
parts of the neutral components of the Higgs and singlets fields do not
mix, leading to three physical CP--odd Higgs bosons and five CP--even states.
The Goldstone states that are absorbed by the $Z$ and $Z^\prime$ bosons are
mixtures of the imaginary parts of $H_d^0$, $H_u^0$, $S$ and $\overline{S}$,
\begin{equation}
  \begin{aligned}
    G &= \sqrt{2} ( \Imag H_d^0 \cos \beta - \Imag H_u^0 \sin\beta ) , \\
    G^\prime &= \sqrt{2} (\Imag S \cos \theta - \Imag \overline{S} \sin \theta )
    \cos \gamma - \sqrt{2} ( \Imag H_u^0 \cos\beta + \Imag H_d^0 \sin \beta )
    \sin \gamma ,
  \end{aligned} \label{eq:pseudoscalar-goldstones}
\end{equation}
where
\begin{equation} \label{eq:gamma-definition}
  \tan \gamma = \frac{v}{2 s}\sin 2\beta .
\end{equation}
For phenomenologically viable scenarios with $s \gg v$, $\tan\gamma$ goes
to zero.  Expressed in terms of the field basis $(P_1, P_2, P_3)$, where
\begin{equation}
  \begin{aligned}
    P_1 &= \sqrt{2} (\Imag H_u^0 \cos\beta + \Imag H_d^0 \sin\beta ) \cos \gamma
    + \sqrt{2} (\Imag S \cos\theta - \Imag \overline{S} \sin\theta )
    \sin \gamma, \\
    P_2 &= \sqrt{2} (\Imag S \sin\theta + \Imag \overline{S} \cos\theta), \\
    P_3 &= \sqrt{2} \Imag \phi ,
  \end{aligned} \label{eq:pseudoscalar-basis}
\end{equation}
the pseudoscalar mass matrix $\tilde{M}^2$ has elements
\begin{equation}
  \begin{aligned}
    \tilde{M}_{11}^2 &= \frac{\sqrt{2} s}{\sin 2\beta \cos^2 \gamma}
    \left ( T_\lambda \cos\theta - \frac{\lambda \sigma \varphi}{\sqrt{2}}
    \sin\theta \right ) , \\
    \tilde{M}_{12}^2 &= \tilde{M}_{21}^2 = \frac{v}{\sqrt{2} \cos\gamma}
    \left ( T_\lambda \sin\theta + \frac{\lambda \sigma \varphi}{\sqrt{2}}
    \cos\theta \right ) , \\
    \tilde{M}_{13}^2 &= \tilde{M}_{31}^2 = \frac{\lambda \sigma v s}{2
      \cos\gamma} \sin\theta , \\
    \tilde{M}_{22}^2 &= \frac{2 \sigma \varphi}{\sin 2\theta} \left (
    \frac{\kappa_\phi}{2} \varphi + \frac{\mu_\phi}{\sqrt{2}}
    + \frac{\Lambda_F}{\varphi} \right ) + \frac{v^2 \sin 2\beta}{\sqrt{2}
      s \sin 2\theta} \left ( T_\lambda \sin^3 \theta - \frac{\lambda
      \sigma \varphi}{\sqrt{2}} \cos^3\theta \right ) \\
    & \quad {} + \frac{\sqrt{2} T_\sigma \varphi}{\sin 2\theta} , \\
    \tilde{M}_{23}^2 &= \tilde{M}_{32}^2 = \frac{T_\sigma s}{\sqrt{2}} -
    \sigma s \left ( \kappa_\phi \varphi + \frac{\mu_\phi}{\sqrt{2}} \right ) -
    \frac{\lambda \sigma}{4} v^2 \sin 2\beta \cos\theta , \\
    \tilde{M}_{33}^2 &= \frac{T_\sigma s^2}{2\sqrt{2} \varphi} \sin 2\theta
    - 2 \mu_\phi B_\phi - \frac{3 T_{\kappa_\phi}}{\sqrt{2}} \varphi
    - \frac{\sqrt{2}}{\varphi} \left (\mu_\phi \Lambda_F + \Lambda_S \right )
    + \sigma \kappa_\phi s^2 \sin 2\theta - \frac{\kappa_\phi \mu_\phi}
    {\sqrt{2}} \varphi \\
    & \quad {} - 4 \kappa_\phi \Lambda_F + \frac{\sigma \mu_\phi s^2}{2\sqrt{2}
      \varphi} \sin 2\theta - \frac{\lambda \sigma s}{4\varphi} v^2 \sin\theta
    \sin 2\beta .
  \end{aligned} \label{eq:pseudoscalar-mass-matrix}
\end{equation}
In the parameter space of interest here, the structure of the full
$3 \times 3$ matrix is such that it can be approximately diagonalized
analytically.  Because $M_{Z^\prime}, M_S \gg M_Z$ and we restrict our attention
to small values of $\lambda$, the mixings between $P_1$ and $P_2$, $P_3$ are
rather small and may be safely neglected.  In this approximation,
the mass of one CP--odd state is set by $\tilde{M}_{11}$. Thus it has almost
the same mass as the charged Higgs states. The masses of two other
CP--odd states are set by $\tilde{m}_{+}$ and $\tilde{m}_{-}$
which are given by
\begin{equation}
\begin{aligned}
 \left ( \tilde{m}_{\pm}^{\text{\DRbar}} \right )^2 \approx
\frac{1}{2} \left \{\tilde{M}_{22}^2 + \tilde{M}_{33}^2 \pm
\sqrt{(\tilde{M}_{22}^2 - \tilde{M}_{33}^2)^2 + 4 \tilde{M}_{23}^4}
\right \} .
\end{aligned}
\label{eq:heavy-pseudoscalar-masses}
\end{equation}
As follows from Eq.~(\ref{eq:heavy-pseudoscalar-masses}) in some cases
$\tilde{m}_{-}$ can be rather small so that the lightest CP--odd state $A_1$
becomes the lightest particle in the spectrum. This happens, for example, in
the limit $\kappa_\phi, \mu_\phi, \Lambda_F, \Lambda_S \to 0$, when
$m_{A_1}^{\text{\DRbar}}$ vanishes and the superpotential possesses a global
$U(1)_{PQ}$ PQ symmetry which is spontaneously broken by the
VEVs $s_1$, $s_2$ and $\varphi$.  For small but non-vanishing $U(1)_{PQ}$
violating couplings, the state $A_1$ is a light pseudo-Goldstone boson of the
approximate PQ symmetry and can be lighter than the SM-like Higgs.  In this
case, the decay $h_1 \to A_1\,A_1$ is kinematically allowed and
can in principle lead to non-negligible branching fractions for non-standard
decays of the SM Higgs \cite{Athron:2014pua}.  Even for larger values of the
couplings $\kappa_\phi$, $\mu_\phi$, $\Lambda_F$ and $\Lambda_S$, $m_{A_1}$ may
be small provided that the remaining parameters in
Eq.~(\ref{eq:heavy-pseudoscalar-masses}) are tuned so that $\tilde{m}_{-}\to 0$.
It is important to note that in either case, the vanishing of
$m_{A_1}^{\text{\DRbar}}\approx \tilde{m}_{-}$ does not also require that the
lightest neutralino mass becomes small, as occurs for example in the
PQ-symmetric NMSSM.  Indeed, from Eq.~(\ref{eq:heaviest-neutralino-masses}) and
Eq.~(\ref{eq:susy-scale-neutralino-masses}) it is clear that the singlino
dominated states should remain heavy, while $m_{\tilde{\chi}_1^0}$ is governed
by the values of the gaugino masses and $\mu_{\text{eff}}$.  This means that by
varying the other Lagrangian parameters for fixed $M_{1/2}$ and
$\mu_{\text{eff}}$, the value of $m_{A_1}$ can be chosen independently of
$m_{\tilde{\chi}_1^0}$.  In particular, for a given $m_{\tilde{\chi}_1^0}$ this
allows for the possibility of resonant annihilations
$\tilde{\chi}_1^0\, \tilde{\chi}_1^0 \to A_1\to \mbox{\it SM particles}$
with $m_{A_1} \approx 2 m_{\tilde{\chi}_1^0}$, leading
to regions of parameter space in which the well-known $A$-funnel mechanism
is responsible for setting the DM relic density \cite{Drees:1992am,
  Roszkowski:2001sb,Djouadi:2005dz}.

The real parts of $H_d^0$, $H_u^0$, $S$, $\overline{S}$ and $\phi$ form
five physical CP--even Higgs states, $h_i$, related by the unitary
transformation
\begin{equation}\label{eq:Hmixing}
  \begin{pmatrix}
    h_1 \\ h_2 \\ h_3 \\ h_4 \\ h_5
  \end{pmatrix} =
  U_h \begin{pmatrix}
    \Real H_d^0 \\ \Real H_u^0 \\ \Real S \\ \Real \overline{S} \\ \Real \phi
  \end{pmatrix} ,
\end{equation}
where $U_h$ diagonalizes the CP--even Higgs mass matrix, $M^2$.  In the basis
$(S_1, S_2, S_3, S_4, S_5)$, where
\begin{equation}
  \begin{aligned}
    \sqrt{2} \Real S &= S_1 \cos \theta + S_2 \sin \theta + s_1 , \\
    \sqrt{2} \Real \overline{S} &= -S_1 \sin\theta + S_2 \cos\theta
    + s_2  , \\
    \sqrt{2} \Real \phi &= S_3 + \varphi , \\
    \sqrt{2} \Real H_d^0 &= S_5 \cos\beta - S_4 \sin\beta + v_1 , \\
    \sqrt{2} \Real H_u^0 &= S_5 \sin\beta + S_4 \cos\beta + v_2 ,
  \end{aligned} \label{eq:scalar-basis}
\end{equation}
and using the EWSB conditions Eq.~(\ref{eq:ewsb-conditions}) to
eliminate the soft Higgs masses, this has elements
\begin{align}
  M_{11}^2 &= g_1'^2 Q_S^2 s^2 - \frac{\sigma^2 s^2}{2}
  \sin^2 2\theta + \sqrt{2} T_\sigma \varphi \sin 2\theta +
  \left ( \kappa_\phi \sigma \varphi^2 + \sqrt{2} \sigma \mu_\phi \varphi
  + 2 \sigma \Lambda_F \right ) \sin 2\theta \nonumber \\
  & \quad {} + \frac{T_\lambda}
  {2\sqrt{2} s} v^2 \cos\theta \sin 2\beta - \frac{\lambda \sigma
    \varphi}{4s} v^2 \sin\theta \sin 2\beta , \nonumber \\
  M_{12}^2 &= M_{21}^2 = \frac{\sigma^2 s^2}{4} \sin 4\theta -
  \sqrt{2} T_\sigma \varphi \cos 2\theta - \left ( \kappa_\phi \sigma \varphi^2
  + \sqrt{2} \sigma \mu_\phi \varphi + 2 \sigma \Lambda_F \right ) \cos 2\theta
  \nonumber \\
  & \quad {} + \frac{T_\lambda}{2\sqrt{2} s} v^2 \sin\theta \sin 2\beta +
  \frac{\lambda \sigma \varphi}{4s} v^2 \cos\theta \sin 2\beta , \nonumber\\
  M_{13}^2 &= M_{31}^2 = \sigma^2 \varphi s \cos 2\theta -
  \frac{\lambda \sigma}{4} v^2 \sin\theta \sin 2\beta ,
  \nonumber \\
  M_{14}^2 &= M_{41}^2 = \frac{g_1'^2}{2} Q_S (Q_{H_u}
  - Q_{H_d}) s v \sin 2\beta - \frac{T_\lambda}{\sqrt{2}} v
  \cos\theta \cos 2\beta \nonumber \\
  & \quad {} - \frac{\lambda \sigma}{2} \varphi v \sin\theta
  \cos 2\beta , \nonumber \\
  M_{15}^2 &= M_{51}^2 = g_1'^2 Q_S ( Q_{H_d} \cos^2\beta
  + Q_{H_u} \sin^2\beta) s v - \frac{T_\lambda}{\sqrt{2}} v
  \cos\theta \sin 2\beta + \lambda^2 v s \cos^2\theta \nonumber \\
  & \quad {} - \frac{\lambda \sigma}{2} \varphi v \sin\theta \sin 2\beta
  , \nonumber \\
  M_{22}^2 &= \frac{\sigma^2 s^2}{2} \sin^2 2\theta + \frac{\sqrt{2}
    T_\sigma \varphi}{\sin 2\theta} \cos^2 2\theta + \left ( \kappa_\phi
  \sigma \varphi^2 + \sqrt{2} \sigma \mu_\phi \varphi + 2 \sigma
  \Lambda_F \right )\frac{\cos^2 2\theta}{\sin 2\theta} \nonumber \\
  & \quad {} + \frac{T_\lambda v^2}{2\sqrt{2} s \cos\theta}\sin^2\theta
  \sin 2\beta - \frac{\lambda \sigma \varphi v^2}{4 s \sin\theta} \cos^2\theta
  \sin 2\beta , \nonumber \\
  M_{23}^2 &= M_{32}^2 = -\frac{T_\sigma}{\sqrt{2}} s + \sigma^2 \varphi s
  \sin 2\theta - \sigma s \left (\kappa_\phi \varphi + \frac{\mu_\phi}{\sqrt{2}}
  \right ) \label{eq:scalar-mass-matrix} \\
  & \quad {} + \frac{\lambda \sigma}{4} v^2 \cos\theta \sin 2\beta ,
  \nonumber \\
  M_{24}^2 &= M_{42}^2 = \left (-\frac{T_\lambda}{\sqrt{2}} v \sin\theta
  + \frac{\lambda \sigma}{2} \varphi v \cos \theta \right )
  \cos 2\beta , \nonumber \\
  M_{25}^2 &= M_{52}^2 = \frac{\lambda^2}{2} s v \sin 2\theta +
  \left (-\frac{T_\lambda}{\sqrt{2}} v \sin \theta + \frac{\lambda
    \sigma}{2} \varphi v \cos\theta \right )\sin 2\beta ,
  \nonumber \\
  M_{33}^2 &= \frac{T_\sigma s^2}{2\sqrt{2} \varphi} \sin 2\theta -
  \sqrt{2} \frac{\Lambda_S}{\varphi} + \frac{T_{\kappa_\phi}}{\sqrt{2}} \varphi
  + \mu_\phi \left (\frac{\sigma s^2}{2\sqrt{2} \varphi} \sin 2\theta
  + 3\frac{\kappa_\phi \varphi}{\sqrt{2}} - \frac{\sqrt{2} \Lambda_F}{\varphi}
  \right ) \nonumber \\
  & \quad {} + 2\kappa_\phi^2 \varphi^2 - \frac{\lambda \sigma s}{4 \varphi}
  v^2 \sin\theta \sin 2\beta , \nonumber \\
  M_{34}^2 &= M_{43}^2 = \frac{\lambda \sigma}{2} s v \sin\theta \cos 2\beta
  , \nonumber \\
  M_{35}^2 &= M_{53}^2 = \frac{\lambda \sigma}{2} s v \sin\theta \sin 2\beta
  , \nonumber \\
  M_{44}^2 &= \frac{\sqrt{2} s}{\sin 2\beta} \left ( T_\lambda \cos\theta
  - \frac{\lambda \sigma \varphi}{\sqrt{2}} \sin\theta \right ) +
  \left ( \frac{\bar{g}^2}{4} - \frac{\lambda^2}{2} \right ) v^2
  \sin^2 2\beta \nonumber \\
  & \quad {} + \frac{g_1'^2}{4} ( Q_{H_u} - Q_{H_d})^2 v^2
  \sin^2 2\beta , \nonumber \\
  M_{45}^2 &= M_{54}^2 = \left ( \frac{\lambda^2}{4} - \frac{\bar{g}^2}{8}
  \right ) v^2 \sin 4\beta + \frac{g_1'^2}{2} v^2 ( Q_{H_u} -
  Q_{H_d} ) \nonumber \\
  & \quad {} \times ( Q_{H_d} \cos^2\beta + Q_{H_u}
  \sin^2 \beta) \sin 2\beta , \nonumber \\
  M_{55}^2 &= \frac{\lambda^2}{2} v^2 \sin^2 2\beta + \frac{\bar{g}^2}{4}
  v^2 \cos^2 2\beta + g_1'^2 v^2 ( Q_{H_d} \cos^2\beta +
  Q_{H_u} \sin^2\beta )^2 . \nonumber
\end{align}
With the exceptions of $M_{45}^2$, $M_{54}^2$ and $M_{55}^2$, the size
of the mass matrix elements is determined by the singlet VEVs $s$ and $\varphi$.
For small values of $\lambda$ such that $\lambda s \sim \sigma s \sim
\sigma \varphi \sim M_S$, it is therefore expected that all but the lightest
state have masses of the order of the SUSY scale or heavier.  In particular,
for $\lambda \sim \sigma \to 0$ the element $M_{11}^2 \sim M_{Z^\prime} \gg
M_S$, while all other matrix elements are substantially smaller.
Thus the mass of the heaviest CP--even state is approximately degenerate with
the $Z^\prime$ mass.  After neglecting all terms which are proportional to
$\lambda v$ in Eqs.~(\ref{eq:scalar-mass-matrix}) it is easy to see that in the
limit $M_S\gg M_Z$ the mass of another CP--even state is set by $M_{44}$, i.e.,
this state is almost degenerate with the charged Higgs states, while the masses
of two other CP--even states are determined by $m_{+}$ and $m_{-}$,
\begin{equation}
\begin{aligned}
 \left (m_{\pm}^{\text{\DRbar}} \right )^2 \approx
\frac{1}{2} \left \{M_{22}^2 + M_{33}^2 \pm
\sqrt{(M_{22}^2 - M_{33}^2)^2 + 4 M_{23}^4}
\right \}\,.
\end{aligned}
\label{eq:heavy-scalar-masses}
\end{equation}

The mass of the lightest state, on the other hand, is bounded from above by
the smallest element $M_{55}^2$, i.e., the tree-level lightest CP--even Higgs
mass satisfies
\begin{equation} \label{eq:higgs-mass-upper-bound}
  \left ( m_{h_1}^{\text{\DRbar}} \right )^2 \leq
  \frac{\lambda^2}{2} v^2 \sin^2 2\beta + \frac{\bar{g}^2}{4}
  v^2 \cos^2 2\beta + g_1'^2 v^2 ( Q_{H_d} \cos^2\beta +
  Q_{H_u} \sin^2\beta )^2 .
\end{equation}
Consequently $h_1 \approx S_5$ is always light, and for $M_S \gg M_Z$ is SM-like
in its interactions.  While the upper bound
Eq.~(\ref{eq:higgs-mass-upper-bound}) is larger than in the MSSM, it is still
the case that radiative corrections are important for reaching $m_{h_1} \approx
125$ GeV.  Moreover, the by-now very precise measurement of the Higgs mass,
$m_{h_1}^{\text{exp.}} = 125.09 \pm 0.21 \pm 0.11$ GeV \cite{Aad:2015zhl},
strongly constrains the parameter space of SUSY models and necessitates a
reliable calculation of the physical Higgs mass.  In principle, the physical
Higgs masses can be determined from the poles in the propagator after including
the one-loop self-energies by solving
\begin{equation} \label{eq:higgs-pole-masses}
  \det \left [ p_i^2 \bm{1} - M^2(M_S) + \Sigma_h(p_i^2) \right ] = 0
\end{equation}
with $m_{h_i}^2 = \Real (p_i^2)$ and where $M^2(M_S)$ is the tree-level Higgs
mass matrix, evaluated here at $M_S$, and $\Sigma_h(p^2)$ denotes the
self-energies.  The required one-loop self-energies are automatically
included in the tools used in our numerical studies, described below.
However, for large values of $M_S \gg M_Z$ this strategy leads to large
logarithmic contributions to the Higgs masses due to heavy states, which should
be resummed to get an accurate estimate for the Higgs mass.  In our case, the
discussion above indicates that the SUSY spectrum is split, containing many
heavy scalars, notably the MSSM sfermions and the exotic scalars, as well as
light neutralinos and exotic fermions.  Such scenarios are well handled by
an effective field theory (EFT) approach to calculate the lightest Higgs mass,
in which the large logarithms are resummed.  In the MSSM, the largest of these
contributions is usually associated with the third generation sfermions, and in
particular the stops.  In the SE$_6$SSM, there are also contributions from the
heavy exotic scalars that should be accounted for.  Because the exotic Yukawa
couplings $\tilde{\lambda}_{\alpha \beta}$ and $\kappa_{ij}$ are very small in
the models we consider, these logarithmic corrections to the Higgs mass are
very small and can be neglected compared to the contributions from the stops
and other MSSM sfermions\footnote{We have confirmed numerically that the
  contributions in Eq.~(\ref{eq:higgs-pole-masses}) from the exotic states are
  negligible compared to those from the stops and sbottoms.}.  In our results
below, to obtain the light CP--even Higgs mass we therefore make use of the
known EFT calculation in the MSSM, which includes the dominant contributions to
the Higgs mass.  While a complete EFT calculation including the exotic states
would be more accurate\footnote{Such a calculation has been presented very
  recently \cite{Athron:2016fuq}, but this was not available when the scans
  presented here were performed.}, we expect that in this case the accuracy of
our calculation should not be significantly reduced due to the small size of the
exotic contributions.

\section{Results} \label{sec:results}
\subsection{Scan Procedure}
To study scenarios in the CSE$_6$SSM that are able to account for the
observed relic DM density with a MSSM-like DM candidate, a dedicated CSE$_6$SSM
spectrum generator was created using \FSv\ \cite{Athron:2014yba,Athron:2014wta}
and \SARAHv\ \cite{Staub:2008uz,Staub:2010jh,Staub:2012pb,Staub:2013tta}.  The
generated code\footnote{All of the code used in the following analysis is
  made available at \url{https://doi.org/10.5281/zenodo.215628}.},
which internally also relies on some routines from SOFTSUSY
\cite{Allanach:2001kg,Allanach:2013kza}, provides a precise
determination of the mass spectrum by making use of the full two-loop RGEs
and one-loop self-energies for all of the masses.  Leading two-loop
contributions to the CP--odd and CP--even Higgs masses taken from the known
NMSSM \cite{Degrassi:2009yq} and MSSM \cite{Degrassi:2001yf,Brignole:2001jy,
  Dedes:2002dy,Brignole:2002bz,Dedes:2003km} expressions were initially also
included\footnote{While full two-loop corrections to the Higgs masses, in
  the gaugeless limit, can be calculated for a general model in \SARAH\
  \cite{Goodsell:2014bna,Goodsell:2015ira}, this capability was not
  available in \FS\ at the time our numerical study was done.}, since the
additional contributions from new states are expected to be small by virtue
of their small couplings.

However, as noted above, for the solutions presented below this fixed order
Higgs mass calculation suffers from the effects of large logarithmic
contributions that are not resummed and so a MSSM EFT calculation is
employed to predict $m_{h_1}$ instead.  To do so, at the SUSY scale defined by
$M_S = \sqrt{m_{\tilde{t}_1}^\text{\DRbar} m_{\tilde{t}_2}^{\text{\DRbar}}}$,
with $m_{\tilde{t}_\alpha}^{\text{\DRbar}}$ given by Eq.~(\ref{eq:stop-masses}),
we performed a simple tree-level matching to the MSSM.  In this simple matching
procedure, the \DRbar\ MSSM soft scalar masses $m_{Q_{ii}}^2$, $m_{u^c_{ii}}$,
$m_{d^c_{ii}}$, $m_{L_{ii}}$, $m_{e^c_{ii}}$, gaugino masses $M_1$, $M_2$, $M_3$
and soft trilinear $A_t \equiv T^U_{33} / y^U_{33}$ are set at $M_S$ to
their values obtained in the CSE$_6$SSM after running from $M_X$.  The MSSM
$\mu$ parameter is set to its effective value at $M_S$,
Eq.~(\ref{eq:mu-eff-defn}), while an effective MSSM pseudoscalar mass,
$(m_A)_{\text{eff}}$, is obtained from the effective soft bilinear
\begin{equation} \label{eq:Bmu-eff-defn}
  (B\mu)_{\text{eff}} = \frac{T_\lambda s}{\sqrt{2}} \cos\theta - \frac{\lambda
    \sigma}{2} s \varphi \sin \theta.
\end{equation}
The lightest CP--even Higgs mass was then calculated using \SUSYHDv\
\cite{Vega:2015fna} to obtain a more accurate estimate for the SM-like
Higgs mass.  The remaining heavy CP--even Higgs masses were computed using
the ordinary fixed order approach.

As mentioned above, for the purposes of studying the MSSM-like DM candidate
it is most convenient to directly vary the parameters $M_{1/2}$ and
$\mu_{\text{eff}}$.  For this reason, we implemented a solver algorithm in \FS\
that makes use of the semi-analytic solutions to the RGEs.  A similar algorithm
has previously been used in studies of the constrained E$_6$SSM, where it was
described in Ref.~\cite{Athron:2009bs}.  The main advantage of this algorithm
over the standard two-scale fixed point iteration is that by expanding
all of the soft parameters at low-energies using the semi-analytic solutions,
the EWSB conditions can be used to fix a subset of the input high-scale
parameters in terms of the remaining input parameters.  In particular, the
low-energy soft Higgs and singlet masses can be written in the form
\begin{equation} \label{eq:soft-scalar-mass-expansion}
  m_\Phi^2(M_S) = a_\Phi (M_S) m_0^2 + b_\Phi (M_S) M_{1/2}^2
  + c_\Phi(M_S) M_{1/2} A_0 + d_\Phi(M_S) A_0^2 ,
\end{equation}
for $\Phi = H_d, H_u, S, \overline{S}, \phi$.  Imposing the EWSB conditions
Eq.~(\ref{eq:ewsb-conditions}) then allows $m_0$ to be fixed, as desired,
along with $\tan\theta$, $\varphi$, $\Lambda_F(M_X)$ and $\Lambda_S(M_X)$.
The parameters $\lambda$ and $M_{1/2}$ remain free parameters that can be varied
to set the mass and composition of $\tilde{\chi}_1^0$.

To satisfy the limits on the $Z^\prime$ mass, we take advantage of the mechanism
described below Eq.~(\ref{eq:singlet-vevs-magnitude}) to set $M_{Z^\prime}$ well
above the current limits, and so we set $M_{Z^\prime} \approx 240$ TeV.  This
requires a very large value of $s = 650$ TeV at the SUSY scale.  Acceptably
small values of $\mu_{\text{eff}} \lesssim 1$ TeV for reproducing the DM relic
density are then achieved for very small $|\lambda|$, though $\mu_{\text{eff}}$
is still large enough to evade limits from LEP.  In this study we focus on
scenarios in which the LSP is either a mixed bino-Higgsino or pure Higgsino
dark matter candidate.  To do so, we considered
$|\lambda(M_X)| = 9.15181 \times 10^{-4}$ and
$|\lambda(M_X)| = 2.4 \times 10^{-3}$, for both $\lambda < 0$ and
$\lambda > 0$.  Because $\tan\theta \approx 1$ for such large values of $s$,
this corresponds to $|\mu_{\text{eff}}(M_X)| \approx 347$ GeV and
$|\mu_{\text{eff}}(M_X)| \approx 898$ GeV, giving values at the SUSY scale of
$|\mu_{\text{eff}}(M_S)| \approx 417$ GeV and
$|\mu_{\text{eff}}(M_S)| \approx 1046$ GeV, respectively\footnote{
  The values of $|\mu_{\text{eff}}|$ given are the mean values over
  all of the obtained valid solutions.  The exact values of
  $|\mu_{\text{eff}}(M_S)|$ and $|\mu_{\text{eff}}(M_X)|$ vary over
  the parameter space scanned, since $\tan\theta$ varies slightly over
  the scanned region, as it is an EWSB output parameter, and the RG
  evolution also changes slightly due to sparticle threshold
  corrections.  For the smaller value of $|\lambda(M_X)|$, the
  solutions we present have $409 \text{ GeV} \leq
  |\mu_{\text{eff}}(M_S)| \leq 425$ GeV, and $344 \text{ GeV} \leq
  |\mu_{\text{eff}}(M_X)| \leq 349$ GeV.  For the larger
  $|\lambda(M_X)|$ value we obtain solutions with $1032 \text{ GeV}
  \leq |\mu_{\text{eff}}(M_S)| \leq 1063$ GeV, and $892 \text{ GeV}
  \leq |\mu_{\text{eff}}(M_X)| \leq 903$ GeV.}.

To prevent tachyonic states in the exotic sector, the exotic couplings cannot
be too large, and for our scans we chose fixed values satisfying
$\tilde{\lambda}_{\alpha\beta}(M_X),\, \kappa_{ij}(M_X) \leq 3 \times
10^{-3}$.  Additionally, to simplify our analysis we took these
couplings to be family universal with
$\tilde{\lambda}_{\alpha\beta}(M_X) = \tilde{\lambda}_0
\delta_{\alpha\beta}$ and $\kappa_{ij}(M_X) = \kappa_0 \delta_{ij}$.
A SUSY scale somewhat below $M_{Z^\prime}$ was obtained by choosing
small $\sigma(M_X) = 2 \times 10^{-2}$.  Light inert singlinos in the
spectrum were ensured by choosing extremely small values for the
couplings $\tilde{f}_{i\alpha}$ and $f_{i\alpha}$, while for
simplicity we set the couplings $\tilde{\sigma}(M_X)$,
$\mu_\phi(M_X)$, $g_{ij}^D(M_X)$ and $h^E_{i\alpha}(M_X)$ to zero.  We stress
that the impact of the latter two sets of couplings on the quantities we
investigate is numerically negligible.  We have checked that their values
could also be increased to satisfy constraints on the exotic lifetimes without
altering our results.  We also chose $\kappa_\phi(M_X) = 10^{-2}$, and
$\mu_L(M_X) = 10$ TeV.  While the above fixed couplings impact the mass
spectrum, they do not play a significant role in the predictions for dark
matter, for the scenarios considered here in which the dark matter candidate is
the lightest MSSM-like neutralino, and hence we do not scan over them.

For each parameter point in the scan, the GUT scale
$M_X$ at which these values are set is defined to be the scale at which
$g_1(M_X) = g_2(M_X)$.  This condition is solved iteratively, as described in
Ref.~\cite{Athron:2014yba}.  We do not require that $g_3(M_X)$ is also
unified, but this will be approximately fulfilled due to the inclusion of the
$\hat{L}_4$ and $\hat{\overline{L}}_4$ states.  This is similar to what
occurs in the E$_6$SSM \cite{King:2007uj}.

For $\lambda \ll \bar{g}$, the tree-level upper bound on the SM-like
Higgs mass is maximized for large $\tan\beta$.  We took
$\tan\beta(M_Z) = 10$ to saturate this limit.  As in the CMSSM, the
transformation $M_{1/2} \to -M_{1/2}$, $A_0 \to -A_0$, $B_0 \to -B_0$ and
$\mu_{\text{eff}} \to -\mu_{\text{eff}}$ leaves our results invariant.  We use
this symmetry to fix $M_{1/2} \geq 0$.  Setting $B_0 = 0$, we
scanned over $M_{1/2}$ and $A_0$ by uniformly sampling in the
intervals $[0 \text{ TeV, } 20 \text{ TeV}]$ and $[-20 \text{ TeV, }
  20 \text{ TeV}]$, respectively, to find solutions with the correct
Higgs mass and an allowed DM relic density.  The relic density and direct
detection cross section were calculated numerically with
\MICROMEGASv\ \cite{Belanger:2001fz,Belanger:2004yn,
  Belanger:2006is,Belanger:2008sj,Belanger:2010gh,Belanger:2013oya,
  Belanger:2014vza}, using \CALCHEP\ \cite{Belyaev:2012qa} model files
automatically generated with \SARAH.  The values of the CSE$_6$SSM
parameters used are summarized in
\tabref{tab:cse6ssm-parameter-values}.

\begin{table*}[ht]
  \centering
  \begin{tabular}{ccc}
    \toprule
    & $\lambda(M_X) = \pm 9.15181 \times 10^{-4}$ &
    $\lambda(M_X) = \pm 2.4 \times 10^{-3}$ \\
    \midrule
    $\sigma(M_X)$ & $2 \times 10^{-2}$ & $2 \times 10^{-2}$ \\
    $\kappa_\phi(M_X)$ & $10^{-2}$ & $10^{-2}$ \\
    $\tilde{\lambda}_{\alpha\beta}(M_X) = \tilde{\lambda}_0
    \delta_{\alpha\beta}$
    & $10^{-3}$ & $3 \times 10^{-3}$ \\
    $\kappa_{ij}(M_X) = \kappa_0 \delta_{ij}$ & $10^{-3}$
    & $1.4 \times 10^{-3}$, $3 \times 10^{-3}$ \\
    $\tilde{f}_{11}(M_X)$, $\tilde{f}_{22}(M_X)$, $\tilde{f}_{31}(M_X)$
    & $10^{-7}$ & $10^{-7}$ \\
    $f_{11}(M_X)$, $f_{22}(M_X)$, $f_{32}(M_X)$ & $10^{-7}$ & $10^{-7}$ \\
    $\mu_L(M_X)$ [TeV] & $10$ & $10$ \\
    $s(M_S)$ [TeV] & $650$ & $650$ \\
    $M_{1/2}$ [TeV] & $[0, 20]$ & $[0, 20]$ \\
    $A_0$ [TeV] & $[-20, 20]$ & $[-20, 20]$ \\
    $\tan\beta(M_Z)$ & $10$ & $10$ \\
    \bottomrule
  \end{tabular}
  \caption{Summary of the fixed parameter values and allowed ranges used in the
    CSE$_6$SSM for the two values of $|\lambda(M_X)|$ considered.  The free
    parameters $\tilde{\sigma}(M_X)$, $\mu_\phi(M_X)$, $B_0$, $g_{ij}^D(M_X)$,
    $h^E_{i\alpha}(M_X)$ and the $\tilde{f}_{i\alpha}(M_X)$, $f_{i\alpha}(M_X)$
    not shown are set to zero in both cases.  The parameters
    $m_0$, $\tan\theta$, $\varphi$, $\Lambda_F$ and $\Lambda_S$ are fixed by the
    requirement of correct EWSB.  In the CMSSM, the same ranges are taken for
    $M_{1/2}$ and $A_0$ for the comparison scans with $\mu(M_S) = \pm 417$
    GeV and $\mu(M_S) = \pm 1046$ GeV, and we set $\tan\beta(M_Z)
    = 10$ as well.  The EWSB conditions are used to fix $m_0$ and $B_0$ in the
    CMSSM.}
  \label{tab:cse6ssm-parameter-values}
\end{table*}

For this choice of parameters the lightest neutralino is expected to be
MSSM-like in its composition and couplings.  At the same time, the spectrum and
the RG flow of couplings in the CSE$_6$SSM is very different to that in the
CMSSM.  While the two models may in this limit make very similar
predictions concerning DM, the ranges of parameter space in which this occurs
and their collider signatures can therefore be quite distinct.  This makes it
interesting to compare the CSE$_6$SSM and CMSSM directly.  To do this
comparison, we also generated a CMSSM spectrum generator using \FS\ and \SARAH\
as described above, and modified it to make use of the semi-analytic solver
algorithm.  The MSSM EWSB conditions were used to fix the common soft scalar
mass $m_0$ and soft breaking bilinear $B_0$ at the GUT scale, and $M_{1/2}$ and
$A_0$ were scanned over the same ranges as in the CSE$_6$SSM.  This was done
for values of $\mu(M_S)$ fixed to the mean values obtained in the CSE$_6$SSM,
that is, $|\mu(M_S)| = 417$ GeV and $|\mu(M_S)| = 1046$ GeV, respectively.
The same fixed value of $\tan\beta(M_Z) = 10$ was used.  In this way we are
able to present a more direct comparison of the two models, in which
analogous parameters are approximately matched between the two\footnote{We
  emphasise that our approach in the CMSSM differs from the conventional
  approach in the literature, in which $\mu$ would be determined by the EWSB
  conditions and $m_0$ is an input parameter.}.  The CMSSM solutions that we
obtained have a heavy SUSY scale as well, so that we again used \SUSYHD\ to
compute the lightest Higgs mass.  The predicted DM relic density and direct
detection cross section were calculated in \MICROMEGAS\ using model files
generated by \SARAH\footnote{We checked that the results obtained this way were
  in very good agreement with those found from using the MSSM implementation
  already available in \MICROMEGAS, provided some care was taken to define the
  quark mass parameters consistently in the calculation of the direct detection
  cross sections.}.

In both models, valid points were selected by imposing the theoretical
constraints that the point should have a valid spectrum with correct EWSB and
no tachyonic states.  We required that all couplings remain perturbative up
to the GUT scale.  Since we perform only a na\"{\i}ve matching to the MSSM
in the EFT calculation, we allowed for an uncertainty of $\pm 3$ GeV in the
result for $m_{h_1}$, which is somewhat larger than is reported by \SUSYHD.  For
the CSE$_6$SSM we accepted points with calculated light Higgs masses satisfying
$122 \text{ GeV } \leq m_{h_1} \leq 128$ GeV, and for comparison we allowed the
same range of Higgs masses in the CMSSM.  A point predicting a relic density
$\Omega h^2$ greater than that determined by Planck observations
\cite{Ade:2015xua},
\begin{equation} \label{eq:planck-relic-density}
  (\Omega h^2)_{\text{exp.}} = 0.1188 \pm 0.0010 ,
\end{equation}
is effectively ruled out if one assumes a standard cosmological history.
Points with a predicted relic density that does not exceed this value are not
ruled out in the same way, though in this case additional contributions to DM
are required.  In our scans we excluded all points that have a predicted relic
density $(\Omega h^2)_{\text{th.}} > (\Omega h^2)_{\text{exp.}}$.

To make a clear comparison of the impact of collider bounds on the
CSE$_6$SSM and CMSSM, model specific limits should be applied to each.
However in the CSE$_6$SSM the RGEs drive the sfermions to masses which
are substantially larger than the gaugino masses, creating a
hierarchical spectrum that persists even with the decoupling of the
$Z^\prime$ mass from the rest of the spectrum.  This means that
typically LHC collider limits come from the gaugino sector, especially
the gluino which is produced through strong interactions.  The gluino
decays in an MSSM-like manner and as a result the gluino mass limit
set in the CMSSM in the heavy sfermion limit should, to a reasonable
approximation, apply to the gluino in the CSE$_6$SSM also\footnote{A
  more thorough treatment involves reinterpreting existing searches,
  for which a variety of tools, such as
  \checkmate\ \cite{Drees:2013wra}, \madanalysis\ \cite{Conte:2012fm},
  \smodels\ \cite{Kraml:2013mwa} or \fastlim\ \cite{Papucci:2014rja} are
  available.  However since the situation is fairly simple in this case, with
  very heavy sfermions, we consider this unnecessary here and beyond
  the scope of our analysis.}.  So that the reader can see where
current and future collider limits should constrain the models we will
show explicit gluino mass contours in each model, along with contours
for the physical first generation squark mass, $m_{\tilde{u}_6}$.  Note that
this is approximately degenerate with the remaining first and second generation
squark masses, i.e., $m_{\tilde{q}_{1,2}} \approx m_{\tilde{u}_6}$.

\subsection{Mixed Bino-Higgsino Dark Matter}
\begin{figure}[h!]
  \includegraphics[width=0.5\textwidth]{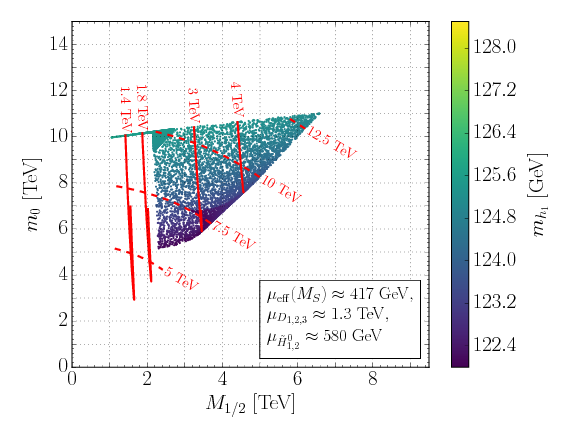}
  \includegraphics[width=0.5\textwidth]{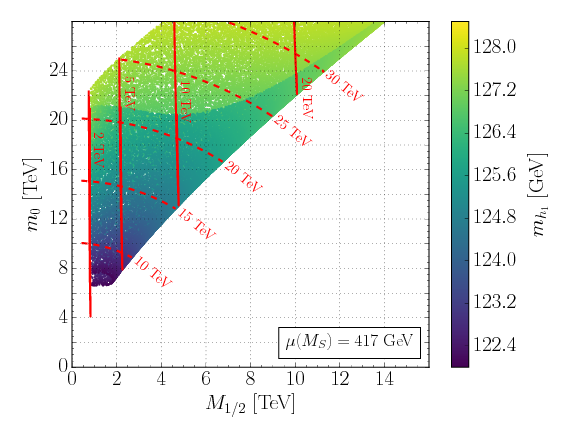}
  \includegraphics[width=0.5\textwidth]{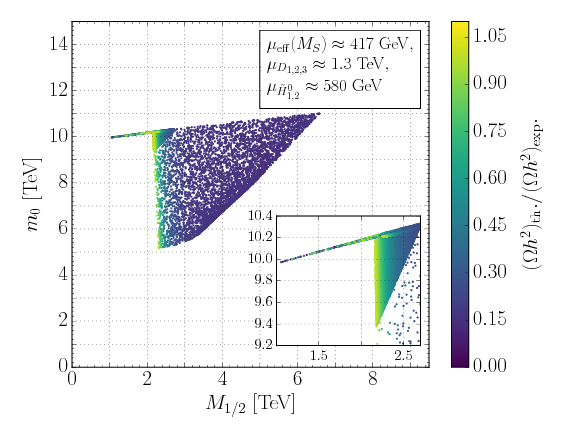}
  \includegraphics[width=0.5\textwidth]{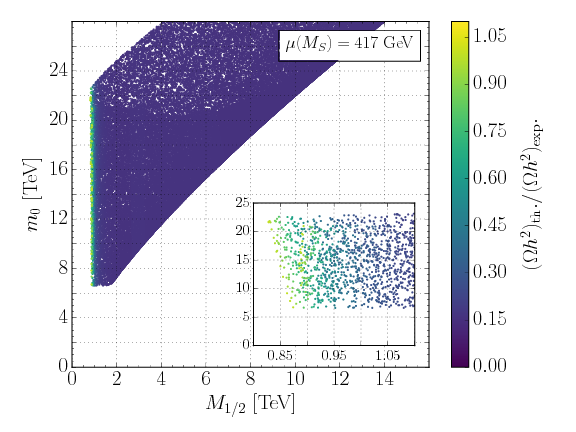}
  \includegraphics[width=0.5\textwidth]{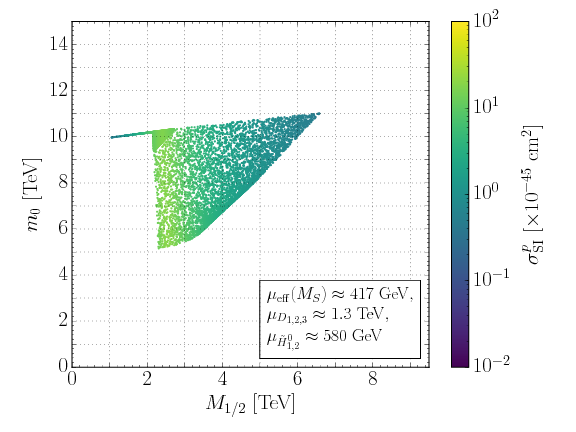}
  \includegraphics[width=0.5\textwidth]{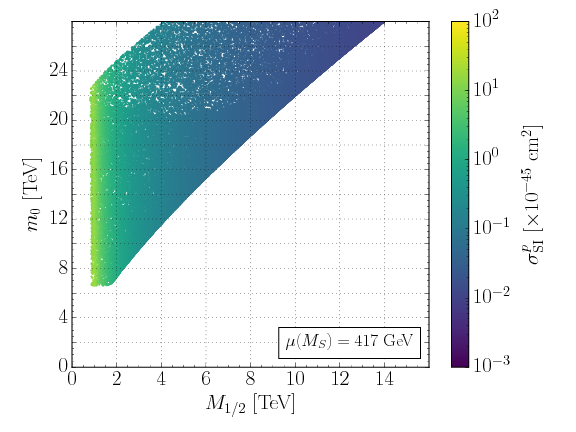}
  \caption{Contour plots in the $M_{1/2} - m_0$ plane of the lightest CP--even
    Higgs mass (top row), DM relic density (middle row) and proton SI cross
    section (bottom row) in the CSE$_6$SSM with $\mu_{\text{eff}}(M_X) \approx
    347$ GeV (left column) and CMSSM with $\mu(M_S) = 417$ GeV (right
    column).  In the top row, we also show contours of the gluino (solid lines)
    and squark (dashed lines) masses.  At large values of $M_{1/2}$, where the
    $\tilde{\chi}_1^0$ is a light Higgsino, the relic density saturates
    with $(\Omega h^2)_{\text{th.}} / (\Omega h^2)_{\text{exp.}} \approx 0.15$.}
  \label{fig:mupos400GeV-m12-m0-planes}
\end{figure}

\begin{figure}[h!]
  \includegraphics[width=0.5\textwidth]{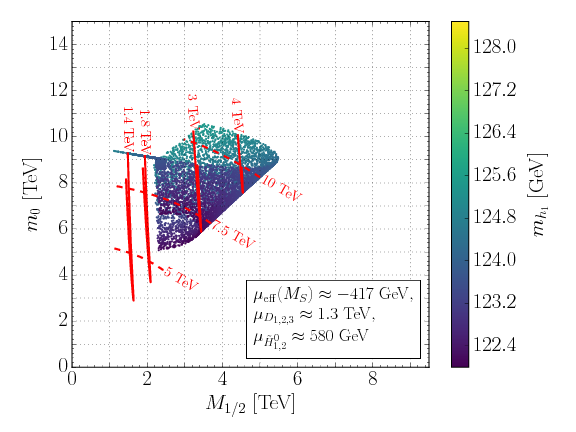}
  \includegraphics[width=0.5\textwidth]{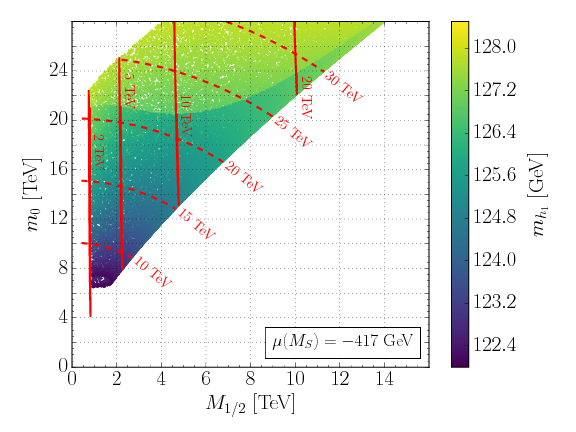}
  \includegraphics[width=0.5\textwidth]{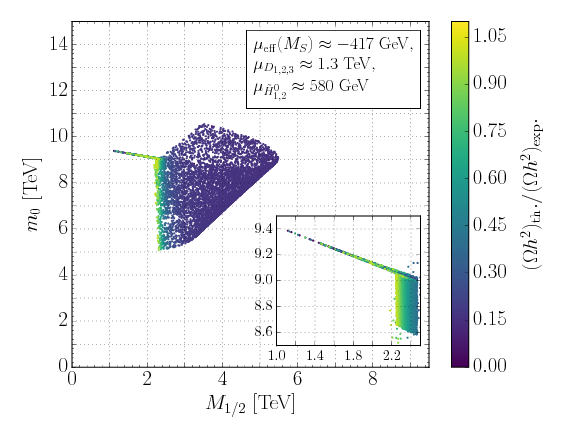}
  \includegraphics[width=0.5\textwidth]{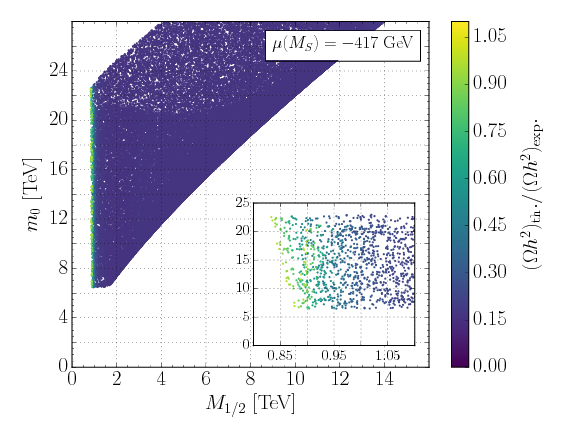}
  \includegraphics[width=0.5\textwidth]{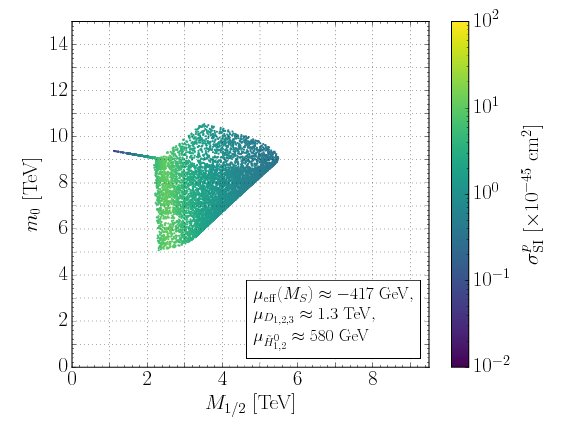}
  \includegraphics[width=0.5\textwidth]{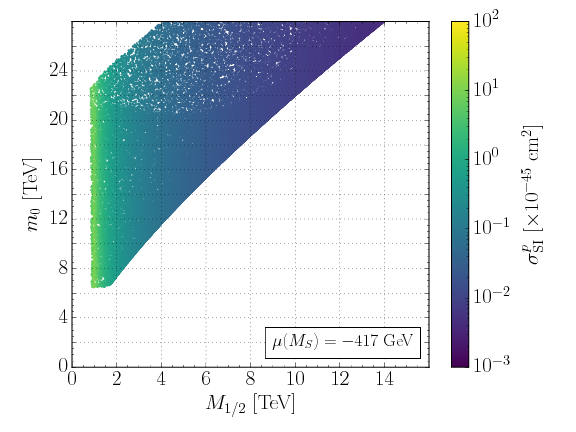}
  \caption{Contour plots in the $M_{1/2} - m_0$ plane of the lightest CP--even
    Higgs mass (top row), DM relic density (middle row) and proton SI cross
    section (bottom row) in the CSE$_6$SSM with $\mu_{\text{eff}}(M_X) \approx
    -347$ GeV (left column) and CMSSM with $\mu(M_S) = -417$ GeV (right
    column).  In the top row, we show contours of the gluino (solid lines)
    and squark (dashed lines) masses.  As for the positive $\mu_{\text{eff}}$
    case, at large $M_{1/2}$ the relic density reaches a limiting value of
    $(\Omega h^2)_{\text{th.}} / (\Omega h^2)_{\text{exp.}} \approx 0.15$.}
  \label{fig:muneg400GeV-m12-m0-planes}
\end{figure}

We first consider cases with a light Higgsino mass term of
$|\mu_{(\text{eff})}(M_S)| \approx 417$ GeV.  The results obtained in the
CSE$_6$SSM and the CMSSM for this value of
$|\mu_{(\text{eff})}|$ are compared in \figref{fig:mupos400GeV-m12-m0-planes}
and \figref{fig:muneg400GeV-m12-m0-planes}.

In the top row of \figref{fig:mupos400GeV-m12-m0-planes} we compare the mass
of the SM-like Higgs in the two models.  In both we find solutions
consistent with $m_{h_1} \approx 125$ GeV, but the allowed regions in the
$M_{1/2} - m_0$ plane clearly differ quite substantially.  For such large values
of $s$ and small values of $\lambda$ the tree-level mass of the lightest
CP--even Higgs in the SE$_6$SSM is approximately the same as it is in the MSSM,
$(m_{h_1}^{\text{\DRbar}})^2 \approx M_Z^2 \cos^2 2\beta$, as follows from
approximately diagonalizing the mass matrix in
Eq.~(\ref{eq:scalar-mass-matrix}).  Without substantial tree-level contributions
from the additional $F$- and $D$-terms, a $125$ GeV Higgs is achieved with large
radiative corrections in the CSE$_6$SSM as well as in the CMSSM.  In principle,
large enough loop corrections result from either large sparticle masses,
particularly stop masses, or large stop mixing.  However, increasing $A_0$ or
$M_{1/2}$ to generate large mixings for fixed $\mu_{(\text{eff})}$ leads to the
value of $m_0$ increasing as needed to satisfy the EWSB conditions.  As a
result in the solutions we obtain $m_0 > A_0, M_{1/2}$ and large enough
radiative corrections must arise from sufficiently heavy sparticle masses
instead.  The effect of the Higgs mass constraint can be clearly seen in the
top row of \figref{fig:mupos400GeV-m12-m0-planes} and
\figref{fig:muneg400GeV-m12-m0-planes}, where the requirement $m_{h_1} \geq 122$
GeV imposes the lower bound on $m_0$ for small values of $M_{1/2}$.

The right-most boundary of the solution region is a consequence of determining
$m_0^2$ from the EWSB conditions.  When the soft masses and SUSY scale are
large and $|\mu_{(\text{eff})}| \ll M_{1/2}$, as is the
case here, the resulting function for $m_0^2(M_{1/2}, A_0)$ has a minimum at
each $M_{1/2}$ with $m_{0,\text{min}}^2(M_{1/2}) > 0$\footnote{
  For example, at tree-level and neglecting small $D$-term contributions
  the EWSB conditions lead to an expression of the form $m_0^2 = \xi_1 M_{1/2}^2
  + \xi_2 M_{1/2} A_0 + \xi_3 A_0^2 + \xi_0 |\mu_{(\text{eff})}|^2$ where the
  coefficients $\xi_1, \xi_3 > 0$ and $\xi_0, \xi_2 < 0$ for $\tan\beta = 10$
  are set by the RG flow.  Because $\xi_1 - \xi_2^2 / (4 \xi_3) > 0$,
  for fixed $|\mu_{(\text{eff})}| \ll M_{1/2}$ it is easy to show that there is
  a non-trivial lower bound on the value of $m_0^2$.}.  Hence
when $\mu_{(\text{eff})}$ is fixed, we do not find points with values of $m_0$
below this boundary for each given value of $M_{1/2}$. This can be contrasted
  with the usual procedure in the CMSSM, where lower values of $m_0^2$ can be
  found by varying $|\mu|$ and $B\mu$.

\begin{figure}[h]
  \includegraphics[width=0.5\textwidth]{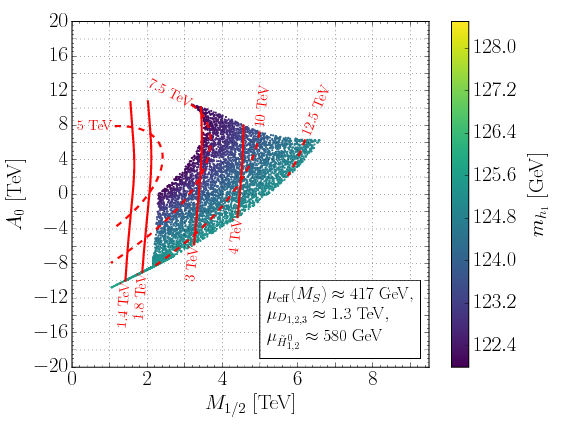}
  \includegraphics[width=0.5\textwidth]{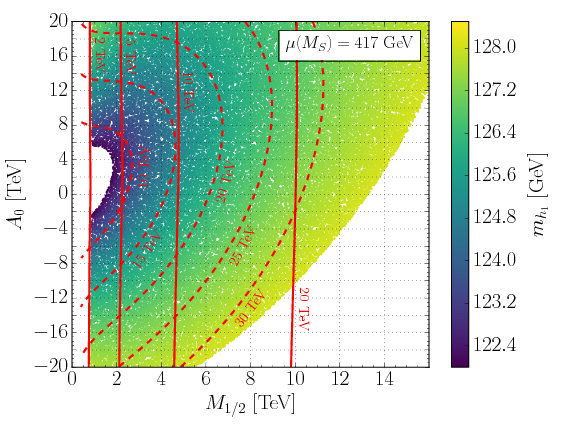}
  \caption{Contour plots of the lightest CP--even Higgs mass in the
    $M_{1/2} - A_0$ plane in the CSE$_6$SSM with $\mu_{\text{eff}}(M_X)
    \approx 347$ GeV (left) and the CMSSM with $\mu(M_S) = 417$ GeV (right).
    Also shown are contours of the gluino (solid lines) and squark (dashed
    lines) masses for both models.}
  \label{fig:mupos400GeV-m12-A0-plane}
\end{figure}

In the CMSSM, the Higgs mass constraint $m_{h_1} \leq 128$ GeV also puts an
upper bound on the possible values of $M_{1/2}$.  This is shown
in \figref{fig:mupos400GeV-m12-A0-plane}, where we plot $m_{h_1}$ in the
$M_{1/2} - A_0$ plane in both models for $\mu_{(\text{eff})} > 0$.  The upper
bound on $m_{h_1}$ cuts off the solution region at large values of $M_{1/2}$ in
\figref{fig:mupos400GeV-m12-A0-plane} in the CMSSM.  In comparison, in the
CSE$_6$SSM the region at large $M_{1/2}$ is ruled out by the presence of
tachyonic states.  The lower right region of the CSE$_6$SSM $M_{1/2} - A_0$
plane in \figref{fig:mupos400GeV-m12-A0-plane} is excluded by tachyonic
pseudoscalars $A_i$, while the uppermost boundary is due to tachyonic
CP--even Higgs states.  This corresponds to the much more restrictive upper
bound on $m_0$ in the CSE$_6$SSM in \figref{fig:mupos400GeV-m12-m0-planes}
compared to the CMSSM.  The same is true for $\mu_{(\text{eff})} < 0$ in
\figref{fig:muneg400GeV-m12-m0-planes}, though the position of the boundary
is modified, leading to the much smaller range of acceptable $m_0$ values in
the CSE$_6$SSM for this value of $|\mu_{\text{eff}}|$.  It should be noted,
however, that these results are obtained for a single value of $s$.  It is
expected that if $s$ and $\lambda$ are allowed to vary while maintaining fixed
$\mu_{\text{eff}}$, additional solutions would be obtained, as is found in the
constrained E$_6$SSM \cite{Athron:2009bs,Athron:2012sq}.  It is important to
emphasize that in the CSE$_6$SSM there is still additional parameter space
available, and that the constraints shown here apply only for a single value of
$M_{Z^\prime}$ in the model.

The large values of $m_0$ required result in a large SUSY scale and all scalars
except the SM-like Higgs $h_1$, and the lightest pseudoscalar $A_1$ in
the CSE$_6$SSM, are very heavy.  In the top row of
\figref{fig:mupos400GeV-m12-m0-planes} and
\figref{fig:muneg400GeV-m12-m0-planes} we show contours of the gluino and
first and second generation squark masses.  The viable solutions that we
find in the CSE$_6$SSM all have squark masses $m_{\tilde{q}_{1,2}} \geq 5.4$
TeV, while in our CMSSM solutions $m_{\tilde{q}_{1,2}} \geq 6.5$ TeV,
so that these states are not observable at the LHC.  On the other
hand, the small exotic couplings lead to light exotic fermions.  For
$|\mu_{\text{eff}}(M_X)| \approx 347$ GeV, the choice of $\kappa_0 = 10^{-3}$
leads to exotic $D$ fermion masses of $\approx 1.3$ TeV.  Similarly, setting
$\tilde{\lambda}_0 = 10^{-3}$ leads to inert Higgsinos with masses
$\approx 580$ GeV.  Both sets of states are therefore light enough to be
produced at the LHC and would be detectable via the signatures
discussed in \secref{sec:particle-spectrum}.  Given the increasingly large
SUSY scale required by LHC searches in constrained models, this makes searches
targeting the exotic spin-$1/2$ leptoquark and inert Higgsino states attractive
for still being able to probe the CSE$_6$SSM parameter space.  Because the
exotic couplings cannot be too large in the scenarios considered here, improved
limits on these states would strongly constrain the solutions we have found
with very small values of $|\mu_{\text{eff}}|$.

In addition to the restriction on the allowed values of $m_0$, there is also
a lower bound on $M_{1/2}$ in both models, which is determined by the relic
density constraint.  The behaviour in the CMSSM in this case is well understood.
When $M_1$ is sufficiently large, $\tilde{\chi}_1^0$ is a nearly pure, light
Higgsino that is underabundant \cite{Edsjo:1997bg}.  The opposite
limit, with small $M_{1/2}$ and $M_1 \lesssim \mu$, leads to an almost pure bino
LSP that is overabundant, due to its small annihilation cross section.
Therefore requiring $\Omega h^2 \leq 0.1188$ amounts to placing a lower
bound on $M_{1/2}$ for fixed $\mu$.

Since $\mu_{(\text{eff})}$ is small in this case, an acceptable relic density is
achieved with relatively low values of $M_{1/2}$.  The minimal allowed value of
$M_{1/2}$ in the CMSSM, $M_{1/2} \approx 0.85$ TeV, leads to $M_1 \approx \mu$
and the LSP is a so-called ``well-tempered'' highly mixed bino-Higgsino state
\cite{ArkaniHamed:2006mb} that saturates the relic density.  This region is
evident in the middle rows of \figref{fig:mupos400GeV-m12-m0-planes} and
\figref{fig:muneg400GeV-m12-m0-planes} as an extremely narrow strip at the
minimum value of $M_{1/2}$ (shown in greater detail in the insets) where
$(\Omega h^2)_{\text{th.}} \approx 0.1188$, while for larger $M_{1/2}$ the
Higgsino DM candidate leads to $(\Omega h^2)_{\text{th.}} \ll 0.1188$.
From comparing the left and right panels in the middle rows of
\figref{fig:mupos400GeV-m12-m0-planes} and
\figref{fig:muneg400GeV-m12-m0-planes} it is clear that similar behaviour
occurs for the $Z_2^E = +1$ DM candidate in the CSE$_6$SSM.  From
Eq.~(\ref{eq:cse6ssm-gaugino-mass-relation}) and
Eq.~(\ref{eq:cmssm-gaugino-mass-relation}) it follows that the necessary
value of $M_1$ occurs for smaller values of $M_{1/2}$ in the CMSSM.

The low allowed values of $M_{1/2}$ imply that  in the light
$\mu_{(\text{eff})}$ scenario the gluino as well as the ordinary neutralino and
chargino states can be light.  Though the location of the well-tempered strip
differs in the two models, the masses of the gluino, neutralino and charginos
are rather similar.  For example, in both models in this strip
$m_{\tilde{\chi}_1^0} \approx 370$ GeV.  In the CMSSM, we find that
$m_{\tilde{g}} \gtrsim 2.1$ TeV, the minimum value occurring in the
well-tempered region.  A very similar result can be
seen in the CSE$_6$SSM, with $m_{\tilde{g}} \gtrsim 2$ TeV except for a narrow
line of solutions where the gluino can be as light as $m_{\tilde{g}} \approx 1$
TeV.

For these solutions, the bino DM candidate is viable due to
the $A$-funnel mechanism.  In the CMSSM, $m_A$ is only light enough so that
$m_A \approx 2 m_{\tilde{\chi}_1^0}$ at large $\tan\beta \gtrsim 50$
\cite{Roszkowski:2001sb}.  Because we only considered $\tan\beta(M_Z) =
10$ in our scans, $m_A > 6$ TeV is always very heavy in our CMSSM results and
the $A$-funnel region is not accessible.  In the CSE$_6$SSM, for a given
value of $\tan\beta$ and $M_{1/2}$ one can make $m_{A_1} \approx
2 m_{\tilde{\chi}_1^0}$ light by fine tuning $A_0$ appropriately.  This
corresponds to the lower boundary of the solution region in
\figref{fig:mupos400GeV-m12-A0-plane}.  Therefore even for $\tan\beta(M_Z) = 10$
light bino DM can satisfy the relic density constraint in the CSE$_6$SSM.  This
does, however, imply a substantial fine tuning; in our scans, additional points
were sampled from this region to overcome this.

In either the bulk or $A$-funnel regions, the gluino is thus observable
at run II or at the high luminosity LHC (HL-LHC); indeed, gluino masses under
$2$ TeV are already rather close to the limits based on the most recent
$\sqrt{s} = 13$ TeV data and so LHC searches will soon be probing this part of
the parameter space.  Similarly, both models also predict light neutralinos and
charginos with masses of a few hundred GeV.  To be precise, our CMSSM
solutions satisfy $366 \text{ GeV} \leq m_{\tilde{\chi}_1^0} \leq 452$ GeV,
$428 \text{ GeV} \leq m_{\tilde{\chi}_2^0} \leq 453$ GeV and $419 \text{ GeV}
\leq m_{\tilde{\chi}_1^\pm} \leq 453$ GeV, while in the CSE$_6$SSM the ranges
are $182 \text{ GeV} \leq m_{\tilde{\chi}_1^0} \leq 426$ GeV, $335 \text{ GeV}
\leq m_{\tilde{\chi}_2^0} \leq 438$ GeV, and $335 \text{ GeV} \leq
m_{\tilde{\chi}_1^\pm} \leq 431$ GeV.  This suggests the neutralinos and
charginos could also be discoverable at the HL-LHC
\cite{runii-neutralinos-3lepton} in the small $\mu_{(\text{eff})}$ case.  The
overall picture for the solutions presented with $|\mu(M_S)| \approx 417$ GeV is
of a split spectrum, with unobservably heavy scalars but light exotic
fermions and EW-inos, as well as a sufficiently light gluino.  This scenario
would therefore predict interesting collider phenomenology in tandem with
accounting for the observed DM relic density.

However, while small values of $\mu_{(\text{eff})}$ permit the neutralinos and
gluino to be observable at the LHC, models with a highly mixed bino-Higgsino DM
candidate are strongly constrained by null results from direct detection
experiments.  In the bottom rows of \figref{fig:mupos400GeV-m12-m0-planes} and
\figref{fig:muneg400GeV-m12-m0-planes} we show the $\tilde{\chi}_1^0$-proton
SI cross section for each sign of $\mu_{(\text{eff})}$.  In the region where
$(\Omega h^2)_{\text{th.}}$ matches the observed value, the
direct detection cross section peaks at $\sim 10^{-45} - 10^{-44}$ cm$^2$ and is
above the 90\% exclusion limits set by \lux\ \cite{Akerib:2015rjg,
  Akerib:2016vxi}.  In both the CSE$_6$SSM and CMSSM, the SI cross section in
this part of the parameter space is dominated by $t$-channel exchange of the
lightest CP--even Higgs $h_1$.  Thus in the leading approximation the SI part
of $\chi_1^0$--nucleon cross section takes the form
\begin{equation}
  \begin{aligned}
    \sigma_{SI} &= \frac{4 m^2_r m_N^2}{\pi v^2 m^4_{h_1}}
    |g_{h_1 \chi_1 \chi_1} F^N|^2 \,, \\
    m_r &= \frac{m_{\chi^0_1} m_N}{m_{\chi^0_1} + m_N} \,,\qquad\qquad\qquad
    F^N = \sum_{q=u,d,s} f^N_{Tq} + \frac{2}{27} \sum_{Q=c,b,t} f^N_{TQ}\,,
  \end{aligned} \label{higgs-chi1-xs}
\end{equation}
where
\begin{equation}
m_N f^N_{Tq} = \langle N | m_{q}\bar{q}q |N \rangle\,, \qquad\qquad
f^N_{TQ} = 1 - \sum_{q=u,d,s} f^N_{Tq}\,,
\end{equation}
while\footnote{The values of these hadronic matrix elements are the default
  values used in \MICROMEGAS, as determined in Ref.~\cite{Belanger:2013oya}
  from lattice results.  A review of some recent determinations of the
  required sigma terms $\sigma_{\pi N}$ and $\sigma_s$ has been given in
  Ref.~\cite{Thomas:2012tg}, while an extraction of these quantities
  from phenomenological inputs using chiral effective field theory has been
  presented in Refs.~\cite{Alarcon:2011zs,Alarcon:2012nr}.}
$f^N_{Tu}\simeq 0.0153$, $f^N_{Td}\simeq 0.0191$ and $f^N_{Ts}\simeq 0.0447$.
The size of the cross section in Eq.~(\ref{higgs-chi1-xs}) is set by the
$h_1 \, \tilde{\chi}_1^0 \, \tilde{\chi}_1^0$ coupling $g_{h_1 \chi_1 \chi_1}$,
which is given by
\begin{equation} \label{eq:leading-SI-xs-contribution}
  g_{h_1 \chi_1 \chi_1} = \frac{1}{2} \left ( \sqrt{\frac{3}{5}} g_1 N_{14}
  - g_2 N_{13} \right ) \left [ N_{11} (U_h)_{11} - N_{12} (U_h)_{12}
    \right ]\,,
\end{equation}
where the neutralino mixing matrix elements $N_{ij}$ are
defined\footnote{Note that in this convention $N_{11}$ and $N_{12}$ specify
  the Higgsino mixing, while $N_{13}$ and $N_{14}$ give the wino and bino
  mixing respectively.} in Eq.~(\ref{eq:neutralino-diagonalisation}) and the
Higgs mixing matrix $U_h$  is defined by Eq.~(\ref{eq:Hmixing}).  In the
CSE$_6$SSM, the contributions to this coupling involving the singlet mixing
components $N_{1j}$, $j = 5, 6, 7, 8$, are negligible in our case and can be
ignored.  In the highly mixed case with $|\mu| \approx M_1$ and
$N_{13} \lesssim N_{14}$, the products $N_{11} N_{14}$ and $N_{12} N_{14}$ that
appear above are large and the SI cross section is enhanced
\cite{Crivellin:2015bva}.  Therefore points with a mixed bino-Higgsino DM
candidate that saturates the relic abundance are excluded, for
both\footnote{For $\mu_{(\text{eff})} < 0$ the SI cross section is slightly
  smaller, due to a cancellation between the contributions from the up- and
  down-type Higgsinos, but this is not significant enough to evade the current
  limits.} signs of $\mu_{(\text{eff})}$.  As $M_{1/2}$ is increased (decreased)
so that $\tilde{\chi}_1^0$ has a smaller (larger) bino component, the SI cross
section decreases as $N_{14} \to 0$ ($N_{11}, N_{12} \to 0$).  Additionally, the
reduction in $\Omega h^2$ for larger values of $M_{1/2}$ implies a reduction in
the local number density of WIMPs and thereby weakens the limits from direct
detection.  We estimate the extent to which this occurs by rescaling the
given limits by the predicted relic abundance, so that a given set of values
$(m_{\tilde{\chi}_1^0}, \sigma_{\text{SI}}^p)$ is not excluded if
\begin{equation} \label{eq:rescaled-dd-limit}
  \sigma_{\text{SI}}^p \leq \frac{(\Omega h^2)_{\text{exp.}}}{
    (\Omega h^2)_{\text{th.}}}
  \sigma_{\text{SI}}^{p,\text{LUX}}(m_{\tilde{\chi}_1^0}),
\end{equation}
where $\sigma_{\text{SI}}^{p,\text{LUX}}(m_{\tilde{\chi}_1^0})$ is the \lux\
limit at the WIMP mass $m_{\tilde{\chi}_1^0}$.  Thus points away from the
well-tempered strip may still avoid the direct detection limits.  In the
CSE$_6$SSM, the presence of the $A$-funnel region also allows for solutions
with $(\Omega h^2)_{\text{th.}} \approx (\Omega h^2)_{\text{exp.}}$ and a
predicted SI cross section below current limits for $\lambda < 0$.
Nevertheless, as discussed below future limits are expected to probe a
substantial portion of the remaining parameter space.  Therefore scenarios
with small $\mu_{(\text{eff})}$ and a mixed bino-Higgsino $\tilde{\chi}_1^0$ are
very tightly constrained.

\subsection{Pure Higgsino Dark Matter}

\begin{figure}[h!]
  \includegraphics[width=0.5\textwidth]{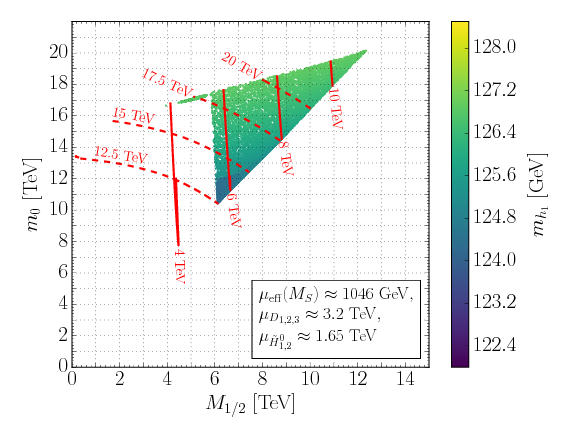}
  \includegraphics[width=0.5\textwidth]{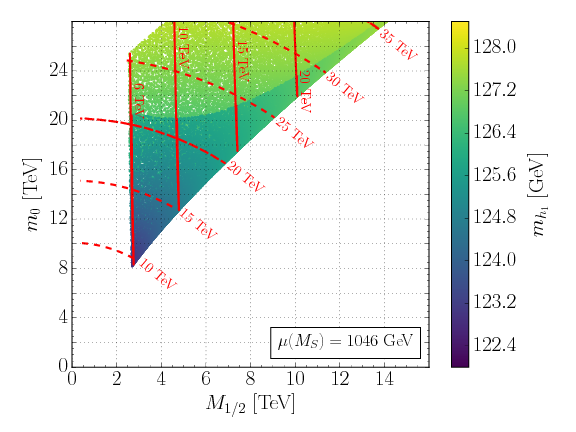}
  \includegraphics[width=0.5\textwidth]{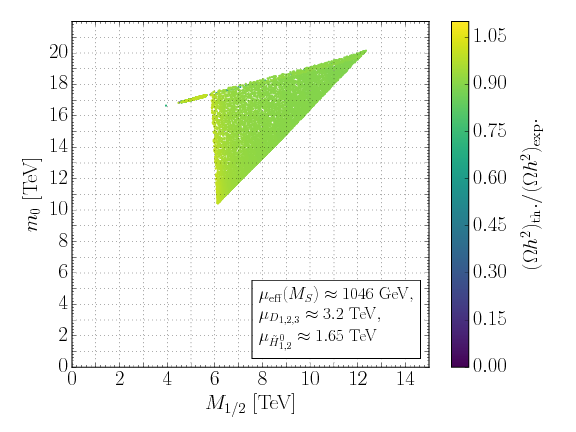}
  \includegraphics[width=0.5\textwidth]{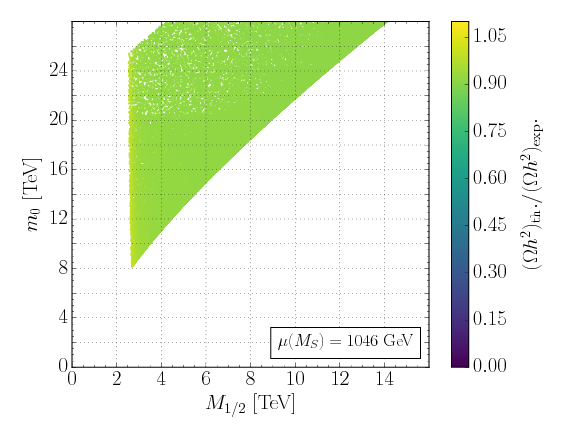}
  \includegraphics[width=0.5\textwidth]{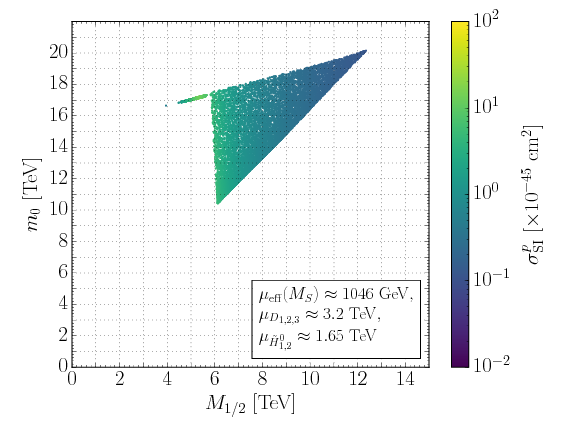}
  \includegraphics[width=0.5\textwidth]{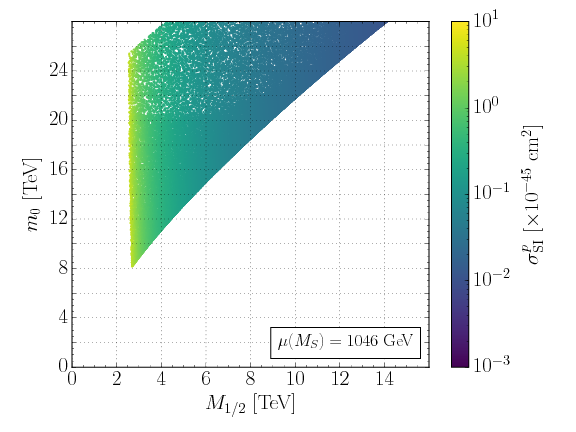}
  \caption{Contour plots in the $M_{1/2} - m_0$ plane of the lightest CP--even
    Higgs mass (top row), DM relic density (middle row) and proton SI cross
    section (bottom row) in the CSE$_6$SSM with $\mu_{\text{eff}}(M_X) \approx
    898$ GeV (left column) and CMSSM with $\mu(M_S) = 1046$ GeV (right column).
    In the top row, we show contours of the gluino (solid lines) and squark
    (dashed lines) masses.}
  \label{fig:mupos1TeV-m12-m0-planes}
\end{figure}

\begin{figure}[h!]
  \includegraphics[width=0.5\textwidth]{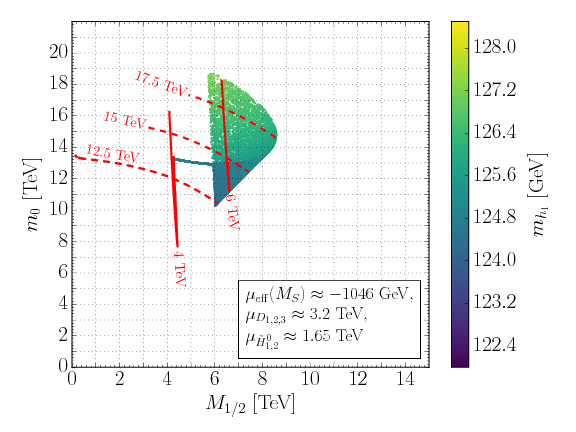}
  \includegraphics[width=0.5\textwidth]{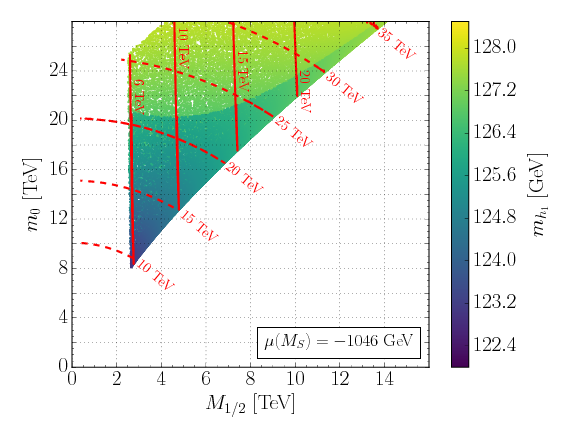}
  \includegraphics[width=0.5\textwidth]{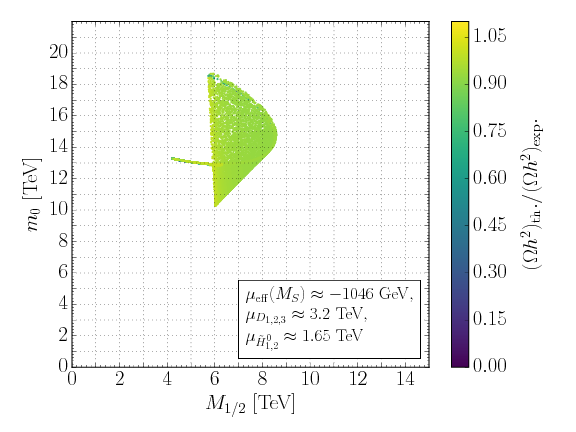}
  \includegraphics[width=0.5\textwidth]{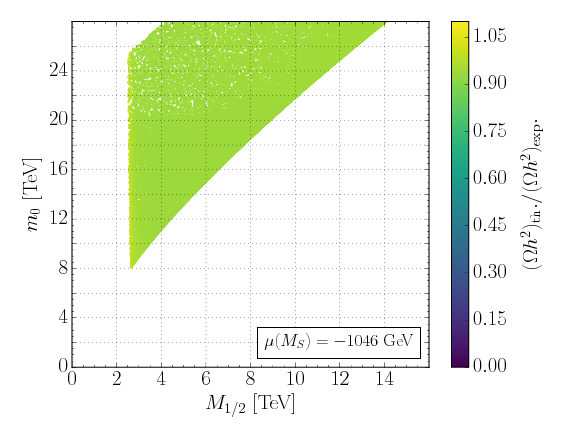}
  \includegraphics[width=0.5\textwidth]{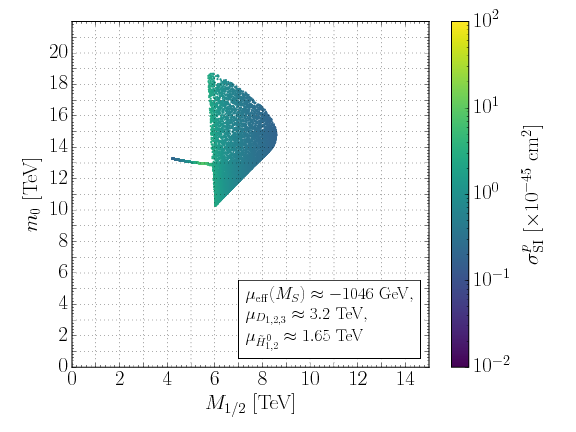}
  \includegraphics[width=0.5\textwidth]{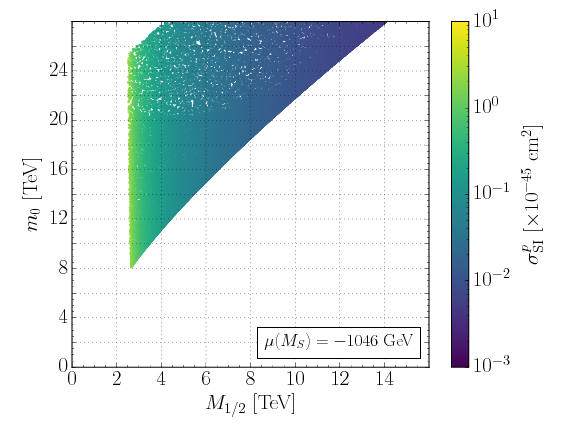}
  \caption{Contour plots in the $M_{1/2} - m_0$ plane of the lightest CP--even
    Higgs mass (top row), DM relic density (middle row) and proton SI cross
    section (bottom row) in the CSE$_6$SSM with $\mu_{\text{eff}}(M_X) \approx
    -898$ GeV (left column) and CMSSM with $\mu(M_S) = -1046$ GeV (right
    column).  In the top row, we show contours of the gluino (solid lines) and
    squark (dashed lines) masses.}
  \label{fig:muneg1TeV-m12-m0-planes}
\end{figure}

Scenarios with a heavy, pure Higgsino DM candidate are less constrained by
direct detection limits due to both the weaker limits at high WIMP masses
and the suppression of the SI scattering cross section for a pure Higgsino
LSP \cite{Cheung:2012qy}.  Analyses of the CMSSM parameter space that also
account for limits from collider searches suggest that this part of the
parameter space is favored by experimental constraints
\cite{Roszkowski:2014wqa}, though scenarios with a relatively light LSP can
still fit the data \cite{Bagnaschi:2015eha}.  To see that this is also true
in the CSE$_6$SSM, in \figref{fig:mupos1TeV-m12-m0-planes} and
\figref{fig:muneg1TeV-m12-m0-planes} we compare the CSE$_6$SSM with
$|\mu_{\text{eff}}(M_S)| \approx 1046$ GeV to the CMSSM with $|\mu(M_S)| = 1046$
GeV.

As in the previous case with small $\mu_{(\text{eff})}$, the region in which we
find solutions in the CSE$_6$SSM is much smaller than in the CMSSM.  The upper
bound on $m_0$ again arises from tachyonic CP--even and CP--odd Higgs states
that occur as $|A_0|$ is increased.  At the same time, the minimum value of
$M_{1/2}$ that satisfies the relic density constraint is much larger.  This is
because a relic density consistent with Eq.~(\ref{eq:planck-relic-density})
requires $\tilde{\chi}_1^0$ to be nearly purely Higgsino with
$m_{\tilde{\chi}_1^0} \approx 1$ TeV, which is achieved for $|M_1| \gtrsim
|\mu_{(\text{eff})}| \approx 1$ TeV.  The condition of universal gaugino masses
at $M_X$ then means that the gluino is now very heavy along with the sfermions.
In the CSE$_6$SSM we find solutions with $m_{\tilde{g}} \geq 3.8$ TeV, compared
to the minimum value of $m_{\tilde{g}} \geq 5.7$ TeV in the CMSSM scan.  The
prospects for an LHC discovery in this scenario are fairly poor in the CMSSM,
as the gluino and all sfermions would be out of reach at run II.

\begin{figure}[h!]
  \includegraphics[width=0.5\textwidth]{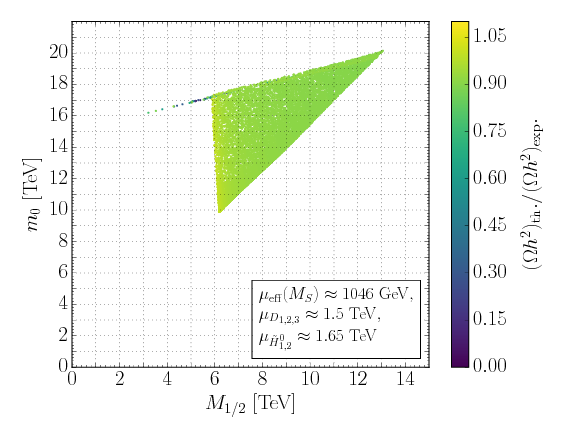}
  \includegraphics[width=0.5\textwidth]{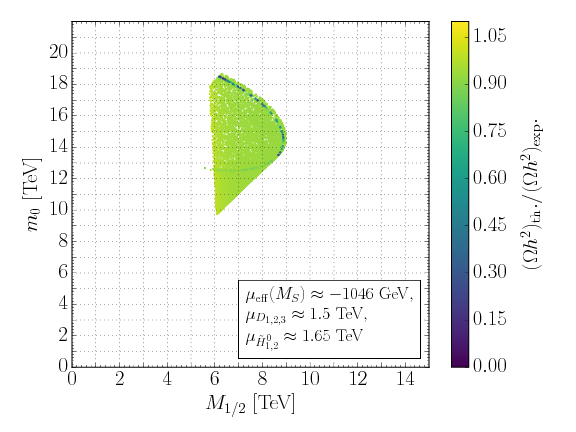}
  \caption{Contour plots in the $M_{1/2} - m_0$ plane of the DM relic density in
    the CSE$_6$SSM with $\mu_{\text{eff}}(M_S) \approx 1046$ GeV (left) and
    $\mu_{\text{eff}}(M_S) \approx -1046$ GeV (right), with reduced values of
    the exotic Yukawa couplings $\kappa_{ij}(M_X)$ such that $\mu_{D_i} \approx
    1.5$ TeV.}
  \label{fig:light-exotics-m12-m0-plane}
\end{figure}

For the CSE$_6$SSM points shown in \figref{fig:mupos1TeV-m12-m0-planes} and
\figref{fig:muneg1TeV-m12-m0-planes} we considered slightly larger
exotic couplings with $\kappa_0 = \tilde{\lambda}_0 = 3 \times
10^{-3}$.  The couplings are required to be large enough to ensure that
$\tilde{\chi}_1^0$ is still the stable second DM candidate, rather
than one of the exotic sector possibilities.  The exotic fermions are
correspondingly heavier, with masses satisfying $3 \text{ TeV} \leq
\mu_{D_i} \leq 3.3$ TeV and $1.63 \text{ TeV} \leq\mu_{\tilde{H}_{I\alpha}^0}
\leq 1.67$ TeV, which also makes them unlikely to be observable at run II or
at the HL-LHC.  Note however that, in addition to being able to vary
$M_{Z^\prime}$, there is also some freedom to vary the exotic couplings to
obtain lighter exotic states.  We illustrate this in
\figref{fig:light-exotics-m12-m0-plane}, where we plot the valid solutions
with $\kappa_0 = 1.4 \times 10^{-3}$, giving $D$ fermion masses of
$\mu_{D_i} \in [1.5 \text{ TeV}, 1.6 \text{ TeV}]$, comparable with the
potential exclusion reach for third generation squarks at the HL-LHC
\cite{Ruhr:2016xsg}.  For fixed $|\lambda(M_X)| = 2.4 \times 10^{-3}$ the
effect of this is to slightly increase the minimum allowed value of $M_{1/2}$
outside of the $A$-funnel region. This is due to an increase in the calculated
$(\Omega h^2)_{\text{th.}}$, which was already rather close to the value from
Planck observations.  The larger value of the relic density in turn arises
because of the increase in $\mu_{\text{eff}}(M_S)$ that results for smaller
values of $\kappa_0$ in the RG running; this can be seen, for example, from
Eq.~(\ref{eq:beta-lambda-one-loop}).  A compensating small reduction in
$\lambda(M_X)$ can be used to maintain the low-energy value of
$\mu_{\text{eff}}$ and therefore $(\Omega h^2)_{\text{th.}}$, in which case the
smaller values of $M_{1/2}$ shown in \figref{fig:mupos1TeV-m12-m0-planes} and
\figref{fig:muneg1TeV-m12-m0-planes} continue to be allowed.  The presence of
light exotics is an important possible signature that allows the model to be
discovered when the SUSY breaking scale is very heavy, as well as
distinguishing the $E_6$ inspired model from the CMSSM.

As can be seen in the middle rows of
\figref{fig:mupos1TeV-m12-m0-planes} and
\figref{fig:muneg1TeV-m12-m0-planes}, and in
\figref{fig:light-exotics-m12-m0-plane}, the prediction for the relic
density in the CSE$_6$SSM remains similar to that in the CMSSM.  In
both models a Higgsino with a mass of approximately $1$ TeV saturates
the observed value in Eq.~(\ref{eq:planck-relic-density}).  The narrow
$A$-funnel region at lower $M_{1/2}$ is again accessible in the
CSE$_6$SSM by tuning $A_0$ to reduce $m_{A_1}$.  As large mixings are
no longer required to reproduce the relic density for
$|\mu_{(\text{eff})}| \approx 1$ TeV, a large fraction of the
solutions found have a predicted SI cross section below the current
\lux\ limits.  Points in both models with $M_{1/2}$ where the LSP
transitions from being pure bino to pure Higgsino, i.e., where $M_1
\approx \mu_{(\text{eff})}$ near the lower bound on $M_{1/2}$, present a
larger cross section that is in excess of the \lux\ limits.  Therefore
even for heavy $\mu_{(\text{eff})}$ in the CMSSM and CSE$_6$SSM
constraints can be put on the parameter space by direct detection
searches.  At larger $M_{1/2}$ (that is, where $M_1$ is significantly
larger than $\mu_{(\text{eff})}$) the models currently evade the SI
direct detection limits, and are very unlikely to be probed by direct
collider searches in the near future if the exotic fermions in the
CSE$_6$SSM are not light.  However, this part of the CSE$_6$SSM, and
CMSSM, parameter space will be constrained by results from \xenon, as
we now discuss in more detail.

\subsection{Impact of Current and Future Searches}
\begin{figure}[h!]
  \includegraphics[width=0.5\textwidth]{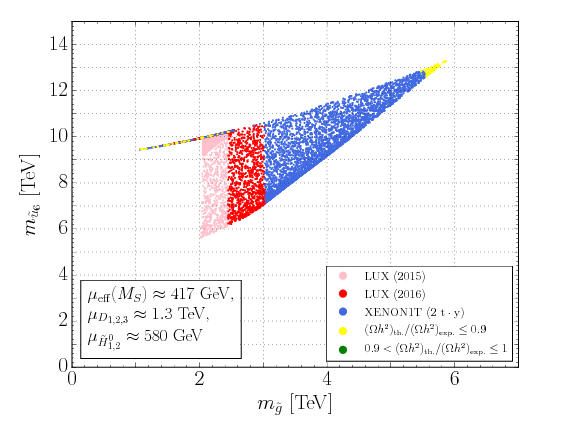}
  \includegraphics[width=0.5\textwidth]{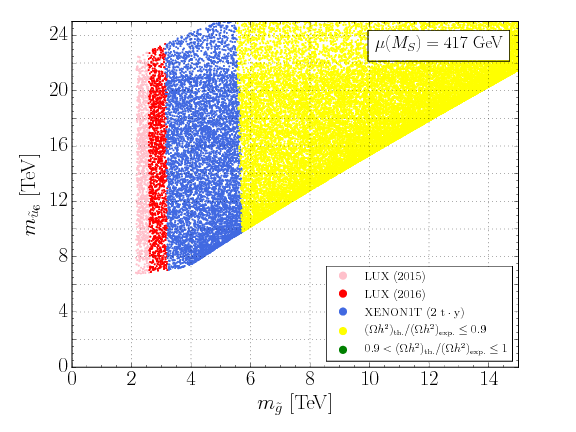}
  \includegraphics[width=0.5\textwidth]{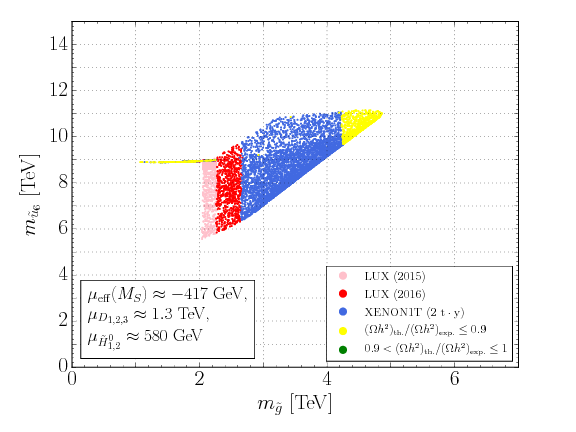}
  \includegraphics[width=0.5\textwidth]{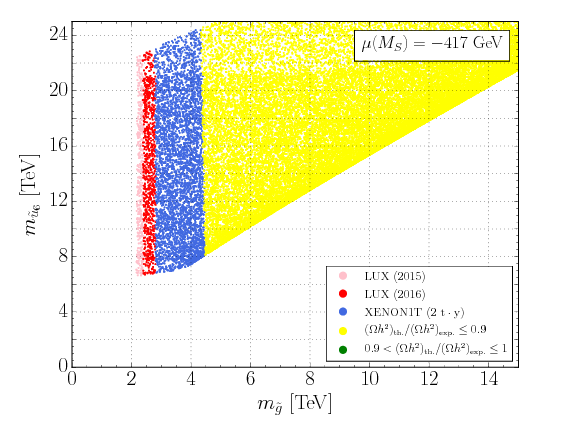}
  \caption{Plots of direct detection and collider constraints in the
    $m_{\tilde{g}} - m_{\tilde{u}_6}$ plane in the CSE$_6$SSM with
    $|\mu_{\text{eff}}(M_X)| \approx 347$ GeV (left column) and the CMSSM with
    $|\mu(M_S)| = 417$ GeV (right column).  In the top row,
    $\mu_{(\text{eff})}(M_X) > 0$, and in the bottom row
    $\mu_{(\text{eff})}(M_X) < 0$.  In each plot, we show points
    that have a SI cross section in excess of the 2015 \cite{Akerib:2015rjg}
    and 2016 \cite{Akerib:2016vxi} \lux\ limits (pink and red, respectively)
    and points that are not currently excluded but are within the projected
    reach \cite{Aprile:2015uzo} of \xenon\ (blue).  In each case, the
    exclusion limit is determined according to Eq.~(\ref{eq:rescaled-dd-limit}).
    Finally, points that are not excluded by any limits but that predict a
    relic density that is less than 90\% of the measured value are shown in
    yellow, while those points with $0.9 < (\Omega h^2)_{\text{th}.} /
    (\Omega h^2)_{\text{exp}.} \leq 1$ are shown in green.}
  \label{fig:mu400GeV-exclusions}
\end{figure}

In \figref{fig:mu400GeV-exclusions} we show the current and future regions
probed by \lux\ and \xenon\ for $|\mu_{(\text{eff})}(M_S)| \approx 417$ GeV in
both models.  As described above, the existing 2015 \lux\ limits already
essentially exclude the well-tempered bino-Higgsino solution region at low
$m_{\tilde{g}}$, i.e., low $M_{1/2}$, where the SI cross section is enhanced by
large mixings.  The effect of the new 2016 limit is to extend this exclusion
to larger gluino masses, despite the reduction in the predicted relic density
and SI cross section.  This is as expected from the results of dedicated MSSM
studies \cite{Cabrera:2016wwr,Baer:2016ucr}.  \xenon\ \cite{Aprile:2015uzo} is
projected to exclude (or discover) even larger values of $m_{\tilde{g}}$.  In
this CMSSM scenario, \xenon\ can potentially exclude $\tilde{g}$ masses up
to $4 - 5$ TeV.

The exclusions set by direct detection searches in the CSE$_6$SSM are to some
extent similar to those in the CMSSM.  In particular, outside of the $A$--funnel
region in the CSE$_6$SSM, the \lux\ limits exclude gluino masses
$m_{\tilde{g}} \lesssim 3$ TeV for $\mu_{(\text{eff})} > 0$ and $m_{\tilde{g}}
\lesssim 2.5$ TeV for $\mu_{(\text{eff})} < 0$ in both models.  Similarly,
\xenon\ will be able to probe gluino masses up to $4 - 5$ TeV in the CSE$_6$SSM
as well.  This accounts for a large fraction of as yet unexcluded solutions in
the CSE$_6$SSM.

\begin{figure}[h!]
  \includegraphics[width=0.5\textwidth]{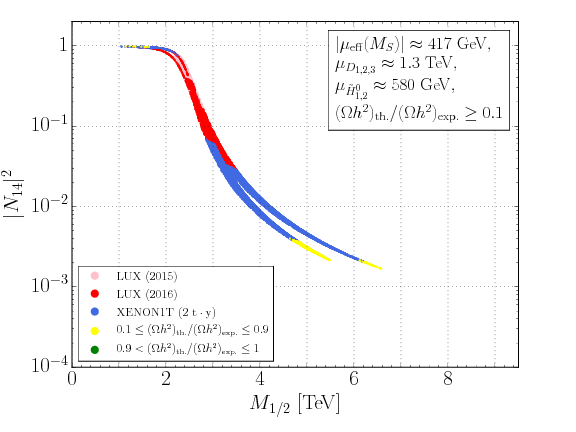}
  \includegraphics[width=0.5\textwidth]{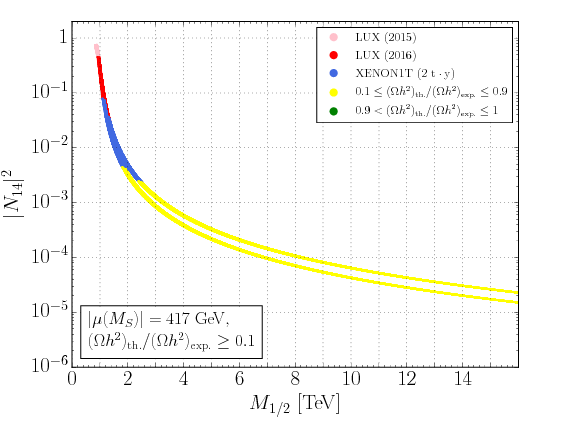}
  \caption{Plots showing points excluded by direct detection constraints
    in the $M_{1/2} - |N_{14}|^2$ plane in the CSE$_6$SSM (left) and CMSSM
    (right) for $|\mu_{(\text{eff})}(M_S)| \approx 417$ GeV, after also
    requiring that the LSP accounts for at least $10$\% of the observed relic
    density.  The scaling of the limits and the colour coding is otherwise
    the same as in \figref{fig:mu400GeV-exclusions}.}
  \label{fig:mu400GeV-mixing-exclusions}
\end{figure}

However, as can be seen from the left column of
\figref{fig:mu400GeV-exclusions}, some points in the $A$-funnel region will
still not be excluded by \lux\ or \xenon.  These points have a suppressed SI
cross section or do not saturate the relic density bound, or both.  This is
also true in both models for those points not excluded at large $m_{\tilde{g}}$.
Points close to the well-tempered region, where the amount of mixing is still
relatively large, only escape being excluded if they lead to an extremely small
relic density.  If it is required that the LSP explains a substantial fraction
of the observed relic abundance, for example $(\Omega h^2)_{\text{th.}} /
(\Omega h^2)_{\text{exp.}} > 0.1$, then these points are removed.  This is
illustrated in \figref{fig:mu400GeV-mixing-exclusions}, where we show the
variation in the bino fraction for points satisfying this criterion.  The
effect of the direct detection limits is to heavily restrict the amount of
mixing allowed.  The surviving points are forced to either be almost pure bino,
at small $M_{1/2}$, or almost pure Higgsino at large $M_{1/2}$ and hence having
a heavy SUSY spectrum.

While the $A$-funnel points will not be observable at \xenon, the fact that
$m_{\tilde{g}} \lesssim 2$ TeV for these solutions means that most are in reach
of LHC searches targeting gluinos.  This highlights the complementary nature
of collider and direct detection searches; similar observations have been made
for the CMSSM (see, for example, Ref.~\cite{Kowalska:2015kaa}).  Given the
similarity of the lightest $Z_2^E = +1$ neutralinos in the CSE$_6$SSM to the
ordinary MSSM neutralino sector, it is not so surprising that this continues
to hold.  In particular, results from \xenon\ will be able to constrain the
CSE$_6$SSM (and CMSSM) at much higher SUSY scales than are expected to be
reached at the LHC.  We conclude from this that direct detection searches, if
no WIMPs are observed, will be able to place indirect limits on the sparticle
masses much higher than can be achieved at run II, when the neutralino does not
annihilate via special mechanisms such as the $A$-funnel.  Thus direct
detection limits are a particularly strong constraint on the CSE$_6$SSM
parameter space.

\begin{figure}[h!]
  \includegraphics[width=0.5\textwidth]{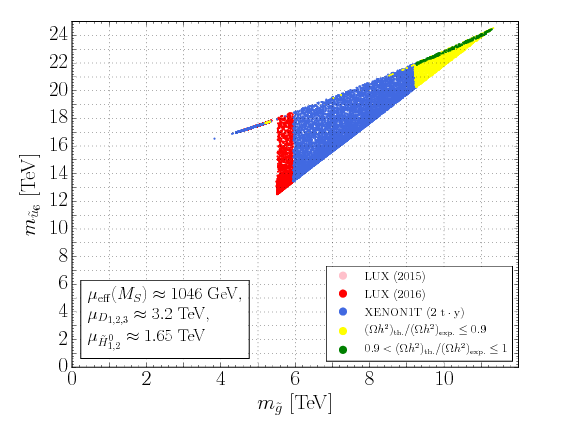}
  \includegraphics[width=0.5\textwidth]{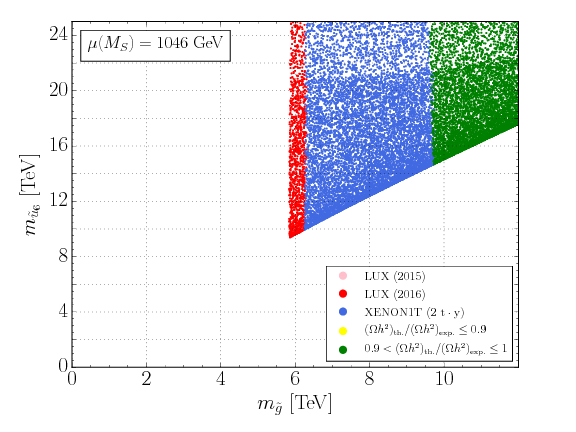}
  \includegraphics[width=0.5\textwidth]{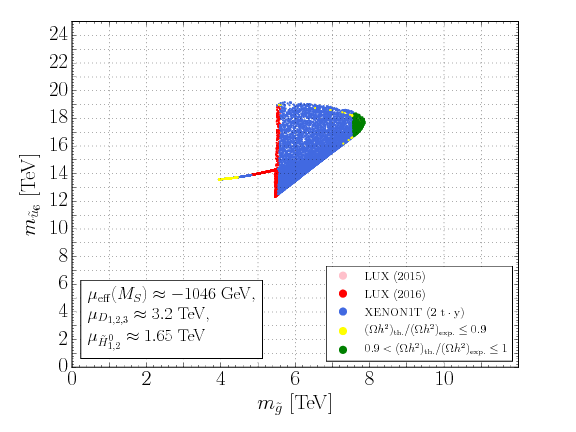}
  \includegraphics[width=0.5\textwidth]{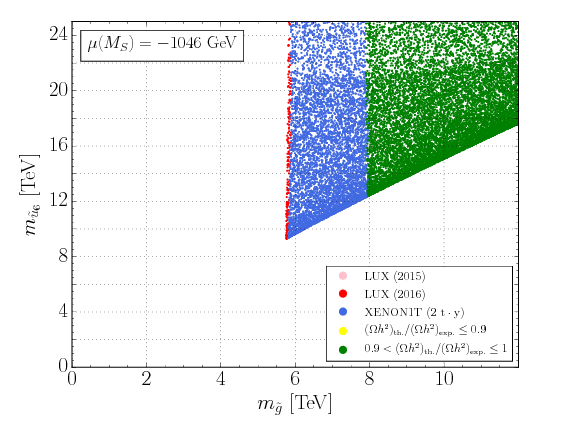}
  \caption{Plots of constraints in the $m_{\tilde{g}} - m_{\tilde{u}_6}$
    plane in the CSE$_6$SSM with $|\mu_{\text{eff}}(M_X)| \approx 898$ GeV
    (left column) and the CMSSM with $|\mu(M_S)| = 1046$ GeV (right column).
    In the top row, $\mu_{(\text{eff})}(M_X) > 0$, and in the bottom row
    $\mu_{(\text{eff})}(M_X) < 0$.  The color coding is the
    same as in \figref{fig:mu400GeV-exclusions}.}
  \label{fig:mu1TeV-exclusions}
\end{figure}

The solutions that we find with a heavy Higgsino DM candidate lead to gluino
and MSSM sfermion masses beyond the exclusion reach at run II. This is shown in
\figref{fig:mu1TeV-exclusions}.  Consequently there are effectively no
constraints on this part of parameter space coming from collider limits, at
least in the CMSSM.  In the CSE$_6$SSM, the possibility of light exotic
fermions, as in \figref{fig:light-exotics-m12-m0-plane}, would allow for the
model to be discovered even if all MSSM-like states and exotic scalars are
heavy.  However, if these states are also heavy then limits from direct
detection searches are much more effective at constraining the parameter space.

\begin{figure}[h!]
  \includegraphics[width=0.5\textwidth]{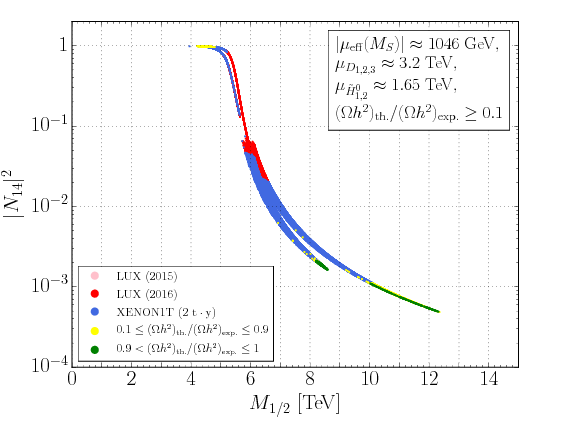}
  \includegraphics[width=0.5\textwidth]{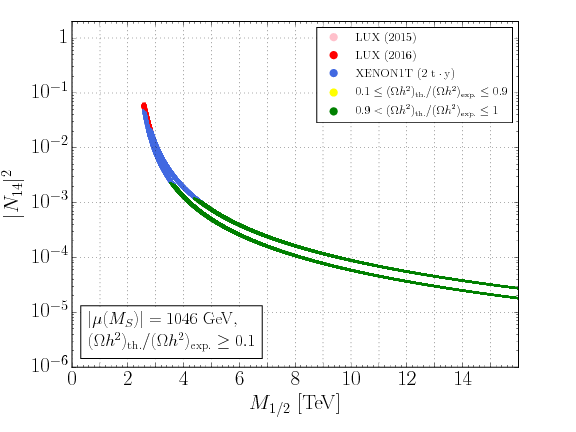}
  \caption{Plots showing points excluded by direct detection constraints
    in the $M_{1/2} - |N_{14}|^2$ plane in the CSE$_6$SSM (left) and CMSSM
    (right) for $|\mu_{(\text{eff})}(M_S)| \approx 1046$ GeV, after also
    requiring that the LSP accounts for at least $10$\% of the observed
    relic density.  The scaling of the limits and the colour coding is the
    same as in \figref{fig:mu400GeV-exclusions}.}
  \label{fig:mu1TeV-mixing-exclusions}
\end{figure}

Prior to the most recent \lux\ limits, all of our solutions with heavy
$|\mu_{(\text{eff})}|$ were consistent with direct detection limits.  This is no
longer true for the new 2016 \lux\ limits, which now exclude points with
$M_1 \approx \mu_{(\text{eff})}$.  Therefore the current direct detection limits
are already probing the heavy $|\mu_{(\text{eff})}|$ parameter space.
Scenarios with a highly mixed bino-Higgsino $\tilde{\chi}_1^0$ accounting for
at least $10$\% of the relic abundance are again all excluded by the
current limits.  This is shown in \figref{fig:mu1TeV-mixing-exclusions}.  Thus
in the case that the LSP is relevant for addressing the DM problem, direct
detection limits place stringent constraints on the allowable bino-Higgsino
admixture.  More extensive coverage of the valid, low mixing regions will
require results from \xenon, however.

It is clear that in the CSE$_6$SSM, results from \xenon\ will place very strong
constraints on the parameter space, as it should be possible to cover almost
all of the allowed region.  As for the previous small $|\mu_{(\text{eff})}|$
case, the surviving regions are the $A$-funnel region and at very
large $m_{\tilde{g}}$.  In this scenario the $A$-funnel region cannot be
searched for directly at the LHC; from the left column of
\figref{fig:mu1TeV-exclusions} it can be seen that the gluino mass
is always greater than $\approx 4$ TeV.  An interesting question is to what
extent indirect DM detection experiments or results from flavor physics can
constrain the CSE$_6$SSM here; we leave this for a future study.  On the other
hand, for very heavy spectra without light exotic fermions neither collider
searches nor results from \xenon\ will constrain the CSE$_6$SSM or the CMSSM.
Even more sensitive direct detection experiments, such as results from LZ,
will be required to directly search for these scenarios.

It should be noted that the large number of solutions for which
$(\Omega h^2)_{\text{th.}}$  is indicated as being less than $90$\% of the
Planck value in \figref{fig:mu1TeV-exclusions} still account for a very large
fraction of the observed relic abundance.  Small changes in $\lambda(M_X)$, or
$\mu(M_X)$ in the CMSSM, are enough to closely reproduce the value in
Eq.~(\ref{eq:planck-relic-density}) without significantly changing any other
results, unlike in the light Higgsino case where the DM candidate is severely
underabundant assuming a standard freeze-out scenario.  At large $M_{1/2}$ the
relic density is still fully accounted for by the Higgsino DM candidate.
Unfortunately, while these scenarios can explain the observed DM density
entirely, the expected collider phenomenology is rather uninteresting as all
states are too heavy to be observable.

\section{Conclusions} \label{sec:conclusions}
We have studied dark matter and LHC phenomenology implications in both
the CMSSM and a constrained version of an $E_6$ inspired model (CSE$_6$SSM).
The SE$_6$SSM is a string inspired alternative to the MSSM, where the
break down of the $E_6$ gauge group leads to a discrete $R$--parity and
a $U(1)_N$ gauge extension surviving to the TeV scale that forbids the
$\mu$-term of the MSSM.  The charges allow the standard see-saw
mechanism for neutrino masses and a leptogenesis explanation of the
matter-anti-matter asymmetry.  The model contains exotic states at low
energies needed to fill three generations of complete $\bm{27}$-plet
representations of $E_6$ and ensure anomaly cancellation, and can give
rise to spectacular collider signatures.  A single additional discrete
symmetry which commutes with $E_6$ is imposed to forbid FCNCs and this
along with $R$--parity leads to multiple dark matter candidates.  In this paper
we focused on scenarios where the lightest exotic particle is an extremely
light singlino which forms hot dark matter, but contributes negligibly to
the relic density.  We showed that the relic density can instead be
explained entirely by the lightest MSSM-like neutralino.

We have performed a detailed exploration of the parameter space of
both the CMSSM and CSE$_6$SSM and compared the results.  We find that
in both models one may fit the observed relic density with a pure
Higgsino neutralino that has a mass around $1$ TeV.  Alternatively this
can be achieved with a mixed bino-Higgsino dark matter candidate,
requiring a fine tuning of $M_1$ and $\mu_{\text{(eff)}}$ to obtain
the well-tempered strip and this can work for lighter neutralino
masses ($\approx 400$ GeV in our example).  However recent direct detection
results have placed strong limits on this mixing, placing a significant
tension between fitting the observed relic density and evading direct
detection limits.  Indeed we find that the recent \lux\ 2016 direct detection
limits constrain Higgsino-bino mixing such that it rules out this
well-tempered strip for both models for light and heavy neutralinos.

However we also found that the CSE$_6$SSM can have special $A$-funnel
solutions where the correct relic density can be achieved for lighter
$M_{1/2}$, a scenario that is only possible in the CMSSM for a much
larger $\tan \beta$ than is considered here.  Such scenarios exist for
both the heavier and lighter Higgsino masses considered. For lighter
Higgsino masses this $A$-funnel region, which can escape direct
detection limits even from the future results of \xenon, will be
probed by the LHC run II.  This demonstrates an important
complementarity between collider searches and experiments for the
direct detection of dark matter.

Such special regions aside however it is now rather difficult to
explain dark matter in the lighter scenarios.  Nonetheless if one
requires only that the relic density is not too large then many
scenarios are still viable and have phenomenology that will be probed
with run II of the LHC.  Since the sfermions will still be very heavy
the main signatures arise from the production of gluinos, charginos
and neutralinos, with MSSM-like signatures.  On the other hand the
leptoquarks in the CSE$_6$SSM can be light enough to detect even when
the SUSY scale is very heavy.  These exotic states would lead to
considerable enhancement of $p\,p \to
t\,\bar{t}\,\tau^+\,\tau^-+E_t^{miss}+X$ and $p\,p \to b\,\bar{b}\,
\tau^+\,\tau^- + E_T^{miss} + X$, where $X$ stands for any light quark
or gluon jets.

Heavier scenarios with a Higgsino dark matter candidate of
around $1$ TeV are also not currently constrained so much by direct
detection and it is possible to fit the relic density in both the
CMSSM and CSE$_6$SSM for a wide range of the parameter space.  These
scenarios have a rather heavy spectrum which is not accessible to the
LHC, however they will be probed by future direct detection
experiments, such as \xenon\ which will be able to probe most of the
viable solutions we have found in the CSE$_6$SSM.  Therefore the
future impact of \xenon\ on these models will be very significant.

\section*{Acknowledgements}
PA acknowledges helpful discussions with Tom Steudner and Dominik
St\"ockinger regarding the Higgs mass calculation. This work was
supported by the University of Adelaide, Monash University and the Australian
Research Council through the ARC Centre of Excellence for Particle Physics at
the Terascale (CoEPP) (CE110001104).

\appendix

\numberwithin{equation}{section}
\renewcommand{\theequation}{\Alph{section}.\arabic{equation}}

\section{RGEs}
\label{app:rges}
In our analysis, the SUSY preserving and soft SUSY breaking parameters at $M_S$
are obtained from the GUT scale boundary conditions by running them using
two-loop RGEs.  These RGEs were automatically derived using \SARAHv, which
makes use of the general results given in Refs.~\cite{Martin:1993zk,
  Fonseca:2011vn,Goodsell:2012fm}.  For completeness, in this appendix we
summarize the complete set of RGEs used to obtain our results.  For a general
parameter $p$, the RG equation for $p$ is expressed in terms of the one- and
two-loop $\beta$ functions, $\beta_p^{(1)}$ and $\beta_p^{(2)}$ respectively,
according to
\begin{equation} \label{eq:two-loop-rge}
  \frac{\mathrm{d} p(t)}{\mathrm{d} t} = \beta_p =
  \frac{\beta_p^{(1)}}{(4\pi)^2} + \frac{\beta_p^{(2)}}{(4\pi)^4} ,
\end{equation}
where $t = \ln Q / M_X$ gives the scale at which $p$ is evaluated.

\subsection{Gauge Couplings}
In general, kinetic mixing of the $U(1)_Y$ and $U(1)_N$ leads to a set of
RGEs for the Abelian gauge couplings involving a set of off-diagonal gauge
couplings.  In the triangle basis of
Eq.~(\ref{eq:off-diagonal-gauge-couplings}), these RGEs can be written
\begin{equation} \label{eq:abelian-gauge-coupling-rge}
  \frac{\mathrm{d} G}{\mathrm{d}t} = G \times B ,
\end{equation}
where the matrix of $\beta$ functions is
\begin{equation}
  B = \begin{pmatrix}
    \beta_{g_1} g_1^2 & 2 g_1 g_1' \beta_{g_{11}} + 2 g_1 g_{11} \beta_{g_1} \\
    0 & g_1'^2 \beta_{g_1'} + 2 g_1' g_{11} \beta_{g_{11}}
    + g_{11}^2 \beta_{g_1}
  \end{pmatrix} .
\end{equation}
The off-diagonal $\beta$ function $\beta_{g_{11}}$ is rather small, with
$\beta_{g_{11}}^{(1)} = - \sqrt{6} / 5$ at one-loop.  As discussed in
\secref{sec:se6ssm}, the effects of kinetic mixing are therefore small if
$g_{11}$ vanishes at the GUT scale, and so we neglect it.  When this is done,
the two-loop RGEs for the diagonal Abelian gauge couplings are
\begin{align}
  \beta_{g_1}^{(1)} &= \frac{48}{5} g_{1}^{3} , \label{eq:beta-g1-one-loop} \\
  \beta_{g_1}^{(2)} &= g_{1}^{3} \Bigl [ \frac{234}{25} g_1^2 +
    \frac{81}{25} g_1'^2 + \frac{54}{5} g_2^2 + 24 g_3^2
    - \frac{6}{5} |\lambda|^2 - \frac{6}{5} |\tilde{\sigma}|^2
    - \frac{26}{5} \trace{y^U y^{U \dagger}} \nonumber \\
  & \quad {} - \frac{14}{5} \trace{y^D y^{D \dagger}}
    - \frac{18}{5} \trace{y^E y^{E \dagger}}
    - \frac{4}{5} \trace{\kappa \kappa^{\dagger}}
    - \frac{6}{5} \trace{\tilde{\lambda} \tilde{\lambda}^{\dagger}}
    \nonumber \\
  & \quad {} - \frac{6}{5} \trace{f f^{\dagger}}
    - \frac{6}{5} \trace{\tilde{f} \tilde{f}^{\dagger}}
    - \frac{14}{5} \trace{g^D g^{D \dagger}}
    - \frac{18}{5} \trace{h^E h^{E \dagger}} \Bigr ] ,
    \label{eq:beta-g1-two-loop} \\
  \beta_{g_1^\prime}^{(1)} &= \frac{213}{20} g_1'^3 ,
    \label{eq:beta-g1p-one-loop} \\
  \beta_{g_1^\prime}^{(2)} &= g_1'^3 \Bigl [ \frac{81}{25} g_1^2
    + \frac{2457}{200} g_1'^2 + \frac{51}{5} g_2^2 + 24 g_3^2
    - \frac{19}{5} |\lambda|^2 - \frac{5}{2} |\sigma|^2 \nonumber \\
  & \quad {} - \frac{4}{5} |\tilde{\sigma}|^2
    - \frac{9}{5} \trace{y^U y^{U \dagger}}
    - \frac{21}{5} \trace{y^D y^{D \dagger}}
    - \frac{7}{5} \trace{y^E y^{E \dagger}} \nonumber \\
  & \quad {} - \frac{57}{10} \trace{\kappa \kappa^{\dagger}}
    - \frac{19}{5} \trace{\tilde{\lambda} \tilde{\lambda}^{\dagger}}
    - \frac{19}{5} \trace{f f^{\dagger}} \nonumber \\
  & \quad {} - \frac{19}{5} \trace{\tilde{f} \tilde{f}^{\dagger}}
    - \frac{21}{5} \trace{g^D g^{D \dagger}}
    - \frac{7}{5} \trace{h^E h^{E \dagger}}
    \Bigr ] . \label{eq:beta-g1p-two-loop}
\end{align}
The $\beta$ functions for the $SU(2)_L$ and $SU(3)_C$ gauge couplings are
the same irrespective of whether or not the kinetic mixing is taken into
account.  They are
\begin{align}
  \beta_{g_2}^{(1)} &= 4 g_2^3, \label{eq:beta-g2-one-loop} \\
  \beta_{g_2}^{(2)} &= g_2^3 \Bigl [ \frac{18}{5} g_1^2 + \frac{17}{5} g_1'^2 +
    46 g_2^2 + 24 g_3^2 - 2 |\lambda|^2 - 2 |\tilde{\sigma}|^2
    - 6 \trace{y^U y^{U \dagger}}
    \nonumber \\
  & \quad {} - 6 \trace{y^D y^{D \dagger}} - 2 \trace{y^E y^{E \dagger}}
    - 2 \trace{\tilde{\lambda} \tilde{\lambda}^{\dagger}}
    - 2 \trace{f f^{\dagger}}
     \nonumber \\
   & \quad {} - 2 \trace{\tilde{f} \tilde{f}^{\dagger}}
     - 6 \trace{g^D g^{D \dagger}} - 2 \trace{h^E h^{E \dagger}}
    \Bigr ] ,
  \label{eq:beta-g2-two-loop} \\
  \beta_{g_3}^{(1)} &= 0 , \label{eq:beta-g3-one-loop} \\
  \beta_{g_3}^{(2)} &= g_3^3 \Bigl [ 3 g_1^2 + 3 g_1'^2 + 9 g_{2}^{2}
    + 48 g_3^2 - 4 \trace{y^U y^{U \dagger}} - 4 \trace{y^D y^{D \dagger}}
    \nonumber \\
    & \quad {} -2 \trace{\kappa \kappa^{\dagger}} - 4 \trace{g^D g^{D \dagger}}
    \Bigr ] .
  \label{eq:beta-g3-two-loop}
\end{align}

\subsection{Superpotential Trilinear Couplings}
When gauge kinetic mixing is neglected, the running of the dimensionless
superpotential couplings is described by the following two-loop $\beta$
functions:


\subsection{Superpotential Bilinear and Linear Couplings}
The $\beta$ functions of the bilinear superpotential parameters
$\mu_\phi$ and $\mu_L$ read
\begin{align}
  \beta_{\mu_\phi}^{(1)} &= 2 \mu_{\phi} \Bigl ( 2 |\kappa_{\phi}|^2
    + 2 |\tilde{\sigma}|^2 + |\sigma|^2 \Bigr ) ,
    \label{eq:beta-muphi-one-loop} \\
  \beta_{\mu_\phi}^{(2)} &= \mu_{\phi} \Bigl \{
    -16 |\kappa_{\phi}|^4 - 8 |\tilde{\sigma}|^4 - 4 |\sigma|^4
    + |\tilde{\sigma}|^2 \Bigl [ \frac{12}{5} g_1^2
    + \frac{8}{5} g_1'^2 + 12 g_2^2 - 16 |\kappa_{\phi}|^2 \nonumber \\
  & \quad {} - 12 \trace{g^D g^{D \dagger}}
    - 4 \trace{h^E h^{E \dagger}} \Bigr ] + |\sigma|^2 \Bigl [ 5 g_1'^2
    - 4 |\lambda|^2 - 8 |\kappa_{\phi}|^2 \nonumber \\
  & \quad {} - 4 \trace{\tilde{\lambda} \tilde{\lambda}^{\dagger}}
    - 6 \trace{\kappa \kappa^{\dagger}} \Bigr ] \Bigr \} ,
    \label{eq:beta-muphi-two-loop} \\
  \beta_{\mu_L}^{(1)} &= \mu_L \Bigl [ 2 |\tilde{\sigma}|^2
    + 3 \trace{g^D g^{D \dagger}} + \trace{h^E h^{E \dagger}}
    - \frac{3}{5} g_1^2  - \frac{2}{5} g_1'^2 - 3 g_2^2 \Bigr ] ,
    \label{eq:beta-muL-one-loop} \\
  \beta_{\mu_L}^{(2)} &= \mu_L \Bigl \{ \frac{297}{50} g_1^4
    + \frac{217}{50} g_1'^4 + \frac{33}{2} g_2^4 + \frac{18}{25} g_1^2 g_1'^2
    + \frac{9}{5} g_1^2 g_2^2 + \frac{6}{5} g_1'^2 g_2^2 \nonumber \\
  & \quad {} - 2 |\tilde{\sigma}|^2 \Bigl ( 2 |\kappa_{\phi}|^2
    + 3 |\tilde{\sigma}|^2+ |\sigma|^2 \Bigr )
    + \frac{2}{5} g_1^2 \Bigl [ -\trace{g^D g^{D \dagger}}
    + 3 \trace{h^E h^{E \dagger}} \Bigr ] \nonumber \\
  & \quad {} + \Bigl ( \frac{3}{10} g_1'^2 - |\tilde{\sigma}|^2 \Bigr )
    \Bigl [ 3 \trace{g^D g^{D \dagger}} + \trace{h^E h^{E \dagger}} \Bigr ]
    + 16 g_3^2 \trace{g^D g^{D \dagger}} \nonumber \\
  & \quad {} - \trace{\tilde{f} h^{E \dagger} h^E \tilde{f}^{\dagger}}
    - 9 \trace{g^D g^{D \dagger} g^D g^{D \dagger}}
    - 3 \trace{g^D g^{D \dagger} y^{D T} y^{D *}} \nonumber \\
  & \quad {} - 3 \trace{g^D g^{D \dagger} y^{U T} y^{U *}}
    - 3 \trace{g^D \kappa^{\dagger} \kappa g^{D \dagger}}
    - 3 \trace{h^E h^{E \dagger} h^E h^{E \dagger}} \nonumber \\
  & \quad {} - 2 \trace{h^E h^{E \dagger} y^E y^{E \dagger}}
    - \trace{h^E \tilde{\lambda}^{\dagger} \tilde{\lambda} h^{E \dagger}}
    \Bigr \} , \label{eq:beta-muL-two-loop}
\end{align}
while that for the linear superpotential parameter $\Lambda_F$ is
\begin{align}
  \beta_{\Lambda_F}^{(1)} &= \Lambda_F \Bigl ( 2 |\kappa_{\phi}|^2
    + 2 |\tilde{\sigma}|^2 + |\sigma|^2\Big),
    \label{eq:beta-LambdaF-one-loop} \\
  \beta_{\Lambda_F}^{(2)} &= \Lambda_F \Bigl \{ -8 |\kappa_{\phi}|^4
    - 4 |\tilde{\sigma}|^2 - 2 |\sigma|^4  + |\tilde{\sigma}|^2 \Bigl [
    \frac{6}{5} g_1^2 + \frac{4}{5} g_1'^2 + 6 g_2^2 - 8 |\kappa_{\phi}|^2
    \nonumber \\
  & \quad {} - 6 \trace{g^D g^{D \dagger}} - 2 \trace{h^E h^{E \dagger}}
    \Bigr ] + |\sigma|^2 \Bigl [  \frac{5}{2} g_1'^2 - 2 |\lambda|^2
    - 4 |\kappa_{\phi}|^2 \nonumber \\
  & \quad {} - 2 \trace{\tilde{\lambda} \tilde{\lambda}^{\dagger}}
    - 3 \trace{\kappa \kappa^{\dagger}} \Bigr ] \Bigr \} .
    \label{eq:beta-LambdaF-two-loop}
\end{align}

\subsection{Gaugino Masses}
The two-loop $\beta$ functions for the soft gaugino masses are
\begin{align}
  \beta_{M_1}^{(1)} &= \frac{96}{5} g_1^2 M_1 , \label{eq:beta-M1-one-loop} \\
  \beta_{M_1}^{(2)} &= g_1^2 \Bigl [ \frac{936}{25} g_1^2 M_1
    + \frac{162}{25} g_1'^2 ( M_1 + M_1')+ \frac{108}{5} g_2^2 ( M_1 + M_2 )
    + 48 g_3^2 ( M_1 + M_3 ) \nonumber \\
  & \quad {} - \frac{52}{5} \trace{M_1 y^U y^{U \dagger} - y^{U \dagger} T^U}
    - \frac{28}{5} \trace{M_1 y^D y^{D \dagger} - y^{D \dagger} T^D}
    \nonumber \\
  & \quad {} - \frac{36}{5} \trace{M_1 y^E y^{E \dagger} - y^{E \dagger} T^E}
    - \frac{12}{5} \lambda^* \Bigl ( M_1 \lambda - T_{\lambda} \Bigr )
    \nonumber \\
  & \quad {} - \frac{12}{5} \trace{M_1 \tilde{\lambda}
    \tilde{\lambda}^{\dagger}- \tilde{\lambda}^{\dagger} T^{\tilde{\lambda}}}
    - \frac{8}{5} \trace{M_1 \kappa \kappa^{\dagger}
    - \kappa^{\dagger} T^{\kappa}} - \frac{12}{5} \tilde{\sigma}^* \Bigl (
    M_1 \tilde{\sigma} - T_{\tilde{\sigma}} \Bigr ) \nonumber \\
  & \quad {} - \frac{12}{5} \trace{M_1 f f^{\dagger} - f^{\dagger} T^{f}}
    - \frac{12}{5} \trace{M_1 \tilde{f} \tilde{f}^{\dagger}
    - \tilde{f}^{\dagger} T^{\tilde{f}}} \nonumber \\
  & \quad {} - \frac{28}{5} \trace{M_1 g^D g^{D \dagger} -
    g^{D \dagger} T^{g^D}} - \frac{36}{5} \trace{M_1 h^E h^{E \dagger}
    - h^{E \dagger} T^{h^E}} \Bigr ] ,
    \label{eq:beta-M1-two-loop} \\
  \beta_{M_2}^{(1)} &= 8 g_2^2 M_2 , \label{eq:beta-M2-one-loop} \\
  \beta_{M_2}^{(2)} &= g_2^2 \Bigl [ \frac{36}{5} g_1^2 ( M_1 + M_2 ) +
    \frac{34}{5} g_1'^2 ( M_1' + M_2 ) + \frac{184}{5} g_2^2 M_2
    + 48 g_3^2 ( M_2 + M_3 ) \nonumber \\
  & \quad {} - 12 \trace{M_2 y^U y^{U \dagger} - y^{U \dagger} T^U}
    - 12 \trace{M_2 y^D y^{D \dagger} - y^{D \dagger} T^D} \nonumber \\
  & \quad {} - 4 \trace{M_2 y^E y^{E \dagger} - y^{E \dagger} T^E}
    - 4 \lambda^* \Bigl ( M_2 \lambda - T_{\lambda} \Bigr )
    - 4 \trace{M_2 \tilde{\lambda} \tilde{\lambda}^{\dagger}
    - \tilde{\lambda}^{\dagger} T^{\tilde{\lambda}}}\nonumber \\
  & \quad {} - 4 \tilde{\sigma}^* \Bigl ( M_2 \tilde{\sigma}
    - T_{\tilde{\sigma}} \Bigr )
    - 4 \trace{ M_2f f^{\dagger} - f^{\dagger} T^{f}}
    - 4 \trace{M_2 \tilde{f} \tilde{f}^{\dagger}
    - \tilde{f}^{\dagger} T^{\tilde{f}}} \nonumber \\
  & \quad {} - 12 \trace{M_2 g^D g^{D \dagger} - g^{D \dagger} T^{g^D}}
    - 4 \trace{M_2h^E h^{E \dagger} - h^{E \dagger} T^{h^E}} \Bigr ] ,
    \label{eq:beta-M2-two-loop} \\
  \beta_{M_3}^{(1)} &= 0 , \label{eq:beta-M3-one-loop} \\
  \beta_{M_3}^{(2)} &= g_3^2 \Bigl [ 6 g_1^2 ( M_1 + M_3 )
    + 6 g_1'^2 ( M_1'+ M_3 ) + 18 g_2^2 ( M_2 + M_3 ) + 192 g_3^2 M_3
    \nonumber \\
  & \quad {} - 8 \trace{M_3 y^U y^{U \dagger} - y^{U \dagger} T^U}
    - 8 \trace{M_3 y^D y^{D \dagger} - y^{D \dagger} T^D} \nonumber \\
  & \quad {} - 4 \trace{M_3 \kappa \kappa^{\dagger}
    - \kappa^{\dagger} T^{\kappa}}
    - 8 \trace{M_3  g^D g^{D \dagger} - g^{D \dagger} T^{g^D}} \Bigr ] ,
    \label{eq:beta-M3-two-loop} \\
  \beta_{M_1^\prime}^{(1)} &= \frac{213}{10} g_1'^2 M_1^\prime ,
  \label{eq:beta-M1p-one-loop} \\
  \beta_{M_1^\prime}^{(2)} &= g_1'^2 \Bigl [ \frac{162}{25} g_1^2 ( M_1 + M_1' )
    + \frac{2457}{50} g_1'^2 M_1' + \frac{102}{5} g_2^2 ( M_1' + M_2 )
    + 48 g_3^2 ( M_1' + M_3 ) \nonumber \\
  & \quad {} - \frac{18}{5} \trace{M_1' y^U y^{U \dagger} - y^{U \dagger} T^U}
    - \frac{42}{5} \trace{M_1' y^D y^{D \dagger} - y^{D \dagger} T^D}
    \nonumber \\
  & \quad {} - \frac{14}{5} \trace{M_1' y^E y^{E \dagger} - y^{E \dagger} T^E}
    - \frac{38}{5} \lambda^* \Bigl ( M_1' \lambda - T_{\lambda} \Bigr )
    \nonumber \\
  & \quad {} - \frac{38}{5} \trace{M_1' \tilde{\lambda}
    \tilde{\lambda}^{\dagger} - \tilde{\lambda}^{\dagger} T^{\tilde{\lambda}}}
    - \frac{57}{5} \trace{M_1' \kappa \kappa^{\dagger}
    - \kappa^{\dagger} T^{\kappa}}
    - \frac{8}{5} \tilde{\sigma}^* \Bigl ( M_1' \tilde{\sigma}
    - T_{\tilde{\sigma}} \Bigr )\nonumber \\
  & \quad {}
    - 5 \sigma^* \Bigl ( M_1' \sigma - T_{\sigma} \Bigr )
    - \frac{38}{5} \trace{M_1' f f^{\dagger} - f^{\dagger} T^{f}}
    - \frac{38}{5} \trace{M_1' \tilde{f} \tilde{f}^{\dagger}
    - \tilde{f}^{\dagger} T^{\tilde{f}}} \nonumber \\
  & \quad {} - \frac{42}{5} \trace{M_1' g^D g^{D \dagger}
    - g^{D \dagger} T^{g^D}}
    - \frac{14}{5} \trace{M_1' h^E h^{E \dagger} - h^{E \dagger} T^{h^E}}
    \Bigr ] . \label{eq:beta-M1p-two-loop}
\end{align}
As mentioned above, kinetic mixing in this class of $E_6$ inspired models is
small and so we neglect the mixed gaugino mass $M_{11}$.

\subsection{Soft-breaking Trilinear Scalar Couplings}
The two-loop RGEs for the soft scalar trilinear couplings read


\subsection{Soft-breaking Bilinear and Linear Couplings}
The $\beta$ functions for the soft-breaking bilinears are given by
\begin{align}
  \beta_{B_\phi \mu_\phi}^{(1)} &= 2 B_\phi \mu_\phi \Bigl (
    2 |\tilde{\sigma}|^2 + 4 |\kappa_{\phi}|^2 + |\sigma|^2 \Bigr )
    + 4 \mu_\phi \Bigl ( 2 \kappa_{\phi}^* T_{\kappa_\phi}
    + 2 \tilde{\sigma}^* T_{\tilde{\sigma}} + \sigma^* T_{\sigma} \Bigr )
    \nonumber \\
  & \quad {} - 8 \tilde{\sigma}^* \kappa_{\phi} B_L \mu_L ,
    \label{eq:beta-BmuPhi-one-loop} \\
  \beta_{B_\phi \mu_\phi}^{(2)} &= B_\phi \mu_\phi \Bigl \{
    -32 |\kappa_{\phi}|^4 - 8 |\tilde{\sigma}|^4 - 4 |\sigma|^4
    + |\tilde{\sigma}|^2 \Bigl [ \frac{12}{5} g_1^2
    + \frac{8}{5} g_1'^2 + 12 g_2^2 - 32 |\kappa_{\phi}|^2 \nonumber \\
  & \quad {} - 12 \trace{g^D g^{D \dagger}}
    - 4 \trace{h^E h^{E \dagger}} \Bigr ] + |\sigma|^2 \Bigl [ 5 g_1'^2
    - 4 |\lambda|^2 - 16 |\kappa_{\phi}|^2 \nonumber \\
  & \quad {} - 4 \trace{\tilde{\lambda} \tilde{\lambda}^{\dagger}}
    - 6 \trace{\kappa \kappa^{\dagger}} \Bigr ] \Bigr \}
    - \mu_{\phi} \Bigl \{ 16 |\sigma|^2 \sigma^* T_{\sigma}
    + 80 |\kappa_{\phi}|^2 \kappa_{\phi}^* T_{\kappa_\phi} \nonumber \\
  & \quad {} + 32 |\tilde{\sigma}|^2 \tilde{\sigma}^* T_{\tilde{\sigma}}
    + \frac{24}{5} g_1^2 \tilde{\sigma}^* ( \tilde{\sigma} M_1
    - T_{\tilde{\sigma}}) + \frac{16}{5} g_1'^2 \tilde{\sigma}^* (
    \tilde{\sigma} M_1' - T_{\tilde{\sigma}}) \nonumber \\
  & \quad {} + 24 g_2^2 \tilde{\sigma}^* ( \tilde{\sigma} M_2
    - T_{\tilde{\sigma}}) + 10 g_1'^2 \sigma^* (
    M_1' \sigma - T_{\sigma}) + 16 \tilde{\sigma}^* \kappa_\phi^* \Bigl (
    2 \tilde{\sigma} T_{\kappa_\phi} + 3 \kappa_\phi T_{\tilde{\sigma}}
    \Bigr ) \nonumber \\
  & \quad {} + 8 \sigma^* \kappa_{\phi}^* \Bigl ( 2 \sigma T_{\kappa_\phi}
    + 3 \kappa_{\phi} T_{\sigma} \Bigr ) + 8 \lambda^* \sigma^* \Bigl (
    \lambda T_{\sigma} + \sigma T_{\lambda} \Bigr ) \nonumber \\
  & \quad {} + 12 \sigma^* \Bigl [ T_{\sigma} \trace{\kappa \kappa^{\dagger}}
    + \sigma \trace{\kappa^{\dagger} T^{\kappa}} \Bigr ]
    + 8 \sigma^* \Bigl [ T_{\sigma} \trace{\tilde{\lambda}
    \tilde{\lambda}^{\dagger}}
    + \sigma \trace{\tilde{\lambda}^{\dagger} T^{\tilde{\lambda}}} \Bigr ]
    \nonumber \\
  & \quad {} + 24 \tilde{\sigma}^* \Bigl [
    T_{\tilde{\sigma}} \trace{g^D g^{D \dagger}}
    + \tilde{\sigma} \trace{g^{D \dagger} T^{g^D}} \Bigr ] \nonumber \\
  & \quad {} + 8 \tilde{\sigma}^* \Bigl [
    T_{\tilde{\sigma}} \trace{h^E h^{E \dagger}}
    + \tilde{\sigma} \trace{h^{E \dagger} T^{h^E}}\Bigr ] \Bigr \}
    + 8 \kappa_{\phi} \tilde{\sigma}^* B_L \mu_L \Bigl [ 2 |\tilde{\sigma}|^2
    \nonumber \\
  & \quad {} + 3 \trace{g^D g^{D \dagger}}
    + \trace{h^E h^{E \dagger}} - \frac{12}{5} g_1^2
    - \frac{8}{5} g_1'^2 - 12 g_2^2 \Bigr ]
    + 8 \kappa_{\phi} \tilde{\sigma}^* \mu_L \Bigl [
    2 \tilde{\sigma}^* T_{\tilde{\sigma}} \nonumber \\
  & \quad {} + 3 \trace{g^{D \dagger} T^{g^D}} + \trace{h^{E \dagger} T^{h^E}}
    + \frac{12}{5} g_1^2 M_1 + \frac{8}{5} g_1'^2 M_1' + 12 g_2^2 M_2
    \Bigr ] , \label{eq:beta-BmuPhi-two-loop} \\
  \beta_{B_L \mu_L}^{(1)} &=
    B_L \mu_L \Bigl [ 6 |\tilde{\sigma}|^2 + 3 \trace{g^D g^{D \dagger}}
    + \trace{h^E h^{E \dagger}} - \frac{3}{5} g_1^2 - \frac{2}{5} g_1'^2
    - 3 g_2^2 \Bigr ] \nonumber \\
  & \quad {} + \mu_L \Bigl [ 4 \tilde{\sigma}^* T_{\tilde{\sigma}}
    + 6 \trace{g^{D \dagger} T^{g^D}} + 2 \trace{h^{E \dagger} T^{h^E}}
    + \frac{6}{5} g_1^2 M_1 + \frac{4}{5} g_1'^2 M_1' \nonumber \\
  & \quad {} + 6 g_2^2 M_2 \Bigr ]
    - 2 \tilde{\sigma} \kappa_{\phi}^* B_\phi \mu_\phi ,
    \label{eq:beta-BmuL-one-loop} \\
  \beta_{B_L \mu_L}^{(2)} &= B_L \mu_L \Bigl \{ \frac{297}{50} g_1^4
    + \frac{217}{50} g_1'^4 + \frac{33}{2} g_2^4 + \frac{18}{25}g_1^2 g_1'^2
    + \frac{9}{5} g_1^2 g_2^2 + \frac{6}{5} g_1'^2 g_2^2 \nonumber \\
  & \quad {} - 14 |\tilde{\sigma}|^4 + \frac{2}{5} g_1^2 \Bigl [
      -\trace{g^D g^{D \dagger}} + 3 \trace{h^E h^{E \dagger}} \Bigr ]
    + \frac{3}{10} g_1'^2 \Bigl [ 3 \trace{g^D g^{D \dagger}} \nonumber \\
  & \quad {} + \trace{h^E h^{E \dagger}} \Bigr ]
    + 16 g_3^2 \trace{g^D g^{D \dagger}}
    + |\tilde{\sigma}|^2 \Bigl [ \frac{48}{5} g_1^2
    + \frac{32}{5} g_1'^2 + 48 g_2^2 \nonumber \\
  & \quad {} - 2 |\sigma|^2
    - 4 |\kappa_{\phi}|^2 - 15 \trace{g^D g^{D \dagger}}
    - 5 \trace{h^E h^{E \dagger}} \Bigr ]
    - \trace{\tilde{f} h^{E \dagger} h^E \tilde{f}^{\dagger}} \nonumber \\
  & \quad {} - 9 \trace{g^D g^{D \dagger} g^D g^{D \dagger}}
    - 3 \trace{g^D g^{D \dagger} y^{D T} y^{D *}}
    - 3 \trace{g^D g^{D \dagger} y^{U T} y^{U *}} \nonumber \\
  & \quad {} - 3 \trace{g^D \kappa^{\dagger} \kappa g^{D \dagger}}
    - 3 \trace{h^E h^{E \dagger} h^E h^{E \dagger}}
    - 2 \trace{h^E h^{E \dagger} y^E y^{E \dagger}} \nonumber \\
  & \quad {} - \trace{h^E \tilde{\lambda}^{\dagger} \tilde{\lambda}
    h^{E \dagger}} \Bigr \} - \mu_L \Bigl \{ \frac{594}{25} g_1^4 M_1
    + \frac{434}{25} g_1'^4 M_1' + 66 g_2^4 M_2 \nonumber \\
  & \quad {} + \frac{36}{25} g_1^2 g_1'^2 ( M_1 + M_1' )
    + \frac{18}{5} g_1^2 g_2^2 ( M_1 + M_2 )
    + \frac{12}{5} g_1'^2 g_2^2 ( M_1' + M_2 ) \nonumber \\
  & \quad {} - \frac{4}{5} g_1^2 \trace{M_1 g^D g^{D \dagger}
    - g^{D \dagger} T^{g^D}} + \frac{9}{5} g_1'^2 \trace{M_1' g^D g^{D \dagger}
    - g^{D \dagger} T^{g^D}} \nonumber \\
  & \quad {} + 32 g_3^2 \trace{M_3 g^D g^{D \dagger}
    - g^{D \dagger} T^{g^D}} + \frac{12}{5} g_1^2 \trace{M_1 h^E h^{E \dagger}
    - h^{E \dagger} T^{h^E}} \nonumber \\
  & \quad {} + \frac{3}{5} g_1'^2 \trace{M_1' h^E h^{E \dagger}
    - h^{E \dagger} T^{h^E}} + |\tilde{\sigma}|^2 \Bigl [ \frac{48}{5} g_1^2 M_1
    + \frac{32}{5} g_1'^2 M_1' \nonumber \\
  & \quad {} + 48 g_2^2 M_2  + 8 \kappa_{\phi}^* T_{\kappa_\phi}
    + 4 \sigma^* T_{\sigma} + 18 \trace{g^{D \dagger} T^{g^D}}
    + 6 \trace{h^{E \dagger} T^{h^E}} \Bigr ] \nonumber \\
  & \quad {} + \tilde{\sigma}^* T_{\tilde{\sigma}} \Bigl [ 8 |\kappa_{\phi}|^2
    + 4 |\sigma|^2 + 32 |\tilde{\sigma}|^2 + 6 \trace{g^D g^{D \dagger}}
    + 2 \trace{h^E h^{E \dagger}} \Bigr ] \nonumber \\
  & \quad {} + 2 \trace{\tilde{f} h^{E \dagger} T^{h^E}
    \tilde{f}^{\dagger}} + 36 \trace{g^D g^{D \dagger} T^{g^D}
    g^{D \dagger}} + 6 \trace{g^D \kappa^{\dagger} T^{\kappa}
    g^{D \dagger}} \nonumber \\
  & \quad {} + 2 \trace{h^E \tilde{f}^{\dagger} T^{\tilde{f}}
    h^{E \dagger}} + 12 \trace{h^E h^{E \dagger} T^{h^E} h^{E \dagger}}
    + 4 \trace{h^E h^{E \dagger} T^E y^{E \dagger}} \nonumber \\
  & \quad {} + 2 \trace{h^E \tilde{\lambda}^{\dagger} T^{\tilde{\lambda}}
    h^{E \dagger}} + 4 \trace{y^E y^{E \dagger} T^{h^E} h^{E \dagger}}
    + 6 \trace{\kappa g^{D \dagger} T^{g^D} \kappa^{\dagger}} \nonumber \\
  & \quad {} + 2 \trace{\tilde{\lambda} h^{E \dagger} T^{h^E}
    \tilde{\lambda}^{\dagger}} + 6 \trace{g^{D \dagger} y^{D T} y^{D *}
    T^{g^D}} + 6 \trace{g^{D \dagger} y^{U T} y^{U *} T^{g^D}}
    \nonumber \\
  & \quad {} + 6 \trace{y^{D \dagger} T^D g^{D *} g^{D T}}
    + 6 \trace{y^{U \dagger} T^U g^{D *} g^{D T}} \Bigr \} \nonumber \\
  & \quad {} + 4 \tilde{\sigma} \kappa_{\phi}^* \Bigl [ B_\phi \mu_\phi \Bigl (
    2 |\kappa_{\phi}|^2 + 2 |\tilde{\sigma}|^2 + |\sigma|^2 \Bigr )
    \nonumber \\
  & \quad {} + \mu_{\phi} \Bigl (
    2 \kappa_{\phi}^* T_{\kappa_\phi} + 2 \tilde{\sigma}^* T_{\tilde{\sigma}}
    + \sigma^* T_{\sigma} \Bigr ) \Bigr ] .
    \label{eq:beta-BmuL-two-loop}
\end{align}
The two-loop $\beta$ function for the soft-breaking linear coupling $\Lambda_S$
is
\begin{align}
  \beta_{\Lambda_S}^{(1)} &= \Lambda_S \Bigl ( 2 |\kappa_{\phi}|^2
    + 2 |\tilde{\sigma}|^2 + |\sigma|^2 \Bigr )
    + 2 \Lambda_F \Bigl ( 2 \kappa_{\phi}^* T_{\kappa_\phi}
    + 2 \tilde{\sigma}^* T_{\tilde{\sigma}} + \sigma^* T_{\sigma} \Bigr )
    \nonumber \\
  & \quad {} + 2 (B_\phi \mu_\phi) \mu_\phi  \kappa_{\phi}^*
    + 2 (B_\phi \mu_\phi)^* T_{\kappa_\phi}
    - 4 (B_L \mu_L) \mu_{\phi}  \tilde{\sigma}^*
    - 4 (B_L \mu_L)^* T_{\tilde{\sigma}} \nonumber \\
  & \quad {} - 4 \Bigl (m_{L_4}^2 + m_{\bar{L}_4}^2\Bigr ) \tilde{\sigma}
    \mu_L^* + 4 m_{\phi}^2 \kappa_{\phi} \mu_{\phi}^* ,
    \label{eq:beta-LambdaS-one-loop} \\
  \beta_{\Lambda_S}^{(2)} &= \Lambda_S \Bigl \{ |\sigma|^2 \Bigl [
    \frac{5}{2} g_1'^2 - 4 |\kappa_{\phi}|^2 - 2 |\lambda|^2
    - 2 |\sigma|^2 - 3 \trace{\kappa \kappa^{\dagger}}
    - 2 \trace{\tilde{\lambda} \tilde{\lambda}^{\dagger}} \Bigr ]
    \nonumber \\
  & \quad {} + |\tilde{\sigma}|^2 \Bigl [ \frac{6}{5} g_1^2
    + \frac{4}{5} g_1'^2 + 6 g_2^2 - 8 |\kappa_{\phi}|^2 - 4 |\tilde{\sigma}|^2
    - 6 \trace{g^D g^{D \dagger}} - 2 \trace{h^E h^{E \dagger}} \Bigr ]
    \nonumber \\
  & \quad {} - 8 |\kappa_{\phi}|^4 \Bigr \} + \Lambda_F \Bigl \{
    \sigma^* T_{\sigma} \Bigl [ 5 g_1'^2 - 8 |\kappa_{\phi}|^2
    - 4 |\lambda|^2  - 4 |\sigma|^2 - 6 \trace{\kappa \kappa^{\dagger}}
    \nonumber \\
  & \quad {} - 4 \trace{\tilde{\lambda} \tilde{\lambda}^{\dagger}} \Bigr ]
    - |\sigma|^2 \Bigl [5 g_1'^2 M_1'
    + 8 \kappa_{\phi}^* T_{\kappa_\phi} + 4 \lambda^* T_{\lambda}
    + 4 \sigma^* T_{\sigma} + 6 \trace{\kappa^{\dagger} T^{\kappa}} \nonumber \\
  & \quad {} + 4 \trace{\tilde{\lambda}^{\dagger}
    T^{\tilde{\lambda}}} \Bigr ] + \tilde{\sigma}^* T_{\tilde{\sigma}}
    \Bigl [ \frac{12}{5} g_1^2 + \frac{8}{5} g_1'^2
    + 12 g_2^2 - 16 |\kappa_{\phi}|^2 - 8 |\tilde{\sigma}|^2 \nonumber \\
  & \quad {} - 12 \trace{g^D g^{D \dagger}} - 4 \trace{h^E h^{E \dagger}}
    \Bigr ] - |\tilde{\sigma}|^2 \Bigl [ \frac{12}{5} g_1^2 M_1
    + \frac{8}{5} g_1'^2 M_1' + 12 g_2^2 M_2 \nonumber \\
  & \quad {} + 16 \kappa_{\phi}^* T_{\kappa_\phi}
    + 8 \tilde{\sigma}^* T_{\tilde{\sigma}}
    + 12 \trace{g^{D \dagger} T^{g^D}}
    + 4 \trace{h^{E \dagger} T^{h^E}} \Bigr ]
    - 32 |\kappa_{\phi}|^2 \kappa_{\phi}^* T_{\kappa_\phi} \Bigr \} \nonumber \\
  & \quad {} - 4 (B_\phi \mu_\phi)^* \Bigl [ T_{\kappa_\phi} \Bigl (
    4 |\kappa_{\phi}|^2 + 2 |\tilde{\sigma}|^2 + |\sigma|^2 \Bigr )
    + \kappa_{\phi} \sigma^* T_{\sigma}
    + 2 \kappa_{\phi} \tilde{\sigma}^* T_{\tilde{\sigma}} \Bigr ]
    \nonumber \\
  & \quad {} + (B_L \mu_L)^* \Bigl \{ \frac{12}{5} g_1^2 \Bigl (
    \tilde{\sigma} M_1 - T_{\tilde{\sigma}} \Bigr )
    + \frac{8}{5} g_1'^2 \Bigl ( \tilde{\sigma} M_1' - T_{\tilde{\sigma}}
    \Bigr ) + 12 g_2^2 \Bigl ( \tilde{\sigma} M_2 - T_{\tilde{\sigma}} \Bigr )
    \nonumber \\
  & \quad {} + 16 |\tilde{\sigma}|^2 T_{\tilde{\sigma}}
    + 12 \Bigl [ T_{\tilde{\sigma}} \trace{g^D g^{D \dagger}}
    + \tilde{\sigma} \trace{g^{D \dagger} T^{g^D}} \Bigr ]
    + 4 \Bigl [ T_{\tilde{\sigma}} \trace{h^E h^{E \dagger}} \nonumber \\
  & \quad {} + \tilde{\sigma} \trace{h^{E \dagger} T^{h^E}} \Bigr ] \Bigr \}
    - 4 \kappa_{\phi}^* \mu_{\phi} (B_\phi \mu_\phi) \Bigl ( 2 |\kappa_{\phi}|^2
    + 2 |\tilde{\sigma}|^2 + |\sigma|^2 \Bigr ) \nonumber \\
  & \quad {} + \tilde{\sigma}^* \mu_{\phi} B_L \mu_L \Bigl [
    8 |\tilde{\sigma}|^2 + 12 \trace{g^D g^{D \dagger}}
    + 4 \trace{h^E h^{E \dagger}} - \frac{12}{5} g_1^2 - \frac{8}{5} g_1'^2
    \nonumber \\
  & \quad {} - 12 g_2^2 \Bigr ] + \tilde{\sigma}^* \mu_L \mu_{\phi} \Bigl [
    \frac{12}{5} g_1^2 M_1 + \frac{8}{5} g_1'^2 M_1' + 12 g_2^2 M_2
    + 8 \tilde{\sigma}^* T_{\tilde{\sigma}} \nonumber \\
  & \quad {} + 12 \trace{g^{D \dagger} T^{g^D}}
    + 4 \trace{h^{E \dagger} T^{h^E}} \Bigr ]
    - 4 \kappa_{\phi}^* \mu_{\phi}^{2} \Bigl [2 \kappa_{\phi}^* T_{\kappa_\phi}
    + 2 \tilde{\sigma}^* T_{\tilde{\sigma}} + \sigma^* T_{\sigma} \Bigr ]
    \nonumber \\
  & \quad {} - 4 \mu_{\phi}^* \Bigl [ \kappa_{\phi} |\sigma|^2 \Bigl (
    3 m_{\phi}^2 + m_S^2 + m_{\bar{S}}^2 \Bigr )
    + 2 \kappa_{\phi} |\tilde{\sigma}|^2 \Bigl ( 3 m_{\phi}^2 + m_{L_4}^2
    + m_{\bar{L}_4}^2  \Bigr )  \nonumber \\
  & \quad {} + \kappa_\phi \Bigl ( 10 m_{\phi}^2 |\kappa_{\phi}|^2
    + 4 |T_{\kappa_\phi}|^2 + |T_{\sigma}|^2 + 2 |T_{\tilde{\sigma}}|^2 \Bigr )
    + \sigma T_{\sigma}^* T_{\kappa_\phi}
    + 2 \tilde{\sigma} T_{\tilde{\sigma}}^* T_{\kappa_\phi}\Bigr ] \nonumber \\
  & \quad {} + \mu_L^* \Bigl \{ \tilde{\sigma} \Bigl [ 16 |T_{\tilde{\sigma}}|^2
    + 16 m_{L_4}^2 |\tilde{\sigma}|^2 + 16 m_{\bar{L}_4}^2 |\tilde{\sigma}|^2
    + 8 m_{\phi}^2 |\tilde{\sigma}|^2 + 24 m_{L_4}^2 \trace{g^D g^{D \dagger}}
    \nonumber \\
  & \quad {} + 12 m_{\bar{L}_4}^2 \trace{g^D g^{D \dagger}}
    + 8 m_{L_4}^2 \trace{h^E h^{E \dagger}}
    + 4 m_{\bar{L}_4}^2 \trace{h^E h^{E \dagger}} \nonumber \\
  & \quad {} + 12 \trace{T^{{g^D} *} T^{g^D T}} + \trace{T^{{h^E} *} T^{h^E T}}
    + 12  \trace{g^D m_{\bar{D}}^2 g^{D \dagger}} \nonumber \\
  & \quad {} + 12 \trace{g^D g^{D \dagger} m_Q^2}
    + \trace{h^E h^{E \dagger} m_{e^{c}}^{2*}}
    + \trace{h^E m_{H_1}^{2*} h^{E \dagger}}
    \Bigr ] \nonumber \\
  & \quad {} + T_{\tilde{\sigma}} \Bigl [ 12 \trace{T^{{g^D} *} g^{D T}}
    + \trace{T^{{h^E} *} h^{E T}}\Bigr ] \nonumber \\
  & \quad {} -\frac{12}{5} g_1^2 \Bigl (
    \tilde{\sigma} m_{L_4}^2 + \tilde{\sigma} m_{\bar{L}_4}^2
    + 2 \tilde{\sigma} |M_1|^2 - M_1 T_{\tilde{\sigma}} \Bigr ) \nonumber \\
  & \quad {} - \frac{8}{5} g_1'^2 \Bigl (  \tilde{\sigma} m_{L_4}^2
    + \tilde{\sigma} m_{\bar{L}_4}^2 + 2 \tilde{\sigma} |M_1'|^2
    - M_1' T_{\tilde{\sigma}} \Bigr ) \nonumber \\
  & \quad {} - 12 g_2^2 \Bigl ( \tilde{\sigma} m_{L_4}^2
    + \tilde{\sigma} m_{\bar{L}_4}^2 + 2 \tilde{\sigma} |M_2|^2
    - M_2 T_{\tilde{\sigma}} \Bigr ) \Bigr \} .
    \label{eq:beta-LambdaS-two-loop}
\end{align}

\subsection{Soft Scalar Masses}
In writing down the two-loop $\beta$ functions for the soft scalar masses,
the following quantities are defined,
\begin{align}
  \Sigma_{1,1} &= \sqrt{\frac{3}{5}} g_1 \Bigl [ -m_{H_d}^2 - m_{L_4}^2
    + m_{\bar{L}_4}^2 + m_{H_u}^2 + \trace{m_{d^c}^2} - \trace{m_{D}^2}
    \nonumber \\
  & \quad {} + \trace{m_{\bar{D}}^2} + \trace{m_{e^c}^2} - \trace{m_{H_1}^2}
    + \trace{m_{H_2}^2} - \trace{m_L^2} \nonumber \\
  & \quad {} + \trace{m_Q^2} - 2 \trace{m_{u^c}^2} \Bigr ] ,
    \label{eq:soft-traces-1} \\
  \Sigma_{1,4} & = \frac{1}{\sqrt{40}} g_1^\prime \Bigl [
    -6 m_{H_d}^2 + 4 m_{L_4}^2 - 4 m_{\bar{L}_4}^2 - 4 m_{H_u}^2 + 5 m_S^2
    - 5 m_{\bar{S}}^2 \nonumber \\
  & \quad {} + 6 \trace{m_{d^c}^2} - 6 \trace{m_{D}^2}
    - 9 \trace{m_{\bar{D}}^2} + \trace{m_{e^c}^2}
    - 6 \trace{m_{H_1}^2} \nonumber \\
  & \quad {} - 4 \trace{m_{H_2}^2} + 4 \trace{m_L^2} + 6 \trace{m_Q^2}
    + 5 \trace{m_{\Sigma}^2} +3 \trace{m_{u^c}^2} \Bigr ] ,
    \label{eq:soft-traces-2} \\
  \Sigma_{2,11} &= \frac{1}{10} g_1^2 \Bigl [ 3 m_{H_d}^2 + 3 m_{L_4}^2
    + 3 m_{\bar{L}_4}^2 + 3 m_{H_u}^2 + 2 \trace{m_{d^c}^2}
    + 2 \trace{m_{D}^2} \nonumber \\
  & \quad {} + 2 \trace{m_{\bar{D}}^2} + 6 \trace{m_{e^c}^2}
    + 3 \trace{m_{H_1}^2} + 3 \trace{m_{H_2}^2} + 3 \trace{m_L^2} \nonumber \\
  & \quad {} + \trace{m_Q^2} + 8 \trace{m_{u^c}^2} \Bigr ] ,
    \label{eq:soft-traces-3} \\
  \Sigma_{2,14} &= \frac{1}{10} \sqrt{\frac{3}{2}} g_1 g_1^\prime \Bigl [
    3 m_{H_d}^2 - 2 m_{L_4}^2 - 2 m_{\bar{L}_4}^2 - 2 m_{H_u}^2
    + 2 \trace{m_{d^c}^2} + 2 \trace{m_{D}^2} \nonumber \\
  & \quad {} - 3 \trace{m_{\bar{D}}^2} + \trace{m_{e^c}^2}
    + 3 \trace{m_{H_1}^2} - 2 \trace{m_{H_2}^2} - 2 \trace{m_L^2} \nonumber \\
  & \quad {} + \trace{m_Q^2} - 2 \trace{m_{u^c}^2} \Bigr ] ,
    \label{eq:soft-traces-4} \\
  \Sigma_{3,1} &= \frac{1}{40 \sqrt{15}} g_1 \Bigl [ -18 g_1^2 m_{H_d}^2
    - 27 g_1'^2 m_{H_d}^2 - 90 g_2^2 m_{H_d}^2 - 18 g_1^2 m_{L_4}^2
    - 12 g_1'^2 m_{L_4}^2 \nonumber \\
  & \quad {} - 90 g_2^2 m_{L_4}^2 + 18 g_1^2 m_{\bar{L}_4}^2
    + 12 g_1'^2 m_{\bar{L}_4}^2 + 90 g_2^2 m_{\bar{L}_4}^2
    + 18 g_1^2 m_{H_u}^2 + 12 g_1'^2 m_{H_u}^2 \nonumber \\
  & \quad {} + 90 g_2^2 m_{H_u}^2 + 60 \Bigl ( -m_{H_u}^2 + m_{H_d}^2 \Bigr )
    |\lambda|^2 + 60 \Bigl ( - m_{\bar{L}_4}^2 + m_{L_4}^2 \Bigr )
    |\tilde{\sigma}|^2 \nonumber \\
  & \quad {} + 8 g_1^2 \trace{m_{d^c}^2} + 12 g_1'^2 \trace{m_{d^c}^2}
    + 160 g_3^2 \trace{m_{d^c}^2} - 8 g_1^2 \trace{m_{D}^2} \nonumber \\
  & \quad {} - 12 g_1'^2 \trace{m_{D}^2} - 160 g_3^2 \trace{m_{D}^2}
    + 8 g_1^2 \trace{m_{\bar{D}}^2} + 27 g_1'^2 \trace{m_{\bar{D}}^2}
    \nonumber \\
  & \quad {} + 160 g_3^2 \trace{m_{\bar{D}}^2} + 72 g_1^2 \trace{m_{e^c}^2}
    + 3 g_1'^2 \trace{m_{e^c}^2} - 18 g_1^2 \trace{m_{H_1}^2} \nonumber \\
  & \quad {} - 27 g_1'^2 \trace{m_{H_1}^2} -90 g_2^2 \trace{m_{H_1}^2}
    + 18 g_1^2 \trace{m_{H_2}^2} + 12 g_1'^2 \trace{m_{H_2}^2} \nonumber \\
  & \quad {} + 90 g_2^2 \trace{m_{H_2}^2} - 18 g_1^2 \trace{m_L^2}
    - 12 g_1'^2 \trace{m_L^2} -90 g_2^2 \trace{m_L^2} \nonumber \\
  & \quad {} + 2 g_1^2 \trace{m_Q^2} + 3 g_1'^2 \trace{m_Q^2}
    + 90 g_2^2 \trace{m_Q^2} + 160 g_3^2 \trace{m_Q^2} \nonumber \\
  & \quad {} - 64 g_1^2 \trace{m_{u^c}^2} - 6 g_1'^2 \trace{m_{u^c}^2}
    - 320 g_3^2 \trace{m_{u^c}^2} + 60 m_{H_d}^2 \trace{f f^{\dagger}}
    \nonumber \\
  & \quad {} - 60 m_{H_u}^2 \trace{\tilde{f} \tilde{f}^{\dagger}}
    + 180 m_{L_4}^2 \trace{g^D   g^{D \dagger}}
    + 60 m_{L_4}^2 \trace{h^E h^{E \dagger}} \nonumber \\
  & \quad {} + 180 m_{H_d}^2 \trace{y^D y^{D \dagger}}
    + 60 m_{H_d}^2 \trace{y^E y^{E \dagger}}
    - 180 m_{H_u}^2 \trace{y^U y^{U \dagger}} \nonumber \\
  & \quad {} - 60 \trace{f m_{H_2}^{2*} f^{\dagger}}
    + 60 \trace{\tilde{f} m_{H_1}^{2*} \tilde{f}^{\dagger}}
    - 120 \trace{g^D m_{\bar{D}}^2 g^{D \dagger}} \nonumber \\
  & \quad {} - 60 \trace{g^D g^{D \dagger} m_Q^2}
    - 120 \trace{h^E h^{E \dagger} m_{e^{c}}^{2*}}
    + 60 \trace{h^E m_{H_1}^{2*} h^{E \dagger}} \nonumber \\
  & \quad {} + 60 \trace{m_{D}^2 \kappa \kappa^{\dagger}}
    - 60 \trace{m_{\bar{D}}^2 \kappa^{\dagger} \kappa}
    - 60 \trace{m_{H_2}^2 \tilde{\lambda} \tilde{\lambda}^{\dagger}}
    \nonumber \\
  & \quad {} - 120 \trace{y^D y^{D \dagger} m_{d^{c}}^{2*}}
    - 60 \trace{y^D m_Q^{2 *} y^{D \dagger}}
    - 120 \trace{y^E y^{E \dagger} m_{e^{c}}^{2*}} \nonumber \\
  & \quad {} + 60 \trace{y^E m_L^{2*} y^{E \dagger}}
    + 240 \trace{y^U y^{U \dagger} m_{u^{c}}^{2*}}
    - 60 \trace{y^U m_Q^{2 *} y^{U \dagger}} \nonumber \\
  & \quad {} + 60 \trace{\tilde{\lambda} m_{H_1}^{2*} \tilde{\lambda}^{\dagger}}
    \Bigr ] , \label{eq:soft-traces-5} \\
  \Sigma_{2,2} &= \frac{1}{2} \Bigl [ 3 \trace{m_Q^2} + m_{H_d}^2
    + m_{L_4}^2 + m_{\bar{L}_4}^2 + m_{H_u}^2 + \trace{m_{H_1}^2}
    + \trace{m_{H_2}^2} \nonumber \\
  & \quad {} + \trace{m_L^2} \Bigr ] , \label{eq:soft-traces-6} \\
  \Sigma_{2,3} &= \frac{1}{2} \Bigl [ 2 \trace{m_Q^2} + \trace{m_{d^c}^2}
    + \trace{m_{D}^2} + \trace{m_{\bar{D}}^2} + \trace{m_{u^c}^2} \Bigr ]
    , \label{eq:soft-traces-7} \\
  \Sigma_{2,41} &= \frac{1}{10} \sqrt{\frac{3}{2}} g_1 g_1^\prime \Bigl [
    3 m_{H_d}^2 - 2 m_{L_4}^2 - 2 m_{\bar{L}_4}^2 - 2 m_{H_u}^2
    + 2 \trace{m_{d^c}^2} + 2  \trace{m_{D}^2} \nonumber \\
  & \quad {} - 3 \trace{m_{\bar{D}}^2} + \trace{m_{e^c}^2}
    + 3 \trace{m_{H_1}^2} - 2 \trace{m_{H_2}^2} - 2 \trace{m_L^2} \nonumber \\
  & \quad {} + \trace{m_Q^2} - 2 \trace{m_{u^c}^2} \Bigr ] ,
    \label{eq:soft-traces-8} \\
  \Sigma_{2,44} &= \frac{1}{40} g_1'^2 \Bigl [ 18 m_{H_d}^2 + 8 m_{L_4}^2
    + 8 m_{\bar{L}_4}^2 + 8 m_{H_u}^2 + 25 m_S^2 + 25 m_{\bar{S}}^2
    + 12 \trace{m_{d^c}^2} \nonumber \\
  & \quad {} + 12 \trace{m_{D}^2} + 27 \trace{m_{\bar{D}}^2}
    + \trace{m_{e^c}^2} + 18 \trace{m_{H_1}^2} + 8 \trace{m_{H_2}^2}
    \nonumber \\
  & \quad {} + 8 \trace{m_L^2} + 6 \trace{m_Q^2} +25 \trace{m_{\Sigma}^2}
    + 3 \trace{m_{u^c}^2} \Bigr ] , \label{eq:soft-traces-9} \\
  \Sigma_{3,4} &= -\frac{1}{80\sqrt{10}} g_1^\prime \Bigl [
    36 g_1^2 m_{H_d}^2 + 54 g_1'^2 m_{H_d}^2 + 180 g_2^2 m_{H_d}^2
    - 24 g_1^2 m_{L_4}^2 - 16 g_1'^2 m_{L_4}^2 \nonumber \\
  & \quad {} - 120 g_2^2 m_{L_4}^2 + 24 g_1^2 m_{\bar{L}_4}^2
    + 16 g_1'^2 m_{\bar{L}_4}^2 + 120 g_2^2 m_{\bar{L}_4}^2
    + 24 g_1^2 m_{H_u}^2 + 16 g_1'^2 m_{H_u}^2 \nonumber \\
  & \quad {} + 120 g_2^2 m_{H_u}^2 - 125 g_1'^2 m_S^2
    + 125 g_1'^2 m_{\bar{S}}^2 - 40 \Bigl ( 2 m_{H_u}^2 + 3 m_{H_d}^2
    - 5 m_S^2 \Bigr ) |\lambda|^2 \nonumber \\
  & \quad {} + 100 \Bigl ( -m_{\bar{S}}^2 + m_S^2 \Bigr ) |\sigma|^2
    + 80 m_{L_4}^2 |\tilde{\sigma}|^2 - 80 m_{\bar{L}_4}^2 |\tilde{\sigma}|^2
    - 16 g_1^2 \trace{m_{d^c}^2} \nonumber \\
  & \quad {} - 24 g_1'^2 \trace{m_{d^c}^2} - 320 g_3^2 \trace{m_{d^c}^2}
    + 16 g_1^2 \trace{m_{D}^2} + 24 g_1'^2 \trace{m_{D}^2} \nonumber \\
  & \quad {} + 320 g_3^2 \trace{m_{D}^2} +24 g_1^2 \trace{m_{\bar{D}}^2}
    + 81 g_1'^2 \trace{m_{\bar{D}}^2} + 480 g_3^2 \trace{m_{\bar{D}}^2}
    \nonumber \\
  & \quad {} - 24 g_1^2 \trace{m_{e^c}^2} - g_1'^2 \trace{m_{e^c}^2}
    + 36 g_1^2 \trace{m_{H_1}^2} + 54 g_1'^2 \trace{m_{H_1}^2} \nonumber \\
  & \quad {} + 180 g_2^2 \trace{m_{H_1}^2} + 24 g_1^2 \trace{m_{H_2}^2}
    + 16 g_1'^2 \trace{m_{H_2}^2} + 120 g_2^2 \trace{m_{H_2}^2} \nonumber \\
  & \quad {} - 24 g_1^2 \trace{m_L^2} - 16 g_1'^2 \trace{m_L^2}
    - 120 g_2^2 \trace{m_L^2} - 4 g_1^2 \trace{m_Q^2} \nonumber \\
  & \quad {} - 6 g_1'^2 \trace{m_Q^2} - 180 g_2^2 \trace{m_Q^2}
    - 320 g_3^2 \trace{m_Q^2} - 125 g_1'^2 \trace{m_{\Sigma}^2}   \nonumber \\
  & \quad {} - 32 g_1^2 \trace{m_{u^c}^2} - 3 g_1'^2 \trace{m_{u^c}^2}
    - 160 g_3^2 \trace{m_{u^c}^2} - 120 m_{H_d}^2 \trace{f f^{\dagger}}
    \nonumber \\
  & \quad {} - 80 m_{H_u}^2 \trace{\tilde{f} \tilde{f}^{\dagger}}
    + 240 m_{L_4}^2 \trace{g^D g^{D \dagger}}
    + 80 m_{L_4}^2 \trace{h^E h^{E \dagger}} \nonumber \\
  & \quad {} - 360 m_{H_d}^2 \trace{y^D y^{D \dagger}}
    - 120 m_{H_d}^2 \trace{y^E y^{E \dagger}}
    - 240 m_{H_u}^2 \trace{y^U y^{U \dagger}} \nonumber \\
  & \quad {} + 300 m_S^2 \trace{\kappa \kappa^{\dagger}}
    + 200 m_S^2 \trace{\tilde{\lambda} \tilde{\lambda}^{\dagger}}
    + 200 \trace{f f^{\dagger} m_{\Sigma}^2} \nonumber \\
  & \quad {} - 80 \trace{f m_{H_2}^{2*} f^{\dagger}}
    + 200 \trace{\tilde{f} \tilde{f}^{\dagger} m_{\Sigma}^2}
    - 120 \trace{\tilde{f} m_{H_1}^{2*} \tilde{f}^{\dagger}} \nonumber \\
  & \quad {} - 360 \trace{g^D m_{\bar{D}}^2 g^{D   \dagger}}
    + 120 \trace{g^D g^{D \dagger} m_Q^2}
    + 40 \trace{h^E h^{E \dagger} m_{e^{c}}^{2*}} \nonumber \\
  & \quad {} - 120 \trace{h^E m_{H_1}^{2 *} h^{E \dagger}}
    - 120 \trace{m_{D}^2 \kappa \kappa^{\dagger}}
    - 180 \trace{m_{\bar{D}}^2 \kappa^{\dagger} \kappa} \nonumber \\
  & \quad {} - 80 \trace{m_{H_2}^2 \tilde{\lambda} \tilde{\lambda}^{\dagger}}
    + 240 \trace{y^D y^{D \dagger} m_{d^{c}}^{2*}}
    + 120 \trace{y^D m_Q^{2 *} y^{D \dagger}} \nonumber \\
  & \quad {} + 40 \trace{y^E y^{E \dagger} m_{e^{c}}^{2*}}
    + 80 \trace{y^E m_L^{2 *} y^{E \dagger}}
    + 120 \trace{y^U y^{U \dagger}   m_{u^{c}}^{2*}} \nonumber \\
  & \quad {} + 120 \trace{y^U m_Q^{2 *} y^{U \dagger}}
    - 120 \trace{\tilde{\lambda} m_{H_1}^{2 *} \tilde{\lambda}^{\dagger}}
    \Bigr ] , \label{eq:soft-traces-10}
\end{align}
The RGEs are then given by


\bibliography{bibliography}

\end{document}